\documentclass{jfm}

\usepackage{graphicx,psfrag,color}
\usepackage{amsmath}
\usepackage{amsfonts}
\usepackage{mathrsfs}
\usepackage{amssymb}
\usepackage{bm}
\usepackage{url}
\usepackage{color}
\usepackage{natbib}
\usepackage{mdwlist}

\newcommand{\btheta}{\mbox{\boldmath$\hat{\theta}$}}
\newcommand{\bnab}{\mbox{\boldmath$\nabla$}}
\newcommand{\eps}{\epsilon}

\newcommand{\del}{\delta}
\def\half {{\textstyle{1 \over 2}}}
\def\quart {{\textstyle{1 \over 4}}}
\def\dele{{{\delta \over \epsilon^2}}}
\newcommand{\br}{\mathbf{\hat{r}}}
\newcommand{\bhx}{\mathbf{\hat{x}}}
\newcommand{\bhy}{\mathbf{\hat{y}}}
\newcommand{\bhz}{\mathbf{\hat{z}}}
\newcommand{\bu}{\mathbf{u}}
\newcommand{\bU}{\mathbf{U}}
\newcommand{\opL}{{\cal L}}
\newcommand{\dy}{{\bar{y}}}
\newcommand{\dr}{{\bar{r}}}
\newcommand{\ddr}{r^*}
\newcommand{\beq}{\begin{equation}}
\newcommand{\eeq}{\end{equation}}

\renewcommand{\tabcolsep}{9pt}

\begin{document}

\title[Instability Driven by boundary inflow]{Instability driven by boundary inflow across shear: a way to circumvent Rayleigh's stability  criterion in accretion disks?}

\author{R. R. Kerswell}
\affiliation{School of Mathematics, Bristol University, Bristol, BS8 1TW, UK}
\maketitle

\begin{abstract}

We investigate the 2D instability recently discussed by Gallet et al. (2010) and Ilin \& Morgulis (2013) which arises when  a radial crossflow is imposed on a centrifugally-stable swirling flow.
By finding a simpler rectilinear example of the instability - a sheared half plane, the minimal ingredients for the instability are identified and the destabilizing/stabilizing effect of inflow/outflow boundaries clarified. The instability - christened `boundary inflow instability' here - is of critical layer type where this layer is either at the inflow wall and the growth rate is $O(\sqrt{\eta})$ (as found by \cite{ilin13}), or in the interior of the flow and the growth rate is $O(\eta \log 1/\eta)$ where $\eta$ measures the (small) inflow-to-tangential-flow ratio. The instability is robust to changes in the rotation profile even to those which are very Rayleigh-stable and the addition of further physics such as viscosity, 3-dimensionality and compressibility but is sensitive to the boundary condition imposed on the tangential velocity field at the inflow boundary. 
Providing the vorticity is not fixed at the inflow boundary, the instability seems generic and operates by the inflow advecting vorticity present at the boundary across the interior shear.
Both the primary bifurcation to 2D states and secondary bifurcations to 3D states are found to be supercritical. Assuming an accretion flow driven by molecular viscosity only so $\eta=O(Re^{-1})$, the instability is not immediately relevant for accretion disks since the critical threshold is $O(Re^{-2/3})$ and the inflow boundary conditions are more likely to be stress-free than non-slip. However, the analysis presented here does highlight the potential for mass entering a disk to disrupt the orbiting flow if this mass flux possesses vorticity.

\end{abstract}

%
%
\section{Introduction}

Rotating  flows are ubiquitous in nature and industrial applications so understanding their stability continues to be an important and active area of research. The extent of this  activity is perhaps best epitomised by the huge body of work studying Taylor-Couette flow \citep{Couette1888, Mallock1888, Taylor1923} - the flow between two concentric cylinders rotating at different rates - which has become the laboratory paradigm of the subject (e.g. \cite{Tuckerman14,Fardin2014} and references herein). Despite all this work, however, only very recently has it been realised that imposed radial flow can destabilise an otherwise stable rotating flow \citep{gal10,ilin13}. The paper by \cite{ilin13} is particularly revealing because it demonstrates that the (non-dimensionalised) flow
\[
\bU(r, \theta)= -\frac{\eta}{r}\br+\frac{1}{r}\btheta  
\]
between two concentric cylinders at  $r=1$ and $r=a\, >\, 1$ is {\em inviscidly} unstable to infinitesimal 2D oscillatory disturbances  for either inflow $\eta >0$ {\em or} outflow $\eta <0$ providing $\eta$ is not too large.  This is surprising because,  firstly, in the absence of  radial flow, this rotating flow  is (marginally) `Rayleigh-stable' as the angular momentum of the flow nowhere decreases in magnitude radially
(note this is strictly a condition on axisymmetric disturbances: \cite{Rayleigh1917}; \cite{DrazinReid} \S 15.2). Secondly  the flow also fails a requirement for 2D instability - the rotating flow version of Rayleigh's inflexion point theorem (\cite{Rayleigh1880}: \cite{DrazinReid} \S 15.3 and problem 3.2 p 121). Thirdly, it is also somewhat counterintuitive that the instability occurs for both converging and diverging radial flows since the former are stable and the latter generally unstable when the rotation is absent as in Jeffery-Hamel flow (e.g. \cite{Drazin1999}).

 The results of \cite{ilin13} also possess many intriguing features of which four stand out.  Firstly, the existence of an imposed normal flow through the cylinder walls increases the order of the linear operator describing the evolution of small inviscid disturbances from the normal 2 to 3. This means that an extra boundary condition has to be imposed beyond the usual no-normal-flow conditions at either cylinder wall. While it is straightforward to argue that this extra condition must be imposed at the inflow boundary \citep{ilin13}, predicting the effect of a specific choice is less clear: if a no-slip condition is imposed there is instability whereas a vanishing vorticity condition gives stability. Secondly, in the limit of vanishing radial flow ($\eta \rightarrow 0$), \cite{ilin13} find growth rates which scale as $O(\sqrt{\eta})$ rather than the generic $O(\eta)$ one might expect.  This unusual growth rate scaling arises because there are {\em no} discrete modes of the linear problem which can satisfy the usual 2 no-normal velocity conditions for $\eta=0$. This gives rise to a non-standard singular perturbation analysis, the robustness of which is unclear to, say, changes in the rotation profile and/or to the addition of extra physics. Thirdly,  the $\eta \rightarrow 0$ asymptotics presented seems incomplete. The instability described in \cite{ilin13} has a critical layer character but only where this critical layer is actually at the inflow boundary. This suggests the existence of further instabilities with an interior critical layer separated from the boundary. Lastly, the mechanism for the instability is unclear. Crossflow and shear would seem obvious ingredients but rotation or curvature not necessarily so. It is also not apparent  whether the energy to feed the instability comes wholly 'through the boundary' or is extracted at least in part  from  the (interior) rotational energy of the underlying flow.

To keep this study manageable, the focus here is on the $0< \eta \ll 1 $ situation which represents small radial inflow on a predominantly rotating flow. The motivation for this (as in \cite{gal10}) is the accretion disk problem where certainly for cold and hence weakly-ionised disks, the source of inferred disk turbulence remains a hotly contested issue (e.g. \cite{Dubrulle05,Shariff09,bal11, ji13}). As a result there is considerable interest in uncovering robust linear instability mechanisms. Interestingly, the existence of the radial accretion flow is rarely included in theoretical models since it is so small -  $O(1/Re)$ smaller than the azimuthal flow where $Re$ is huge when based on a molecular viscosity (e.g.  \cite{Dubrulle92} quote figures of $10^{14}-10^{26}$) - and presumably its presence only felt over  $O(Re)$  timescales which are far too large to be relevant (e.g \cite{Shariff09} estimates that it would take longer than the age of the universe for molecular viscosity to diffuse momentum  across a typical disk). However, the results of \cite{ilin13} suggests that such a flow could actually drive linear instabilities over a {\em much} shorter $O(\sqrt{Re})$ timescale.

The plan of the paper is to start with perhaps the  simplest example of the instability  which is just a sheared half plane of fluid with imposed inflow. The effect of adding viscosity is discussed as well as the introduction of an outflow (suction) boundary so that the flow domain becomes a channel. Then the discussion  turns to rotating flow with more general profiles, including the  solidly Rayleigh-stable Keplerian profile $\bU=1/\sqrt{r}\btheta$, to examine the robustness of the instability. The effect of further physics in the form of viscosity, 3-dimensionality and compressibility are also broached. Finally, the nonlinear aspects of the instability are probed ranging from a weakly nonlinear analysis around the primary 2D bifurcation through to secondary bifurcations and the ensuing fully 3D finite amplitude solutions.

The findings of the paper, organised  under the various questions posed above, are as follows.

\begin{enumerate}
%
%
\item {\em Is curvature or rotation important for the instability?}. 
No, all the features of the instability are reproduced in a rectilinear flow with inflow described in \S \ref{hp_inviscid}. The instability operates by advecting a source of vorticity at the inflow boundary across shear.
%
%

\item {\em Are there other (non-boundary-layer) modes of instability caused by an inflow boundary?} 
Yes, instabilities exist with interior critical layers distinct  from the inflow boundary: see \S \ref{hp_crit}, \S \ref{swirl_dzdr} and expressions (\ref{crit_growth}), (\ref{hp_vis_crit}) and  (\ref{crit_layer_asym}). Their growth rates are $O(\eta \log(1/\eta)\,)$, while much larger than $O(\eta)$, are smaller than the inviscid boundary-layer modes found by \cite{ilin13} and the equivalent viscous boundary layer modes found by \cite{gal10}.

%
%
\item {\em Given the sensitivity to what is chosen for the extra boundary condition, how generic is the  instability across possible boundary conditions?}
For finite $\eta$ the instability looks generic with only  a no-vorticity boundary condition obviously ensuring stability: see \S \ref{hp_inviscid}.  However for $0< \eta \ll 1$, any restriction on the normal derivative of the tangential velocity effectively kills the instability: see \S \ref{bc_discussion}. The non-slip condition always seems to allow instability to occur \citep{gal10,ilin13}.

%
%
\item {\em How robust is the instability to different rotation profiles?} 
Very robust. The form of the shear is unimportant for the boundary-layer instability  and of secondary importance for the critical-layer instability: see \S \ref{swirl_dzdr}.

%
%
\item {\em What is the effect of adding viscosity?} 
The presence of viscosity introduces a threshold crossflow of $O(Re^{-1})$ in the rectilinear situation where long streamwise wavelengths are permissible (\cite{Nicoud97} and \S \ref{hp_viscous})  or  $O(Re^{-2/3})$ for the rotational situation where the azimuthal wavenumber is an integer (\cite{gal10} and \S \ref{swirl_viscous_bl}).

%
%
\item {\em What is the effect of adding 3 dimensionality?}
Squire's Theorem effectively holds for the boundary layer instabilities since only the shear at the boundary is important and curvature is secondary. As a result, adding 3 dimensionality leads to less unstable disturbances: see \S \ref{section_3D}.

%
%
\item {\em How robust is the instability to adding further physics?}
The instability survives the addition of compressibility (see \S \ref{compress}), 3-dimensionality (see \S \ref{section_3D}) and viscosity (see \S \ref{hp_viscous} and \S \ref{swirl_viscous_bl}). It is also insensitive to the exact shear present as long as it is non-vanishing (see \S \ref{swirl_dzdr}).

%
%
\item {\em Has this instability been seen before \cite{gal10} and \cite{ilin13}?}
Yes, in a rectilinear form by \cite{Nicoud97} who studied `generalised plane Couette flow' where a streamwise pressure gradient is imposed to counterbalance the effects of crossflow in the streamwise momentum balance (see \S \ref{hp}). \cite{Doering00} also saw the instability without an imposed pressure gradient. In this case the introduction of crossflow not only adds a new flow component to the base state but also changes its cross-stream shear. From a stability perspective, how these two effects interact can be subtle and `suction' (as it is typically called) can be found to stabilise or destabilise existing shear instabilities (e.g. \cite{Hains71, Fransson03, Guha10, Deguchi14}). The situation is similar in the Taylor-Couette problem where radial flow can either stabilise or destabilise the well known Taylor vortex instability \citep{Chang67, Bahl70, Min94, Kol97, Kolesov99, Serre08, Martinand09}. The importance of the work of \cite{gal10} and \cite{ilin13} is that they studied the effect of radial flow on centrifugally-stable Taylor-Couette flow. 

%
%
\item {\em Is the instability supercritical or subcritical?}
It is a supercritical instability for bifurcations where the crossflow is fixed and non-slip boundary conditions are applied at the boundaries in the presence of viscosity: see \S \ref{wnl_anal} and Appendix B.

%
%
\item {\em Can secondary bifurcations off the 2D solutions reach crossflow values  which are below the threshold for (primary) instability?}
There is no evidence for this. The six 3D bifurcations found are all supercritical leading to even higher crossflow values.

%
%
\item {\em What is the energy source for the inviscid instability?}  
The instability draws its energy from the underlying shear. The energy of this is replenished by the pressure field, which drives the crossflow, doing work.
%

%
%
\item {\em Is this instability possibly relevant for accretion disks?}  
In a quiescent disk, the (molecular) viscosity-driven accretion flow is $O(Re^{-1})$ whereas the critical threshold for linear instability is a radial flow of $O(Re^{-2/3})$ leaving aside any issues about the exact form of the outermost boundary conditions.  This, together with the fact that no signs of subcriticality have been found in either the primary instability or of any secondary bifurcations, indicates that the instability is not operative in isolation . However, if some other process is able to generate a larger radial flow, then this instability may be triggered as a secondary consequence. The instability primarily derives its energy from the gravitational body force working on the accreting flow. \\

\end{enumerate}

Ilin and Morgulis have themselves continued their investigation to consider the effect of viscosity \citep{ilin15} and 3-dimensionality \citep{ilin15b} albeit from  a complementary perspective: given a crossflow what is the {\em smallest} critical swirling flow for instability? This is equivalent upon rescaling to the problem of fixing the swirling flow and finding when instability disappears as the crossflow is {\em increased} rather than decreased as studied here. That there is a finite range of crossflow for instability when there is both an inflow and outflow boundary appears a generic observation. There is also a tempting general interpretation -  the inflow boundary is responsible for  initially destabilising the flow as the crossflow is increased in magnitude from zero but ultimately the outflow (or more commonly labelled the 'suction') boundary stabilises the flow again when the crossflow becomes large enough. 
The particularly simple half-plane problem treated below helps motivate this simple characterisation. We henceforth refer to the instability first seen by \cite{Nicoud97}, \cite{Doering00}, \cite{gal10} and \cite{ilin13} as  the `boundary inflow instability'.

%
%

\section{2D Instability in Simplest Form: The Half Plane \label{hp} }

The boundary inflow instability operates by advecting vorticity across shear, occurs in 2D and does not need viscosity or underlying vorticity as shown in \cite{ilin13}. Here, we demonstrate that curvature or rotation are  not necessary either by discussing a rectilinear example of the instability. The instability does need a boundary, however,  so cannot be captured by a local analysis.  To see this, consider the simplest possible set-up: a 2D shear flow
\beq
\bU=y \bhx + \eta \bhy
\label{planar_state}
\eeq
where there is a constant pressure gradient $-\eta \bhx$  and $\bU$ conveniently solves both Euler equations and the Navier-Stokes equations (by design\footnote{Removing the pressure gradient means that the base state is no longer a constant shear in the viscous situation.}).
The linearised inviscid equation for the perturbation vorticity $\omega=-\Delta \psi$, where $\psi$ is the streamfunction ($\bu=\bnab \times \psi \bhz$), is
\beq
\biggl( \frac{\partial}{\partial t}+y\frac{\partial}{\partial x} +\eta\frac{\partial}{\partial y} \biggr) \omega=0
\label{inviscid_gov_eqn}
\eeq
which represents just  advection of the vorticity and has solution
\beq 
\omega(x,y,t)=\omega_0 (x-yt+\half \eta t^2,y-\eta t )
\label{expression}
\eeq
where $\omega_0(x,y)$ is the initial vorticity distribution. In an unbounded domain, 
$\omega_0$ can be Fourier transformed so that it is sufficient to consider 
only $\omega_0=\exp(i(kx+ly))$. Then the stream function is 
\beq 
\psi(x,y,t)=\frac{k^2+l^2}{k^2+(l-kt)^2} e^{i(kx+(l-kt)y+\eta [\half k t^2-lt])}
\eeq
which exhibits transient growth but no linear instability just as in  the $\eta=0$ case
\citep{Orr1907,Farrell87}.

%
%
\subsection{Half Plane: Inviscid \label{hp_inviscid}}

A half plane, however, can permit a starting vorticity distribution
which spatially grows towards the boundary. Consider a boundary at $y=0$ and $\eta > 0$ so that
this is an inflow boundary when the fluid domain is $y>0$. Taking the Fourier transform in $x$ of (\ref{expression}) forces 
\beq
\Delta \psi(x,y,t)= f(y-\eta t) e^{ik(x-yt+\half \eta t^2)}
\eeq
where $f$ is an arbitrary function. If $\psi$ is to be a modal
disturbance which depends exponentially on $t$ (and $x$),
i.e. $\psi(x,y,t)=\hat{\psi}_k(y)\exp(i k x+\sigma t)$, then
\beq
f(\xi)=Ae^{-(\sigma \xi+\half ik \xi^2)/\eta}
\eeq
is the only possibility ($A$ is an arbitrary normalisation) and therefore
\beq
\biggl( \frac{d^2}{dy^2}-k^2\biggr)\hat{\psi}_k(y)
= A e^{-(\sigma y+\half iky^2)/\eta}
\label{del_psi}
\eeq
(no modal solution exists for $\eta=0$). We will show that this equation, where the initial vortical distribution decays exponentially as $y \rightarrow \infty$, has  growing modal disturbances (i.e. $\Re e (\sigma)>0$). No instability is possible
if the boundary is an outflow boundary - i.e. $\eta <0$ - as $A$ is forced to be 0 by boundedness. This suggests that the
instability exists providing the boundary is {\em not} a zero-vorticity boundary which would force $A=0$. In what follows, we adopt a non-slip boundary condition as \cite{ilin13} originally did but will revisit this issue in \S \ref{bc_discussion} (\,equation (\ref{del_psi}) is a third order differential equation for $\psi$ integrated once to include an arbitrary constant - hence 3 boundary conditions are needed to specify a unique solution\,).

To confirm there is instability, (\ref{del_psi}) must be solved to derive  the dispersion relation. Using a Green's function approach, setting $A$ to 1 (w.l.o.g.) and 
imposing the 2 (usual inviscid) boundary conditions that $\hat{\psi}_k(0)=0$ 
and $\lim_{y \rightarrow \infty} \hat{\psi}_k(y)=0$, (\ref{del_psi}) has the solution
\beq
\hat{\psi}_k(y)=-
 \int^{y}_0        \frac{e^{-ky   }}{k} \sinh(k \dy) e^{-(\sigma \dy+\half ik \dy^2)/\eta} d \dy
-\int^{\infty}_{y} \frac{e^{-k \dy}}{k} \sinh(k y  ) e^{-(\sigma \dy+\half ik \dy^2)/\eta} d \dy. 
\label{half_plane_soln}
\eeq
A third boundary condition needs to be applied to  give the dispersion relation, and 
with no-slip at the influx boundary ($d\hat{\psi}_k/dy=0$ at $y=0$), this is the relation
\beq
\int^{\infty}_{0} e^{-k \dy-(\sigma \dy+\half ik \dy^2)/\eta} d \dy=0.
\label{dispersion_hp}
\eeq
There is no solution to this for $k=0$ indicating the absence of a 1D instability but there is instability in 2D ($k \neq 0$).
The expression (\ref{dispersion_hp}) is valid for  finite $\eta$ but it is now useful to consider small $\eta$ where this instability can be understood as a critical layer instability. The (special) case where this critical layer is at the (influx) boundary (so it is in fact a boundary layer) was discussed by \cite{ilin13} in their inviscid Taylor-Couette set-up. The growth rates for such modes are $\Re e (\sigma)=O(\sqrt{\eta})$. The other (generic) case when the critical layer is distinct from the inflow boundary has not been discussed before. The growth rate here is smaller  - $\Re e (\sigma)=O(\eta \log(1/\eta))$ (see below) - but is  still larger than the default $O(\eta)$ which would be expected.

The asymptotic form of the dispersion relation (\ref{dispersion_hp}) can be derived as $\eta \rightarrow 0^+$ using standard steepest descent/saddle point ideas. Here we need two parts of the integrand to contribute at the same leading order and precisely cancel. 
This can happen in two ways since there is just one saddle point at $\dy_s=i \sigma/k$:
the contribution from an end point (clearly $\dy=0$) cancels the contribution from the saddle point (the interior critical layer case) or the saddle point and the end point are effectively one and the same asymptotically (the boundary layer case).
In the former case, the leading contributions from the end point at $\dy=0$ (1st term on LHS in (\ref{contributions})) and from the saddle point (2nd term on LHS) must satisfy
\beq
\frac{\eta}{\sigma}+\sqrt{\frac{2 \pi \eta}{ik}}e^{-i \sigma-i \sigma^2/(2 k \eta)} =0
\label{contributions}
\eeq
Now, for the saddle point to be asymptotically separated from the endpoint $\dy=0$, $|\sigma/k|=O(1)$. This together with the fact that the magnitude of the saddle point contribution $O(\sqrt{\eta} \exp( \sigma_r \sigma_i/(k\eta))$ (where $\sigma=\sigma_r+i \sigma_i$) has to be $O(\eta)$ requires {\rm either} $\sigma_r=O(1)$ and then $\sigma_i \ll 1$ or $\sigma_r \ll 1$ and $\sigma_i=O(1)$. The former case is inconsistent because the saddle point is in the wrong part of the $\dy-$complex plane leaving the contribution from the end point $\dy=0$ unbalanced. The latter situation, however, does yield solutions. Defining
\beq
\sigma=i \sigma_i+ \delta(\eta)\tilde{\sigma}_r
\label{crit_scaling}
\eeq
with $\sigma_i$ and $\tilde{\sigma}_r$ both  $O(1)$ quantities and $\delta(\eta) \rightarrow 0$ as $\eta \rightarrow 0$,
then (\ref{contributions}) requires
\beq
\delta(\eta) \tilde{\sigma}_r = -\frac{k \eta}{2\sigma_i} 
\biggl(\log \frac{1}{\eta}+\log \frac{2 \pi \sigma_i^2}{k}\biggr)-k\eta
+O(\eta^2 \log 1/\eta) 
\label{crit_growth}
\eeq
and
\beq
{\rm Arg} 
\biggl[ e^{i(  \sigma_i^2/(2k \eta)+\pi/4 )} \biggr] = O(\eta \log 1/\eta)\qquad 
\label{asym_hp_crit}
\eeq
so there is  instability with asymptotic growth rate $O(\eta \log 1/\eta)$ for a discrete set of frequencies 
\beq
\sigma_i=-\sqrt{\half \pi k \eta (8n-1)}
\label{crit_freq}
\eeq
 where  $n$ is an integer of $O(1/\eta)$. The form of the corresponding eigenfunction is discussed in \S \ref{hp_crit}.

\begin{figure}
\setlength{\unitlength}{1cm}   
\psfrag{u}{{\Large $\Re e(u)$}}
\psfrag{y}{{\Large $y$}}                                                 
\begin{picture}(14,13) 
\put(0,0)   {\includegraphics[width=13cm]{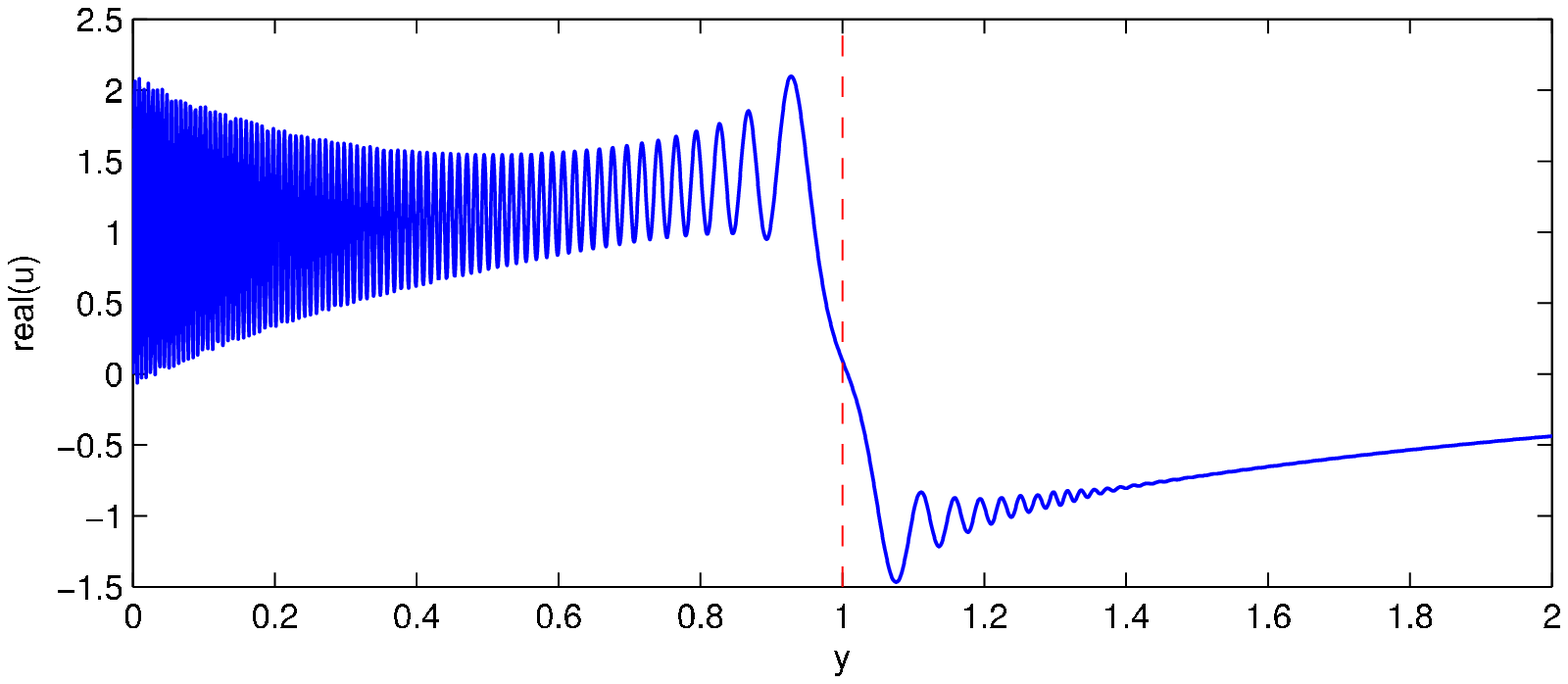}} 
\put(0,6.5)   {\includegraphics[width=13cm]{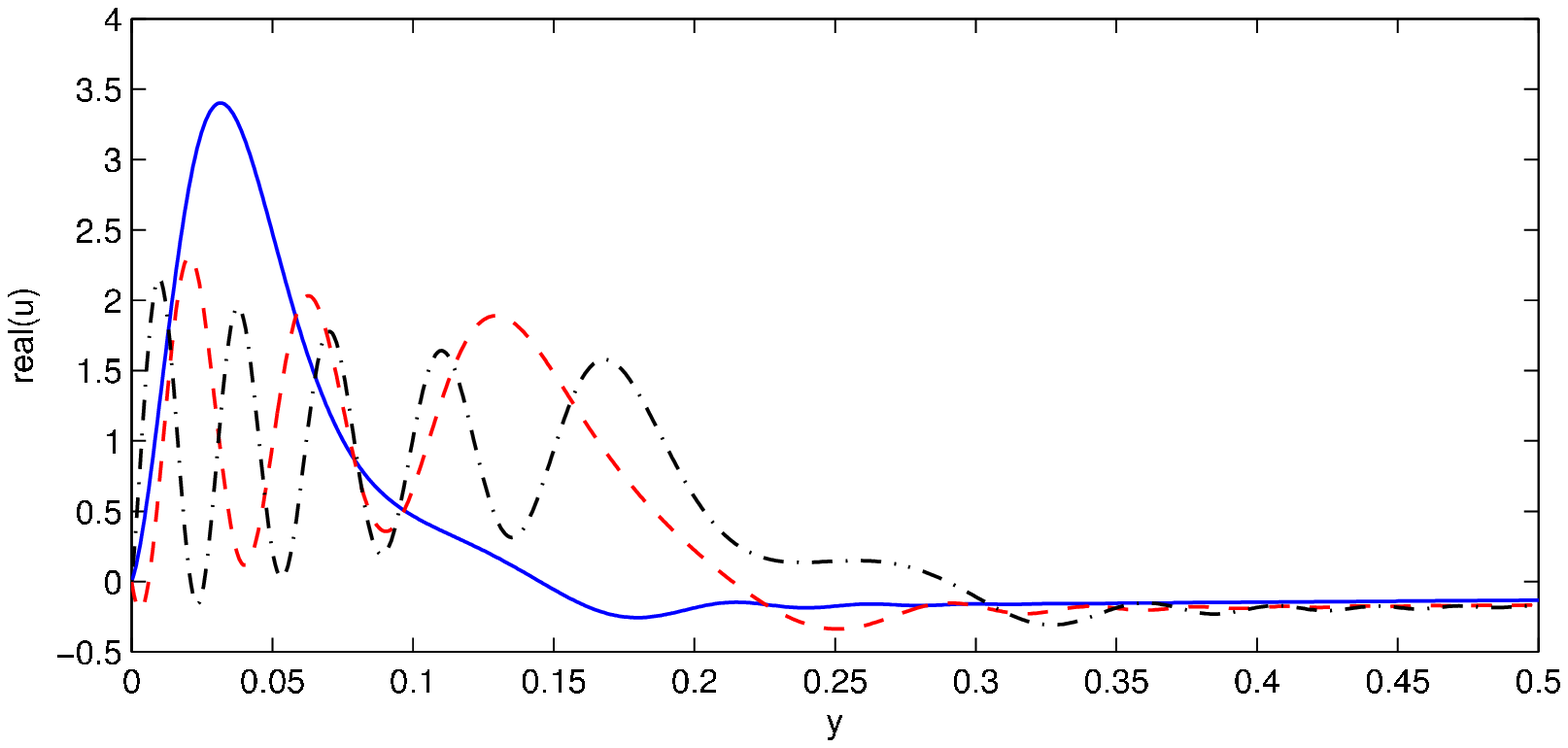}}
\end{picture}       
\caption{Unstable boundary layer eigenfunctions (upper plot: blue solid line is the most unstable, the red dashed line is the third most unstable and the black dash-dot line is the fifth most unstable mode) and an unstable critical layer eigenfunction (lower plot with the critical layer shown as a red dashed line)  for $\eta=10^{-3}$ and $k=1$. In all, the real part of $u=\psi_y$ is shown.  }
\label{eigenplot}
\end{figure}

The other situation - the boundary layer case - arises when the saddle point is within $O(\sqrt{\eta})$ of the endpoint at $\dy=0$, i.e. $|\sigma|=O(\sqrt{\eta})$. At this point, the endpoint and the saddle point contributions cannot be considered separately. Instead $\dy$ and $\sigma$ must be rescaled so that (\ref{dispersion_hp}) becomes
\beq
\int^{\infty}_{0} e^{-\hat{\sigma}z-iz^2} d z=0
\label{dispersion_hp_bl}
\eeq
where $z:=\dy/\sqrt{2\eta/k}$ and 
\beq
\sigma=\sqrt{k\eta/2}\,(\hat{\sigma}_r+i \hat{\sigma}_i)+O(\eta).
\label{bl_scaling}
\eeq
%
This must be solved numerically but a good estimate for the eigenvalues can be found by treating the contributions from the end point and saddle point as if they are separated. This means taking the integer $n$ to be $1 \ll n \ll 1/\eta$ in the frequency expression (\ref{crit_freq}) for the internal critical layer mode and calculating the corresponding growth using (\ref{crit_growth}). This leads to the asymptotic form 
\beq
\hat{\sigma}_i \sim -\sqrt{\pi(8n-1)} \qquad \hat{\sigma}_r \sim \frac{1}{| \hat{\sigma}_i |} \log(\pi \hat{\sigma}_i^2) \qquad {\rm as} \, \, {\rm integer} \, \,n \rightarrow \infty
\label{asym}
\eeq
which performs very well even for the first eigenvalue since $\hat{\sigma}_i$ is already  $< -4.7$ then: see Table 1. Figure \ref{eigenplot} shows a typical critical layer eigenfunction and 3 boundary layer eigenfunctions for $k=1$ and $\eta=10^{-3}$.

%
%
%
%
%
\begin{table}
\begin{center}
\begin{tabular}{@{}rrcrcrll@{}}
  & \multicolumn{3}{c}{asymptotic} & &\multicolumn{3}{c}{actual} \\
 &&&&&& & \\ \hline
 $n$ & $\hat{\sigma}_i$ \quad \qquad& &$ \hat{\sigma}_r $ \quad \qquad  &
  \hspace{1cm}
  &  $\hat{\sigma}_i \quad$ & &\qquad $ \hat{\sigma}_r $ \\
       &                 & &              & &                 &&                \\
100 &  -50.1012 & &0.17909 & & -50.10143  && 0.1790934 \\   
       &                 &  &             & &                 &&                \\
10& -15.7539 & & 0.42268 & & -15.75720 && 0.4226325 \\          
9 & -14.9387 & & 0.43871 & & -14.93867 && 0.4386500 \\          
8 & -14.0684 & & 0.45724 & & -14.07266 && 0.4571595 \\         
7 & -13.1449 & & 0.47903  & & -13.14981 && 0.4789312 \\          
6 &  -12.1513 & & 0.50526  & &-12.15722 && 0.5051295 \\          
5 & -11.0690 & & 0.53781  &  &-11.07620 && 0.5376254 \\
4 & -9.8686  & & 0.57997  &  &-9.877886 && 0.5796846 \\
3 & -8.5004  & & 0.63820  & & -8.513168 && 0.6377301 \\
2 & -6.8647  & & 0.72800  & & -6.884666 && 0.7270530 \\
1 & -4.6894  & & 0.90317 &  & -4.732350 && 0.9003686 \\[6pt]
\end{tabular}
\end{center}
\caption{The 10 most unstable eigenvalues from the dispersion relation (\ref{dispersion_hp_bl}) and asymptotic estimates from (\ref{asym}).}
\label{bl_eigenvalues}
\end{table}

%
%
\subsubsection{Inviscid Asymptotics $0 < \eta \ll 1$ for the Boundary Layer Instability \label{hp_bl} }

Here a modal streamfunction solution of the form $\Psi(x,y,t)=\psi(y) e^{ i k x+\sigma t}$ is sought with a boundary layer of thickness $\sqrt{\eta}$ as first uncovered by \cite{ilin13} (see their \S 3.2). Defining the new boundary layer variable $\xi:= y \sqrt{k/(2 \eta)}$ and splitting  the streamfunction into an expansion of large-scale parts and a boundary layer corrections (hatted variables)
\beq
\psi := (\,\psi_0(y)+\hat{\psi}_0(\xi) \,)+\sqrt{\eta}(\,\psi_1(y)+\hat{\psi}_1(\xi) \,)+ \ldots, 
\eeq
we look for an instability with vanishing frequency (=$k \times\,$ speed of the inflow boundary) to leading order, 
\beq
\sigma:= \sqrt{\eta k/2} \, \hat{\sigma}+O(\eta).
\eeq
The governing equation
\beq
(\sigma +i k y +\eta \partial_y)(\psi_{yy}-k^2 \psi)=0 \label{governing_hp}
\eeq
is then simplified to
\beq
\psi_{0\,yy}-k^2 \psi_0=0
\label{outside}
\eeq
for the leading flow and
\beq
(\hat{\sigma}+2i \xi+\partial_\xi)\hat{\psi}_{0 \xi \xi}=0
\label{bl_asym}
\eeq
for its boundary layer correction.  The interesting observation here is there is no large-scale solution $\psi_0$ which can handle both the $\eta=0$ boundary conditions that $v(x,0,t)=0$ and  $\lim_{y \rightarrow \infty} v(x,y,t)=0$. Crucially, this means that the boundary layer correction $\hat{\psi}_0$ must be $O(1)$ so as to contribute at leading order to fix up the $v(x,0,t)=0$ boundary condition rather than  the usual $O(\sqrt{\eta})$ to satisfy the extra $\eta \neq 0$ tangential boundary condition $u(x,0,t)=0$. As a consequence, there is an $O(1/\sqrt{\eta})$ tangential velocity $u$ in the boundary layer which must vanish at $\xi=0$. Since the large-scale flow cannot contribute at this order, this non-slip condition is solely on the boundary layer flow and is sufficient to determine the growth rate. Integrating (\ref{bl_asym}) twice and incorporating the fact that $\lim_{\xi \rightarrow \infty} \hat{\psi}_{0\, \xi}=0$, leads to
\beq
\hat{\psi}_{0\, \xi}=-A \int^{\infty}_\xi e^{-\hat{\sigma} z-i z^2} \, dz.
\eeq
where $A$ is an arbitrary constant. Imposing $u=0$ at $y=0$ then forces $\hat{\psi}_{0 \xi}(0)=0$ which is precisely condition (\ref{dispersion_hp_bl}).

%
%
\subsubsection{Inviscid Asymptotics $0< \eta \ll 1$ for the Critical Layer Instability \label{hp_crit} }

An interior  critical layer instability is constructed by looking for a mode of O(1) frequency. Adopting the expansion (\ref{crit_scaling}) where again both $\sigma_i$ and $\tilde{\sigma}_r$ are $O(1)$, the critical layer is centred on $y_c:=-\sigma_i/k$ (so $\sigma_i\, <\, 0$) and as in the boundary layer case, has thickness $O(\sqrt{\eta})$. Defining the critical layer variable
\beq
\xi:= \frac{y-y_c}{\sqrt{\eta}}
\eeq
the (leading order) streamfunction in the critical layer, $\hat{\psi}$, satisfies
\beq
(\, \delta \tilde{\sigma}_r/\sqrt{\eta}+ik\xi+\partial_\xi\,) \hat{\psi}_{\xi \xi}=0
\eeq
which can be integrated once to give
\beq
\hat{\psi}_{\xi \xi}= \sqrt{\eta} \, e^{-\delta \tilde{\sigma}_r\xi/\sqrt{\eta}-ik\xi^2/2}
\eeq 
(choosing the normalisation of the mode here for convenience later) and then two further times to give
\begin{align}
\hat{\psi}_{\xi} &=  \sqrt{\eta} \biggl(\,\int^{\xi}_0 e^{-\delta \tilde{\sigma}_r x /\sqrt{\eta}-ikx^2/2} \, dx+A \, \biggr) \hspace{0.25cm}
\sim  \sqrt{\eta} \biggl(\,\pm \sqrt{ \frac{\pi}{2ik} }+A \, \biggr) +h.o.t. \,\,{\rm as} \, \, \xi \rightarrow \pm \infty \label{psi_xi}\\
\hat{\psi}       &= \sqrt{\eta} \biggl(\, \int^{\xi}_0 (\xi -x) e^{-\delta \tilde{\sigma}_r x/\sqrt{\eta}-ik x^2/2} \, dx+A \xi\, \biggr) +B 
                                                                                                              \nonumber \\
                 & \hspace{4.5cm}\sim  \sqrt{\eta} \xi \biggl[\pm \sqrt{ \frac{\pi}{2ik} }+A \biggr]+B+h.o.t.\, \, {\rm as} \, \, \xi \rightarrow \pm \infty \label{psi}
\end{align}
where $A$ and $B$ are $O(1)$ constants. 

Outside the critical layer, the streamfunction consists of a WKB-type solution and simple exponentials which satisfy Laplace's equation,
\begin{align}
y_c< y:          &\quad  \psi_+ = C_{+} e^{-ky} +            & \frac{E_{+}}{k^2(y-y_c)^2}e^{-[\,\delta \tilde{\sigma}_r(y-y_c)+\half i k (y-y_c)^2\,]/\eta}  \label{above}\\
0<y<y_c: & \quad \psi_- = C_{-} e^{-ky}+D_{-} e^{ky}+        & \frac{E_{-}}{k^2(y-y_c)^2}e^{-[\,\delta \tilde{\sigma}_r(y-y_c)+\half i k (y-y_c)^2 \,]/\eta} \label{below}
\end{align}
where $C_{\pm}$, $D_{-}$ and $E_{\pm}$ are constants whose order of magnitude will be set by matching to the critical layer streamfunction. The fact that $\psi_+$ and $d \psi_+/dy$ must vanish as $y \rightarrow \infty$ is imposed by including only the decaying exponential for $y > y_c$: the WKB mode vanishes as $y \rightarrow \infty$ when  $\eta >0$ provided  $\tilde{\sigma}_r > 0$.

The governing equation (\ref{governing_hp}) is third order and so $\psi,\psi_y$ and $\psi_{yy}$ must be continuous everywhere. The oscillatory form of  $\hat{\psi}_{\xi \xi}$ in the critical layer means it must match entirely to the outer WKB-type solution 
\beq
\psi^{WKB}:=\frac{-\eta^{3/2}}{k^2(y-y_c)^2} e^{-[\,\delta \tilde{\sigma}_r(y-y_c)+\half i k (y-y_c)^2\,]/\eta}
\eeq
either side of the critical layer so $E_{\pm}=-\eta^{3/2}$. This means that the  outer WKB solution only contributes at $O(\sqrt{\eta})$ to the tangential velocity $u$ near the critical layer whereas the critical layer streamfunction $\hat{\psi}$ forces an $O(1)$ tangential flow (see (\ref{psi_xi})\,). As a result there must be a large-scale flow of $O(1)$ in both $u$ and $v$ outside the critical layer: this explains the scaling of the integration constants in (\ref{psi}) and means that $C_{\pm}$ and $D_{-}$ are all $O(1)$. There are then six (complex) conditions to be satisfied at leading order  by the five remaining constants and complex eigenvalue $\sigma$.
The first four are matching conditions on $u$ ($\psi_y$) and $v$ ($-ik\psi$) as $y \rightarrow y_c^+$,
\beq
-k C_{+} e^{-ky_c} = \sqrt{ \frac{\pi}{2ik} }+A 
\quad \& \quad 
C_{+} e^{-ky_c} = B
\eeq
and $y \rightarrow y_c^{-}$
\beq
-k C_{-} e^{-ky_c} +k D_{-} e^{ky_c} = -\sqrt{\frac{\pi}{2ik} }+A 
\quad \& \quad
C_{-} e^{-ky_c} +  D_{-} e^{ky_c} = B
\eeq
or eliminating $A$ and $B$, simply two jump conditions across the critical layer
\beq
[u]^{+}_{-}=\sqrt{\frac{2\pi}{ik}} \quad \& \quad [v]^{+}_{-}=0.
\eeq
The two remaining (boundary) conditions, $v(x,0,t)=0$ and $u(x,0,t)=0$,  require 
\beq
C_{-}+D_{-} = 0  \quad \& \quad
-kC_{-}+kD_{-} -\frac{i \sqrt{\eta}}{k y_c} e^{[\,\delta \tilde{\sigma}_r y_c-\half i k y_c^2\,]/\eta} =0  
\eeq
where, while the outer WKB-type solution contributes at $O(\sqrt{\eta})$ to $u$ near the critical layer, it  must contribute at $O(1)$ to $u$ at the inflow boundary.
The resulting dispersion relation is
\beq
\frac{i\sqrt{\eta}}{ky_c}e^{[\,\delta \tilde{\sigma} y_c-\half iky_c^2\,]/\eta}+\sqrt{ \frac{2\pi}{ik} } e^{-k y_c}=0
\label{balance}
\eeq
which leads to the same leading expressions for $\tilde{\sigma}_r$ and $\sigma_i $ given in (\ref{crit_growth}) and (\ref{crit_freq}).
The necessary change in the scaling of the WKB solution contribution gives the dominant $O(\eta \log 1/\eta)$ contribution to $\tilde{\sigma}_r$ and the exact numerical counterbalancing of the large scale tangential flow gives the subdominant $O(\eta)$ contribution.

%
%
\subsubsection{The Need for Shear and  Inflow \label{shear_inflow} }

To emphasize that shear {\em is} a key ingredient of the instability, the above problem can be solved for the shearless flow $\bU=\bhx + \eta \bhy$ leading to the requirement that there must exist $\Re e({\sigma})>0$ with
\beq
\int^{\infty}_{0} e^{-k \dy-\sigma \dy/\eta} d \dy=0
\eeq
which is never satisfied. The presence of an inflow boundary is also crucial: converting the above inflow boundary to an outflow boundary ($\eta \rightarrow -\eta$) removes the instability. This is why the instability discussed here is not relevant to the considerable literature on suction boundary layers where suction is always a stabilizing effect (e.g. \cite{Joslin1998}).

%
%
\subsubsection{The Third Boundary Condition \label{bc_discussion} }

Imposing the parametrised boundary condition $(1-\beta)u+\beta du/dy=0$ as the second boundary condition at $y=0$ gives the modified dispersion relation
\beq
\int^{\infty}_{0} e^{-k \dy-(\sigma \dy+\half ik \dy^2)/\eta} d \dy= \frac{\beta}{1-\beta}  
\label{modified_dispersion}
\eeq
($\beta=0$ recovers non-slip  and $\beta=1$ a no-vorticity or equivalently stress-free condition as $v=0$ along $y=0$). 
For the boundary layer instabilities, the LHS is $O(\sqrt{\eta})$ so this dispersion relation can only be assumed similar to the non-slip relation (\ref{dispersion_hp_bl}) if $\beta \ll \sqrt{\eta}$. In fact,  numerical computations indicate that the boundary layer instability  is suppressed by $\beta \approx 2.6\sqrt{\eta/k} \ll 1$. For the critical layer instabilities, the condition (\ref{contributions}) becomes a possible balance between 3 different terms
\beq
\frac{\eta}{\sigma}+\sqrt{\frac{2 \pi \eta}{ik}}e^{-i \sigma-i \sigma^2/(2 k \eta)} = \frac{\beta}{1-\beta}  
\label{further_contributions}
\eeq
where $\sigma$ now has an $O(1)$ frequency as in (\ref{crit_scaling}).  If $ \beta \gg \eta$, the dominant balance is between the 2nd and 3rd terms (as opposed to the 1st and 2nd for non-slip) and now leads to damped eigenvalues. It is then clear that for instability to occur, there should be no restriction on the normal derivative of the tangential velocity at the inflow boundary. In practice this means  that the instability only really occurs for non-slip boundary conditions when $0 \leq \eta \ll 1$ which incidentally is the one choice which, in concert with the no-normal flow condition, means no disturbance kinetic energy is being advected {\em into} the domain through the inflow boundary.

%
%

\subsection{Half Plane: Viscous \label{hp_viscous} }

The base state (\ref{planar_state}) is unchanged (by design) when viscosity is introduced but the linearised disturbance equation (\ref{inviscid_gov_eqn}) now includes diffusion of vorticity:
%
\beq
\biggl( \frac{\partial}{\partial t}+y\frac{\partial}{\partial x} +\eta\frac{\partial}{\partial y} \biggr) \omega=
\frac{1}{Re}\nabla^2 \omega.
\label{viscous_gov_eqn}
\eeq
Looking for a modal solution $\omega(x,y,t)=\hat{\omega}(y) \exp(ikx+\sigma t)$ leads to the equation
\beq
\frac{d^2 \hat{\omega}}{dy^2} -Re \,\eta \frac{d \hat{\omega}}{dy}-\left[Re(\sigma+iky)+k^2)\right] \hat{\omega}=0
\eeq
which has the bounded solution (as $y \rightarrow \infty$)
\beq
\hat{\omega}= e^{\half Re \,\eta\,y} Ai\biggl[ \frac{Re \sigma+k^2+\quart Re^2\eta^2}{(ikRe)^{2/3}}+(ikRe)^{1/3}y\biggr].
\eeq 
where $Ai(z)$ is the Airy function bounded as $|z| \rightarrow \infty$ with $|\arg(z)| < \pi$.
The dispersion relation is then
\beq
\int^{\infty}_0 e^{(\half Re \,\eta-k) \dy} 
Ai\biggl[ \frac{Re \sigma+k^2+\quart Re^2\eta^2}{(ikRe)^{2/3}}+(ikRe)^{1/3}\dy\biggr] \, d\dy=0.
\label{hp_vis}
\eeq
and instability ($\Re e(\sigma)$ crosses through 0 to become positive ) occurs at a critical inflow 
$\eta_{crit}(k,Re)$. This integral is actually the same as that treated by \citet{gal10} (see their expression (39)) after the transformations
\beq
N_g^{2/3}:=2^{1/3} \frac{Re\, \eta-2k}{(kRe)^{1/3}}, \qquad a_g=-\biggl(\frac{4Re}{N_g k^2} \biggr)^{1/3}(\sigma+k \eta)
\eeq
($N_g$ and $a_g$ from \citet{gal10}).  For the $k$ of interest (\,$\leq O(1)$\,), $k \eta \ll | \sigma |$ and we can reuse their critical value of $N_g$ which has  $a_g$ passing through the imaginary axis to give
\beq
\eta_{crit}  = \biggl(\half N_g^2 k \biggr)^{1/3} Re^{-2/3},\qquad 
\sigma_{crit}= - \biggl(\frac{N_g k^2}{4Re}\biggr)^{1/3}a_g
\eeq
where $N_g=4.57557$ and $a_g=5.63551 i$ (consistent with \citet{gal10} who quote `4.58' and `5.62i' for $N_g$ and $a_g$ respectively). 
This threshold $\eta_{crit}$ tends to zero as $k \rightarrow 0$ albeit with the unstable eigenfunction extending a distance $O((kRe)^{-1/3})$ in the $y$ direction. In terms of connecting this analysis to other problems, there are two notable cases: $k=O(1)$ which is the interesting case in rotating flow where the wavenumber is forced to be an integer by periodicity, and $k=O(Re^{-1})$ which is gives the most unstable disturbance in a domain bounded in $y$ (i.e. a channel see \cite{Nicoud97}). In the former case, $\eta_{crit}=O(Re^{-2/3})$ and the implication from the scaling of the critical frequency is the growth rate away from criticality (i.e. $|\eta-\eta_{crit}|=O(\eta_{crit})$) will be $O(Re^{-1/3})$ or $O(\sqrt{\eta})$ as before.  For $k=O(Re^{-1})$, $\eta_{crit}=O(Re^{-1})$ which is consistent with the numerical findings in \cite{Nicoud97} that the threshold `crossflow' Reynolds number for inflow instability in their plane Couette flow is independent of the shear Reynolds number.

For any given $k$, further boundary-layer-type instabilities exist as $\eta$ increases with the first 6 thresholds listed in Table 2 for $k=1$. Within this `boundary-layer' scaling of $\tilde{\lambda}:=-Re^{1/3} \sigma_i$ and $\tilde{\eta}_{crit}:= Re^{2/3}\eta_{crit}$ for $k=O(1)$, asymptotic predictions for $\tilde{\lambda} \rightarrow \infty$ can be derived following the same route as in the inviscid case. This proves a little more involved leading to two coupled relations 
\begin{align}
\tilde{\eta}_{crit} &=\tilde{\lambda}
\biggl( \frac{3k}{2}\log(2\pi \tilde{\lambda}/k) -\frac{k}{2}\log \biggl[ \frac{2}{3k \log(2\pi \tilde{\lambda}/k)} \biggr]\biggr)^{-1/3}, \label{asym_1} \\
\frac{ \tilde{\lambda}^2 }{ \tilde{\eta}_{crit} } &= \half \pi (8n-1)  \qquad \qquad n=1,2,3, \ldots 
\label{asym_2}
\end{align}
which work reasonably well for $\tilde{\lambda}=O(10)$ given that higher order terms may only be $\log(\tilde{\lambda})$ smaller.

As in the inviscid case, there are also interior critical layer modes excited for even higher inflows: if the critical layer is at $y=-\sigma_i/k=O(1)$, then 
\beq
\eta_{crit}=|\sigma_i| \biggl( \frac{2}{k Re \log Re} \biggr)^{1/3} \gg O(Re^{-2/3}).  
\label{hp_vis_crit}
\eeq

%
%
\begin{table}
\begin{center}
\begin{tabular}{@{}rrrrrrrr@{}}
  & \multicolumn{3}{c}{asymptotic} & &\multicolumn{3}{c}{actual} \\
 &&&&&& & \\ \hline
 $n$ & $\quad \tilde{\lambda}$ \quad & &$ \quad \tilde{\eta}_{crit} $   &
  \hspace{1cm}
     & $\quad \tilde{\lambda}$ \quad & & $ \tilde{\eta}_{crit} $ \\
       &                 & &              & &                 &&                \\
6 &  35.2992  & &  16.8776  &  &  35.183   &&  16.501 \\          
5 &  29.5968  & &  14.2990  &  &  29.482   &&  13.885 \\
4 &  23.8395  & &  11.6711  &  &  23.730   &&   11.201 \\
3 &  18.0097  & &  8.9777   &  & 17.9033  &&    8.4197\\
2 &  12.0754  & &  6.1886   &  & 11.9737  &&    5.4762 \\
1 &   5.9596  & &  3.2301   &  &  5.8938   &&    2.1875 \\[6pt]
\end{tabular}
\end{center}
\caption{The 6 lowest inflow thresholds for instability in the dispersion relation (\ref{hp_vis}) with $k=1$ as  $Re \rightarrow \infty$ and asymptotic estimates from (\ref{asym_1}) \& (\ref{asym_2}).}
\label{hp_vis_eigenvalues}
\end{table}

%
%
\subsection{Inflow and Suction Together: Inviscid Plane Couette flow with Suction \label{hp_suction} }

The half plane system exhibits boundary inflow instability for all $\eta>0$. This can be seen by a simple rescaling of space
\beq
\kappa=\eta^{1+\beta}k, \quad Y=y/\eta^{1+\beta} \quad \&\quad  \eps=1/\eta^{\beta}
\label{rescaling}
\eeq
where $\beta>0$ so that the equation (\ref{governing_hp}) becomes
\beq
(\sigma +i \kappa Y +\eps \partial_Y)(\psi_{YY}-\kappa^2 \psi)=0
\eeq
which is just the original equation with a new small number $\eps$ as $\eta \rightarrow \infty$. However, this is rather artificial since the applied pressure gradient also has to be increased with $\eta$ to maintain the constant shear in $y$. Resorting to a  constant pressure gradient instead now means the shear field decreases as  $\eta$ increases again making it difficult to draw general conclusions for large $\eta$. Introducing another boundary is then the only alternative and this must be an outflow boundary if the resulting base flow is to be steady and spatially non-developing. The simplest  modification is to add an outflow boundary at $y=1$ which is moving at $\bhx$ so that the 
constant-vorticity basic flow (\ref{planar_state}) is still a solution \citep{Nicoud97}.
The equivalent expression to (\ref{half_plane_soln}) is then
\begin{align}
\hat{\psi}_k(y)= &-
 \int^{y}_0   \frac{\sinh (k \dy)}{k\sinh k} \sinh k(1-y) e^{-(\sigma \dy+\half ik \dy^2)/\eta} d \dy \nonumber\\
& \hspace{1cm}-\int^{1}_{y} \frac{\sinh k(1-\dy)}{k \sinh k} \sinh(k y  ) e^{-(\sigma \dy+\half ik \dy^2)/\eta} d\dy 
\end{align}
with the dispersion relation
\beq
\int^{1}_{0}  \sinh k(1-\dy)e^{-(\sigma \dy+\half ik\dy^2)/\eta} d\dy=0.
\label{dispersion_pcf}
\eeq
when non-slip is applied at the influx boundary $y=0$ for $\eta>0$. This is essentially the same as the half plane dispersion relation and will have unstable eigenvalues as there is an inflow boundary. The simple rescaling (\ref{rescaling}), however, is disallowed and instability is lost if $\eta$ becomes too large (see figure 12 of \cite{Nicoud97}). This could be interpreted as the stabilising influence of the newly-introduced suction boundary ultimately overpowering the destabilising inflow boundary. Further evidence for this comes from the pressure-gradient-free version of this flow which is also linearly unstable in the presence of viscosity \citep{Doering00}. In this case, the base state varies exponentially in the cross-stream direction (see eqn 2.13 of \cite{Doering00}) and possesses an area of linear instability in the $(\tan \theta, Re)$ plane (figure 3 of \cite{Doering00} where $\tan \theta$  is their proxy for $\eta$). The lower boundary of this instability region, $\theta_{lower}(Re)$, plausibly scales like $Re^{-1}$ which is the viscous threshold for the instability as described in \S \ref{hp_viscous} whereas the upper boundary has $\theta_{upper}(Re)= \cot^{-1} 54,370$, which appears to be  suction ultimately stabilizing the flow \citep{Hocking74}.

%
%
%

\section{2D Swirling Flow with Radial Inflow \label{swirl}}

We now add curvature to the discussion and consider the basic, steady,
axisymmetric solution 
\beq 
\bU= -\frac{\eta}{r}\br+r \Omega(r) \btheta 
\label{basic}
\eeq 
between two boundaries at $r=1$ and $r=a \,(>1)$ with $\Omega(1)=1$
(i.e. the inner radius $r^*$ and the angular velocity there $\Omega^*$ are used
as length and inverse timescales respectively) and $0< \eta$ so
that there is a radial inflow. The 2D Navier-Stokes equations for the deviation $\bu$ of the 
total flow $\bu_{tot}$ from the basic solution (\ref{basic}), 
\beq
\bu:=\bu_{tot}-\bU=u(r,\theta,t)\br+v(r,\theta,t) \btheta
\eeq 
are 
\begin{align}
\biggl(\frac{\partial}{\partial t}+\Omega \frac{\partial}{\partial \theta} -\frac{\eta}{r} \frac{\partial}{\partial r} 
\biggr) u
+\frac{\eta u}{r^2} -2 \Omega v+\bu.\bnab u &-\frac{v^2}{r}+\frac{\partial p}{\partial r} \nonumber\\ 
&= \frac{1}{Re}\biggl[ \biggl(\nabla^2-\frac{1}{r^2} \biggr)u-\frac{2}{r^2}\frac{\partial v}{\partial \theta} \biggr]
\label{radial}
\end{align}
\begin{align}
\biggl(\frac{\partial}{\partial t}+\Omega \frac{\partial}{\partial \theta} -\frac{\eta}{r} \frac{\partial}{\partial r} 
\biggr) v
+u\frac{d(r\Omega)}{dr}-\frac{\eta v}{r^2}+\Omega u +&\bu .\bnab v+\frac{uv}{r}+
\frac{1}{r}\frac{\partial p}{\partial \theta} \nonumber\\  
&=\frac{1}{Re}\biggl[ \biggl(\nabla^2-\frac{1}{r^2} \biggr)v+\frac{2}{r^2} \frac{\partial u}{\partial \theta}\biggr] 
\label{azimuth}
\end{align}
together with incompressibility
\beq
\frac{1}{r}\frac{\partial (ru)}{\partial r}+\frac{1}{r}\frac{\partial v}{\partial \theta}=0  \label{incompressibility}
\eeq
where
\beq
Re:= \frac{r^{*2} \Omega^*}{\nu}
\eeq
and $\nu$ is the kinematic viscosity. Introducing a streamfunction 
\beq
u=\frac{1}{r} \frac{ \partial \psi}{\partial \theta}, \quad v=-\frac{\partial \psi}{\partial r},
\eeq
reduces the system (\ref{radial})-(\ref{incompressibility}) to
\beq
\biggl( \frac{\partial}{\partial t}+\Omega(r) \frac{\partial}{\partial \theta} -\frac{\eta}{r}\frac{\partial}{\partial r} \biggr) \Delta \psi = \frac{1}{r}\frac{dZ}{dr} \frac{\partial \psi}{\partial \theta}+\frac{1}{Re} \Delta^2 \psi+\frac{1}{r} J(\psi,\Delta \psi)
\label{master}
\eeq
where the Jacobian is defined as
\beq
J(A,B):= \frac{\partial A}{\partial r} \frac{\partial B}{\partial \theta}-
         \frac{\partial A}{\partial \theta} \frac{\partial B}{\partial r}
\eeq
and 
\beq
Z:=\frac{1}{r} \frac{d(r^2 \Omega)}{dr}
\eeq
is the vorticity of the basic flow (\ref{basic}). There are two special cases where the gradient of the vorticity vanishes: uniform rotation ($\Omega=1$ so $Z=2$) and irrotational flow $\Omega=1/r^2$ with $Z=0$ originally considered by \cite{ilin13}.

%
%
\subsubsection{Basic state \label{basic_state}}

There are two usual scenarios: a) the swirl field $\Omega(r)$ is set by the
boundary conditions as in Taylor-Couette flow or b) the swirl field is
determined by an imposed body (gravitational) force as in the
astrophysical context. In the former, the swirl is determined by the
azimuthal momentum equation {\em given} a radial flow with the radial
momentum equation determining the pressure field. For example 
Ilin \& Morgulis (2013) discuss the inviscid, irrotational,  axisymmetric Taylor-Couette-like flow 
\beq
\Omega_{inviscid}:=\frac{1}{r^2} 
\label{inviscid}
\eeq 
which is the only possibility with axisymmetric radial flow  (recall $\Omega(1)=1$ by nondimensionalization). Gallet et al. (2010) consider the viscous Taylor-Couette situation where
possible  base flows form a 1-parameter family 
\beq
\Omega_{TC}=\frac{1-A}{r^2}+Ar^{-\eta Re} 
\label{viscous}
\eeq 
with $A$ a constant set by the motion of the outer cylinder and is
generally rotational ($Z=A(2-\eta Re)r^{-\eta Re}$). In the (latter)
astrophysical context, the acting gravitational (body) force sets the
swirl field (via the radial momentum equation) which then {\em sets}
the radial flow by the need to balance ensuing azimuthal viscous
stresses. The focus here is on the latter situation and we consider the general set of profiles
\beq
\Omega=r^\alpha
\label{astro}
\eeq 
in order to understand how robust the boundary inflow instability is. Profiles with $-2 < \alpha < 0$
have angular momentum increasing with radius and so are 
Rayleigh-stable \citep{Rayleigh1917}.  The gradient of the vorticity
$dZ/dr=\alpha(\alpha+2)\Omega/r$ also does not change sign across the
domain $r \in [1,a]$ so that the flow is inviscidly stable (for
disturbances which vanish at $r=1$ and $a$) by a rotating flow
analogue of Rayleigh's inflexion-point theorem \citep{Rayleigh1880}.
Particularly interesting choices for $\alpha$ are $\alpha=-3/2$ which
is the Keplerian profile for a thin accretion disk due to the radial force balance
\beq
-r \Omega^2 = -\frac{GM}{r^2}
\eeq
(where $G$ is the gravitational constant and $M$ the mass of the central object) and $\alpha=-1$
which is used to model spiral galaxies.  Since we work with deviations away
from the basic state, the exact body force required to maintain the underlying rotation profile
is not explicitly needed in what follows.  In
contrast to the Taylor-Couette situation, the azimuthal component of
the Navier-Stokes equations forces the existence of a small radial
flow 
\beq 
U= \frac{1}{Re} \biggl( \frac{\partial}{\partial r}\log
\left|\frac{\partial}{\partial r}(r^2 \Omega)\right|
-\frac{1}{r}\biggr)=\frac{\alpha}{Re} \frac{1}{r} 
\label{enforced}
\eeq 
which is an inflow if $d\Omega/dr<0$ ($\alpha <0$). 
Studying the consequences of this small accretion flow is the motivation for this work.

%
%
%
%
\begin{figure}
\begin{center} 
\psfrag{X}{{\Large $\hat{\sigma}_r$}}
\psfrag{Y}{{\Large $\sigma_i$}}
\resizebox{1.05\textwidth}{!}{\includegraphics[angle=0]{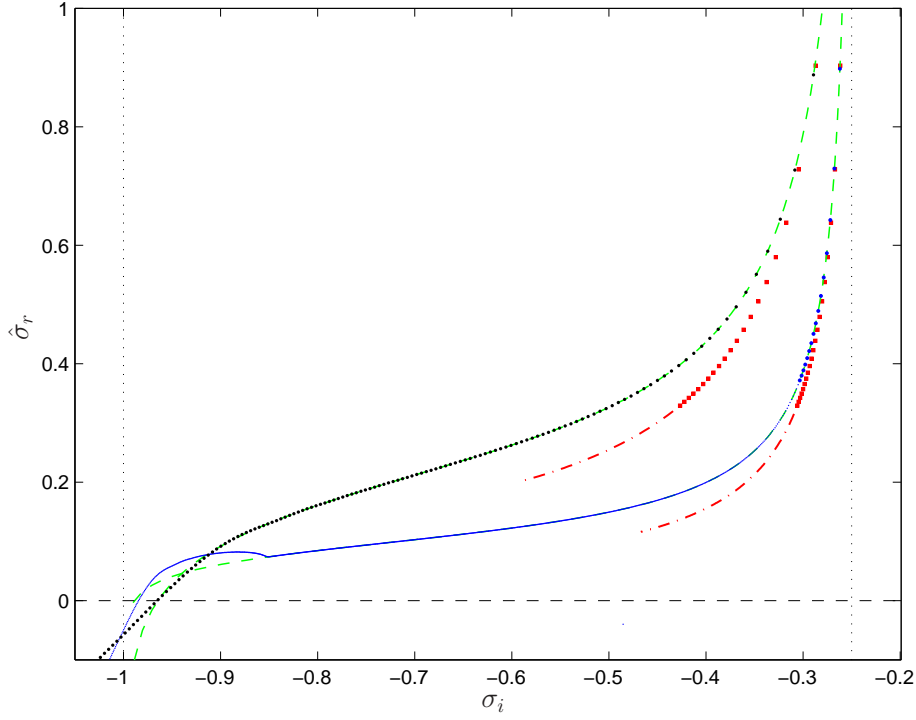}}
\end{center}
\caption{Inviscid 2D instabilities for $\alpha=-2$, $m=1$ and $a=2$ with scaled growth rate $\hat{\sigma}_r=\sigma_r/\sqrt{-m \eta \Omega^{'}(a)/(2a)}$ plotted against frequency $\sigma_i$. The right (left) vertical dashed line is $\sigma_i=-m\Omega(a)$ ($\sigma_i=-m\Omega(1)$).
The eigenvalues from a full 2D eigenvalue calculation are shown as (the upper) black dots for $\eta=10^{-3}$ ($N=1000$) and (the lower) blue dots for $\eta=10^{-4}$ (since $N=4000$ here, the 15 most unstable eigenvalues are marked with a normal-sized dot and the rest by smaller blue dots). The red squares are the 20 most unstable modes from the boundary layer asymptotics (section \ref{swirl_dzdr_0}) with the dash-dot red line tracing the path of the rest (eigenvalues calculated from the boundary layer equation (\ref{bl_asym}) are indistinquishable from the asymptotics at this scale). The green dashed lines are the critical layer asymptotic expression  
(\ref{crit_layer_asym}) with $\eta=10^{-3}$ and $10^{-4}$ plugged in and appropriately rescaled: $
\hat{\sigma}_r=\delta_1 \tilde{\sigma}_r\sqrt{  \frac{2a}{-m \eta \Omega^{'}(a)} }$ 
and $\sigma_i=-m \Omega(R_s)$ with $R_s$ taken over $[1,a]$.  
Notice that at $\eta=10^{-4}$ even $N=4000$ struggles to fully resolve the critical layer eigenvalues near the inner radius - see the hump in the numerical data which breaks the otherwise excellent correspondence with the asymptotic prediction. This hump is delayed to lower $\sigma_i$ if $N$ is increased until it eventually disappears - e.g. the $\eta=10^{-3}$ curve.}
\label{instab_invis_2D}
\end{figure}
%
%

%
%
%
\begin{figure}
\setlength{\unitlength}{1cm}   
\psfrag{u}{{\Large $\Re e(u)$}}
\psfrag{y}{{\Large $y$}}                                                 
\begin{picture}(14,12) 
\put(0,0)   {\includegraphics[width=13cm]{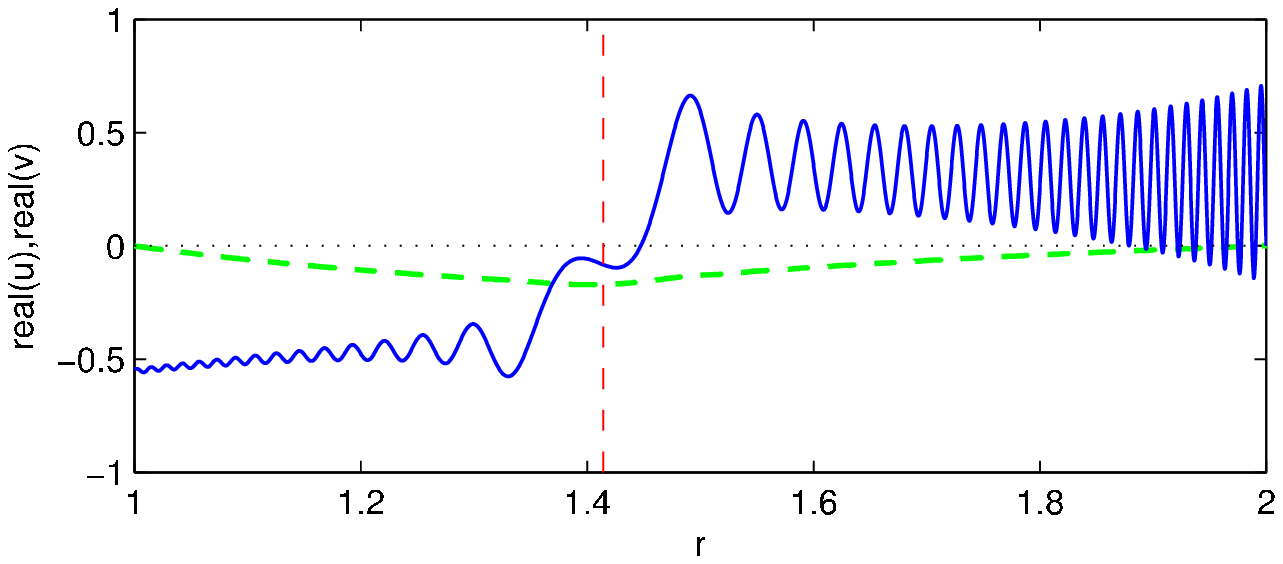}} 
\put(0,6)   {\includegraphics[width=13cm]{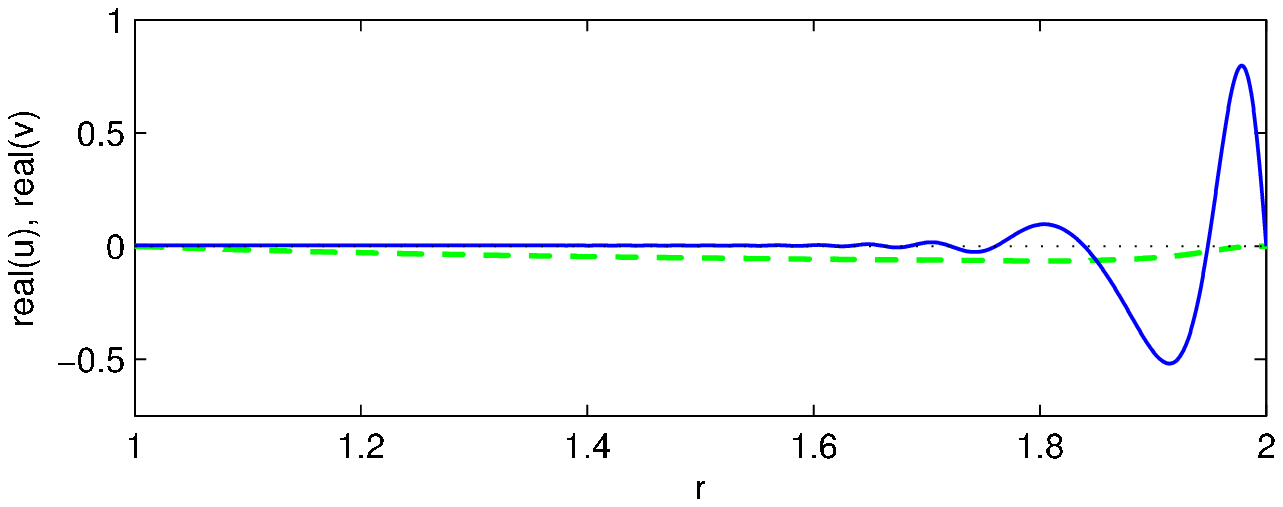}}
\end{picture}       
\caption{The most unstable boundary layer eigenfunction (upper plot) and an unstable critical layer eigenfunction with $\sigma_i \approx -0.5$ (lower plot) for $\eta=10^{-3}$, $m=1$, $a=2$ and $\alpha=-2$. In both, the green dashed line is $\Re e (u)$ and the blue solid line is $\Re e(v)$. The critical layer position is shown as a red dashed line for the critical layer eigenfunction.}
\label{eig_plot}
\end{figure}

%
%
%
%
%
\subsection{2D Inviscid Swirling Flow with $dZ/dr=0$ \label{swirl_dzdr_0} }
%
%
%

We study the simplest case of vanishing vorticity gradient in the base flow first ($dZ/dr=0$), so $\Omega=1/r^2$ or $1$, to show how the analysis mirrors that in the half plane case. The case of uniform rotation $\Omega=1$ initially looks uninteresting because the base flow needs an azimuthal as well as radial body force to maintain it (since $\alpha=0$ in (\ref{enforced})\,).  But it is worth considering as then only the enforced radial inflow creates shear in the base flow and the question is whether this is enough to generate instability.
The inviscid, linearised governing equation (\ref{master}) is just
\beq
\biggl( \frac{\partial}{\partial t}+\Omega(r) \frac{\partial}{\partial \theta} -\frac{\eta}{r}\frac{\partial}{\partial r} \biggr) \omega = 0
\eeq
which is the `curved' analogue of (\ref{inviscid_gov_eqn}) and again 3rd order rather than the usual 2nd order. As before, the solution can be written down using characteristics as
\beq
\omega(r,\theta,t)=\omega_0 
\biggl(r^2+2 \eta t,\theta+\frac{1}{\eta} \int^r_{\sqrt{r^2+2 \eta t}} \Omega(\ddr)\ddr d\ddr \biggr)
\eeq
where $\omega_0(r,\theta)$ is the initial vorticity.
After a discrete Fourier transform, the $e^{im \theta}$ component is
\beq
\biggl( \frac{1}{r} \frac{\partial}{\partial r}r \frac{\partial}{\partial r} -\frac{m^2}{r^2} \biggr)\psi_m(r,t)
=g(r^2+2 \eta t) \exp \biggl(\frac{im}{\eta} \int^r_{\sqrt{r^2+2 \eta t}} \Omega(\ddr)\ddr d\ddr\biggr)
\eeq
where $g$ is an arbitrary function. For modal growth, $\psi_m(r,t)$ should only depend on $t$ through an $\exp(\sigma t)$ factor for some complex constant $\sigma$ which forces
\beq
g(\xi)= B \exp \biggl( \frac{\sigma \xi}{2 \eta} -\frac{im}{\eta} \int^a_{\sqrt{\xi}} \Omega(\ddr)\ddr d\ddr\biggr)
\eeq
($B$ some constant which can be set to $1$ providing the influx boundary conditions don't force $B=0$) and letting $\psi_m(r,t)=\hat{\psi}_m(r) \exp(\sigma t)$ then
\beq
\biggl( \frac{1}{r} \frac{\partial}{\partial r}r \frac{\partial}{\partial r} -\frac{m^2}{r^2} \biggr)\hat{\psi}_m(r)
=\exp \biggl(\frac{\sigma r^2}{2 \eta}-\frac{im}{\eta} \int^a_r \Omega(\ddr)\ddr d\ddr \biggr).
\eeq
This has the solution
\begin{align}
\hat{\psi}_m(r) &=\int^r_1    G_+(r,\dr;a,m)\exp \biggl(\frac{\sigma \dr^2}{2 \eta}-\frac{im}{\eta} \int^a_\dr \Omega(\ddr)\ddr d\ddr \biggr)  d\dr \nonumber\\
&\hspace{1cm}
+\int^a_r    
G_{-}(r,\dr;a,m)\exp \biggl(\frac{\sigma \dr^2}{2 \eta}-\frac{im}{\eta} \int^a_\dr \Omega(\ddr)\ddr d\ddr \biggr)  d\dr
\label{Greens}
\end{align}
where  
\begin{align}
G_{-}(r,\dr;a,m) &:=\frac{\dr}{2m(a^{2m}-1)}
\biggl(\dr^{m}-\frac{a^{2m}}{\dr^m}\biggr) \biggl(  r^m-\frac{1}{r^m} \biggr) \nonumber \\
G_{+}(r,\dr;a,m) &:=\frac{\dr}{2m(a^{2m}-1)}
\biggl(\dr^{m}-\frac{1}{\dr^m} \biggr)\biggl(r^m-\frac{a^{2m}}{r^m} \biggr) \nonumber 
\end{align}
make up the Green's function built around the 2 (usual) no-normal-velocity boundary conditions that $\hat{\psi}_m(1)=0$ 
and $\hat{\psi}_m(a)=0$. Imposing the third boundary condition of non-slip, $d \hat{\psi}_m(a)/dr=0$, gives the dispersion relation
\beq
\int^a_1 
(\dr^{m+1}-\dr^{1-m})
\exp \biggl(\frac{\sigma \dr^2}{2 \eta}-\frac{im}{\eta} \int^a_\dr \Omega(\ddr)\ddr d\ddr \biggr) 
\, d\dr=0. 
\label{dispersion_tc}
\eeq
For $\Omega=1/r^2$ this is exactly expression (5.3) from \cite{ilin13} which they establish has unstable eigenvalues.  
The energy budget for an infinitesimal 2D disturbance,
\beq
\frac{d}{dt}\biggl\langle \half \bu^2 \biggr\rangle = \eta \biggl\langle \frac{v^2-u^2}{r^2} \biggr\rangle-\eta \int^{2\pi}_0 \half v^2 \biggl|_{r=1} \, d\theta- \biggl\langle ruv \frac{d \Omega}{dr} \biggr\rangle
\label{energy_growth}
\eeq
where $\langle \quad \rangle:= \int^{2\pi}_0 \int^a_1 r dr d \theta$, $u=0$ at $r=1$ and $u=v=0$ at $r=a$, clearly shows
that enhanced growth rates of $O(\sqrt{\eta})$ are only possible if $d\Omega/dr$ is non-zero somewhere in the domain, emphasizing the importance of azimuthal shear. In particular, for $\Omega=1$, the last term drops so that any instability can only have a growth rate of $O(\eta)$ at best. In fact, the dispersion relation appears only to have stable solutions, indicating that boundary inflow and just the shear due to the radial inflow are insufficient to drive any instability.

The $\eta \rightarrow 0$ asymptotic analysis for $\Omega=1/r^2$  mirrors that in the half-plane case with the saddle point at 
$\sigma + im\Omega(r_s)=0$. For the boundary layer scenario, adopting the scalings
\begin{align}
\sigma=&\,\,-im\Omega(a)+\sqrt{\eta \chi /a}\,(\hat{\sigma}_r+i \hat{\sigma}_i)+O(\eta)
\nonumber\\
z:=&\,\,\sqrt{a\chi}\,\frac{(a-r)}{\sqrt{\eta}}
\nonumber
\end{align}
where $\chi:=-\half m \Omega^{'}(a)$  (\,$\Omega^{'}(r)<0$ for cases of interest) and $z$ is a boundary layer variable (as before), retrieves (\ref{dispersion_hp_bl}) which has the unstable eigenvalues $\hat{\sigma}$ listed in Table 1: there is instability with growth rates  $O(\biggl(\sqrt{\eta (-\half m \Omega^{'}(a))/a}\biggr)$.

For the critical layer scenario, the appropriate eigenvalue scaling is again (\ref{crit_scaling}) which given 
$\sigma + im\Omega(r_s)=0$, leads to the expansion
\beq
r_s:= R_s+ \frac{i\delta(\eta) \tilde{\sigma}_r}{m \Omega^{'}(R_s)}+\ldots
\eeq
for the saddle point position where $\sigma_i=-m \Omega(R_s)$ defines $R_s$.
The balance between the leading contribution from the saddle point at $r=r_s$ (first term) and the endpoint $r=a$ (second term) is then
\beq
\frac{\sqrt{-i\eta\, \pi}(r_s^{m+1}-r_s^{1-m})}{\sqrt{-\half m r_s \Omega^{'}(r_s)}}
\exp \biggl(\frac{\sigma r_s^2}{2\eta}-\frac{im}{\eta} \int^a_{r_s} \Omega(r^*)r^* dr^*\biggr)
-\frac{i \eta \,(a^{m}-a^{-m})}{( \sigma_i+m\Omega(a) )}\exp \biggl(\frac{\sigma a^2}{2\eta}\biggr)=0
\nonumber
\eeq
which leads to the frequency condition
\beq
{\rm Arg} 
\biggl\{ \exp \biggl[
i \bigg(  \frac{\sigma_i(R_s^2-a^2)}{2\eta}-\frac{m}{\eta} \int^a_{R_s} \Omega(r^*)r^* dr^*+\frac{\pi}{4} 
\biggr) \biggr] \biggr\} = 0 
\eeq
and growth rate
\beq
\delta \tilde{\sigma}_r=\frac{\eta}{(a^2-R_s^2)} 
\biggl[
\log \biggl(\frac{1}{\eta} \biggr)
+ 
2\log \biggl(\frac{ (R_s^{m+1}-R_s^{1-m})(-\sigma_i-m \Omega(a))\sqrt{2\pi}}{(a^{m}-a^{-m})\sqrt{-m R_s \Omega^{'}(R_s)}} \biggr)
+o(1) \biggr]
\label{crit_layer_asym}
\eeq
Figure \ref{instab_invis_2D} shows that this performs well for numerical calculations with $\eta=10^{-3}$ and $10^{-4}$.
The most unstable boundary layer eigenfunction for $\eta=10^{-3}$ is shown in figure \ref{eig_plot} along with a critical layer mode with $\sigma_i \approx -0.5$ which is typical.

%
%
\subsection{2D Inviscid Swirling Flow with $dZ/dr \neq 0$ \label{swirl_dzdr} }

The asymptotic analysis can easily be extended to treat more general rotation profiles $\Omega(r)$ where the gradient of vorticity is non-zero. The idea here is to assume some small viscosity to induce  a radial inflow (via \ref{enforced}) and then to show that the inviscid instability mechanism is robust to the exact form of the azimuthal flow (e.g. whether it has a vorticity gradient or not).
So, taking the linear, inviscid version of perturbation equation (\ref{master}) and inserting  the usual ansatz $\psi(r,\theta,t)=\psi(r)\exp(i m \theta+\sigma t)$ gives
\beq
\biggl( \sigma+im\Omega(r)-\frac{\eta}{r}\frac{d}{d r} \biggr) {\cal L}\psi = \frac{im \psi}{r}\frac{dZ}{dr}
\label{master1}
\eeq
where 
\beq
{\cal L}:= \frac{d^2}{dr^2}+\frac{1}{r} \frac{d}{dr}-\frac{m^2}{r^2}.
\eeq
The $\eta=0$ problem 
\beq
\biggl( \sigma+im\Omega(r) \biggr) {\cal L}\psi = \frac{im \psi}{r}\frac{dZ}{dr}
\eeq
subject to the 2 boundary conditions $\psi(1)=\psi(a)=0$ seems to have no discrete modal solutions for the profiles $\Omega=r^\alpha$ studied here. This is easily proven for the special choices $\alpha=0 \,\, \& \, -2$ (so $dZ/dr=0$) \citep{ilin13} but is only suggested by numerical evidence generally. This observation is significant because it forces a non-standard singular perturbation analysis in which the additional flow component for $\eta \neq 0$ has to contribute at leading order to help satisfy one of these 2 boundary conditions rather than just `fixing up' the third boundary condition. 

The analysis of the boundary layer instability is exactly the same as the $\Omega=1/r^2$ case because the presence of a vorticity gradient does not effect the boundary layer problem  at leading order. So the growth rate $\Re
e(\sigma) $ scales like $\sqrt{- \half m\Omega^{'}(a) \eta/a }$ with
azimuthal wavenumber $m$, local shear at the boundary $-\Omega^{'}(a)>0$, and radial flow $\eta$. The key point is that the
boundary layer instability only depends on the shear at the inflow boundary.

The analysis for the critical layer instability also proceeds in a similar fashion with one key difference: the problem for the large-scale flow is now complicated by a non-vanishing vorticity gradient but this only has a secondary effect: $dZ/dr$ affects the growth rate at $O(\eta)$ rather than at the leading $O(\eta \log 1/\eta)$ level. The starting point is the ansatz 
\beq
\sigma=i \sigma_i +\delta \tilde{\sigma}_r \sqrt{\chi/R_s}
\eeq
where $\sigma_i=-m \Omega(R_s)$ and $R_s$ is the position of the critical layer, $\chi:=-\half m \Omega^{'}(R_s)$ and $\delta \ll \sqrt{\eta}$.
In the critical layer
\beq
(\, -\delta \tilde{\sigma}_r/\sqrt{\eta}+2i\xi+\partial_\xi\,) \hat{\psi}_{\xi \xi}=0
\eeq
to leading order where $\xi:=\sqrt{R_s \chi/\eta}(r-R_s)$. This has the arbitrarily-normalised solution
\beq
\hat{\psi}_{\xi \xi}= e^{\delta \tilde{\sigma}\xi/\sqrt{\eta}-i\xi^2}.
\label{crit_rot}
\eeq 
Matching this to $\psi_{rr}$ outside the critical layer requires the WKB solution
\beq
\psi^{WKB}=\frac{-R_s \chi \, \eta}{m^2r^2[\,\Omega(r)-\Omega(R_s)\,]^2}
\exp\biggl[ 
\frac{im}{\eta} \int^r_{R_s} \dr(\, \Omega(\dr)-\Omega(R_s) \,) d\dr+\sqrt{\frac{\chi}{R_s}}\frac{\delta \tilde{\sigma}(r^2-R_s^2)}{2\eta}
\biggr]
\eeq
to exist there.
There are two other `outer' large-scale solutions either side of critical layer which solve
\beq
[ \, \Omega(r)-\Omega(R_s)\, ] {\cal L} \psi  = \frac{\psi_0}{r}\frac{dZ}{dr}
\eeq
and together accommodate the jump conditions
\beq
[u]^{+}_{-}=0 \quad \& \quad [v]^{+}_{-}=\sqrt{\frac{\pi R_s \chi}{i \eta}}
\eeq
across the critical layer (found by integrating (\ref{crit_rot}) twice) and the no-normal velocity boundary conditions at $r=1$ and $r=a$.
Then, as in section \ref{hp_crit}, balancing the large-scale component of the azimuthal velocity with that from the WKB solution at $r=a$ furnishes the growth rate and dispersion relation for the frequency.  Only the  higher order $O(\eta)$ part of the growth rate 
depends on $dZ/dr$ or indeed $\Omega(r)$ whereas the leading $O(\eta \log 1/\eta)$ part does not (note $\chi$ which contains $\Omega^{'}(R_s)$ needs to be non-zero but otherwise scales out).

%
%
\subsection{Viscous Asymptotics for $0<\eta \ll 1$ for the boundary layer instability \label{swirl_viscous_bl}}

The inviscid asymptotics can be generalised to include viscosity which we do here just for the more dangerous boundary layer instability as this defines the viscous threshold for instability. In the absence of any other physics, the radial inflow is set by the viscosity present via (\ref{enforced}). However, to extract the scaling law for the viscous instability threshold for more generally-maintained radial inflows, we ignore this connection and assume $\eta$ can vary independently of $Re$. It is a simple matter to reinstate this connection later to establish stability or instability for a purely hydrodynamic situation.  On this basis, the linearised disturbance equation for $\psi=\psi(r) e^{i m \theta+\sigma t}$ is  
\beq
\biggl( \sigma+im\Omega(r)-\frac{\eta}{r} \frac{d}{d r} \biggr)\opL \psi= \frac{i m}{r} \frac{dZ}{dr}\psi+\frac{1}{Re} \opL^2 \psi
\eeq
Setting $\sigma+im \Omega(a)=\eps \hat{\sigma}$ with $\hat{\sigma}=O(1)$
and where $\eps$ is the width of the boundary layer at $r=a$, and balancing 
the frequency, inflow and viscous terms \citep{gal10}
\beq 
\hat{\sigma} /\eps \sim \eta/\eps^3 \sim \frac{1}{\eps^4 Re} 
\eeq
leads to $\eps=Re^{-1/3}$ and $\eta=\hat{\eta}\eps^2$ where $\hat{\eta}=O(1)$. 
Defining the boundary layer variable
\beq
y:=\frac{ \sqrt{\chi} }{ \sqrt{\hat{\eta}/a} }\biggl(\frac{a-r}{\eps} \biggr)
\eeq
with, recall,  $\chi:=-\half m \Omega^{'}(a)$ and adopting the standard boundary layer decomposition
$\psi=\psi_0+\hat{\psi}_0+ \eps (\psi_1+\hat{\psi}_1)+\ldots$, leads to the boundary layer equation at $r=a$ (the boundary layer at $r=1$ is very weak and can be ignored at leading order) 
\beq \biggl[
  \frac{d^2}{dy^2}-N\frac{d}{dy}-N(\hat{\sigma}_0 +2 i y)\biggr]
\frac{d^2 \hat{\psi}_0}{dy^2}=0 
\label{bl_eqn}
\eeq 
with 
\beq 
N:= \frac{
  (\hat{\eta}/a)^{3/2}}{\chi^{1/2}} \qquad \& \qquad 
  \hat{\sigma}_0:=\hat{\sigma}/(N \chi^2)^{1/3}.
\eeq 
This is exactly equation (36) in \cite{gal10} if their `a'$:= -\hat{\sigma}_0$. 
The solution which vanishes for $y \rightarrow \infty$ is 
\beq 
\frac{d^2
  \hat{\psi}_0}{dy^2}=e^{\half N y} Ai \biggl[
  \frac{N^{1/3}\hat{\sigma}_0}{(2i)^{2/3}}+\frac{N^{4/3}}{4(2i)^{2/3}}+(2i)^{1/3}
  N^{1/3} y \biggr] 
\eeq where $Ai$ is the Airy function. Imposing the further non-slip condition  $d \hat{\psi}_0/dy=0$ at $y=0$ 
defines the dispersion relation
\beq
\int^{\infty}_0e^{\half N y} Ai \biggl[
  \frac{N^{1/3}\hat{\sigma}_0}{(2i)^{2/3}}+\frac{N^{4/3}}{4(2i)^{2/3}}+(2i)^{1/3}
  N^{1/3} y \biggr]\, dy=0.
\label{dispersion}
\eeq 
The onset of instability is found for $N=N_c$ where $\Re e(\hat{\sigma}_0)$ passes
through zero (in the positive direction).  Numerically, it is better to
solve (\ref{bl_eqn}) directly as an eigenvalue problem rather
than treat this integral equation (see Appendix A).  We find $N_c=4.57557$ and $\hat{\sigma}_0=-5.63551 i$
which is consistent with \cite{gal10}: the threshold radial flow rate for instability is
\beq
\eta_{crit}=a N_c^{2/3} \chi^{1/3} Re^{-2/3}.
\label{eta_crit}
\eeq
Since the viscously-induced radial flow in an accretion disk is $O(Re^{-1})\ll \eta_{crit}$,
the instability is not expected to be triggered unless it is substantially subcritical. Whether this is the case will be considered below in \S \ref{nonlinearity} after we examine the effect of introducing some extra physics to the instability.

%
%
%
%
\section{2D Compressible Swirling Flow with Radial Inflow \label{compress} }
%
%

From the astrophysical perspective, an important ingredient so far missing from the models considered  is compressibility.
Adding this extra physics also provides an opportunity to test the robust of the boundary inflow instability. As the simplest model to include compressibility, we consider the fluid to be an isothermal perfect gas  so that pressure $p$
and density $\rho$ are simply related by $p=c^2\rho$ where $c^2$ is the isothermal speed of sound. The divergence-free flow
\beq 
\bU= -\frac{\eta}{r}\br+ \frac{1}{r}\btheta 
\label{Comp_basic}
\eeq 
is a steady solution of the inviscid 2D momentum equations
\begin{align}
U \frac{dU}{dr}-\frac{V^2}{r}+\frac{1}{\rho_0} \frac{d p_0}{dr} &= -\frac{1}{r^3} \label{Comp_1}\\
U \frac{dV}{dr}+\frac{U V}{r} \hspace{1.3cm}                      &= 0              \label{Comp_2}
\end{align}
where $\rho_0(r)=\rho(r)/\rho(1)$ so the density is non-dimensionalised by its value at the inner radius $r^*$ and $p_0(r)=p(r)/(\rho(1) r^{*2} \Omega^{*2})$. This non-dimensionalisation means $p_0=\delta \rho_0$ where $\delta:=c^2/(r^{*2} \Omega^{*2})$ and $\delta \rightarrow \infty$ is the incompressible limit. In a thin Keplerian disk, however, the basic rotation speed is highly supersonic so  $\delta \ll 1$  (e.g. p87, \cite{Frank}) {\em and} $\eta \sim Re^{-1} \ll \delta$ so the question is whether the instability still operates for $\eta \ll \delta  \ll 1$.
  
In the laboratory, the body force on the right hand side of (\ref{Comp_1}) would be absent leaving only the pressure gradient to drive the centripetal acceleration. In an accretion disk, however, the gravitational attraction to the central object plays this role and the pressure gradient only exists to balance the much smaller radial advection term. To model this latter situation,  a body force is included (albeit here with different radial dependence so $\Omega:=1/r^2$) to support the irrotational basic state needed to satisfy (\ref{Comp_2}) in the absence of viscosity.  The pressure gradient is then only $O(\eta^2)$. With this, the density drifts very slowly as mass balance requires
\beq
\frac{\partial \rho_0}{\partial t}=\frac{\eta}{r} \frac{\partial \rho_0}{\partial r} \sim O\biggl(\frac{\eta^3}{\delta} \biggr) \rho_0.
\eeq
Given the instability being studied here develops over a much shorter $O(1/\sqrt{\eta})$ timescale, both this secular change and the very small pressure gradient can be ignored i.e. it is valid to set $\rho_0=1$ and $p_0=\delta$\footnote{Formally, this is because $\mathbf{u_1}.\nabla \rho_0 \ll \mathbf{U}.\nabla \rho_1$ and $\rho_1/\rho_0^2 \,dp_0/dr \ll 1/\rho_0\, dp_1/dr$.}. After doing this, the 2D linearised disturbance equations are then 
\begin{align}
\biggl(\sigma+im\Omega(r) -\frac{\eta}{r} \frac{d}{d r} \biggr) u_1 &= -\frac{\eta}{r^2} u_1+2 \Omega v_1-\frac{d p_1}{d r}  \label{2D_1}\\
\biggl(\sigma+im\Omega(r)-\frac{\eta}{r} \frac{d}{d r} \biggr) v_1 &= \,\,\, \frac{\eta}{r^2} v_1 -Z u_1-\frac{im}{r}p_1  \label{2D_2}\\
\biggl(\sigma+im\Omega(r)-\frac{\eta}{r} \frac{d}{d r} \biggr) p_1 &=- \delta \nabla \cdot \mathbf{u_1} \label{2D_3}
\end{align}
where the disturbance fields $\mathbf{u_1}$ \& $p_1$ are taken proportional to  $e^{i m \theta+\sigma t}$  ($p_1=\delta \rho_1$). Assuming the boundary layer scalings of \S \ref{swirl_dzdr_0},
\beq
\sigma   = -im \Omega(a)+\sqrt{\frac{-m \eta \Omega^{'}(a)}{2a}} \hat{\sigma}+\ldots, \qquad
\xi      = (a-r)\sqrt{\frac{-ma \Omega^{'}(a)}{2\eta}}, 
\eeq
\beq
(u_1,v_1,p_1)   = \biggl(\sqrt{\eta} \hat{u}, \sqrt{\frac{-m a\Omega^{'}(a)}{a}} \hat{v}, \sqrt{\eta} \hat{p} \biggr)
\eeq
The equations become to leading order in $\eta$
\begin{align}
0 &= 2 \Omega \hat{v}+\frac{d \hat{p}}{d \xi},  \label{2D_new1}\\
-\half m \Omega^{'}(a) \biggl(\hat{\sigma}+2i \xi+\frac{d}{d \xi} \biggr) \hat{v} &= -Z(a) \hat{u}-\frac{im}{a} \hat{p},  \label{2D_new2}\\
\biggl(\hat{\sigma}+2i \xi+\frac{d}{d \xi} \biggr) \hat{p} &=- \frac{\delta}{a \eta} \biggl( -\frac{d \hat{u}}{d \xi}+\frac{im \hat{v}}{a} \biggr), \label{2D_new3}
\end{align}
which can be reduced to the equation
\beq
\biggl( \hat{\sigma}+2i \xi+\frac{d}{d \xi} \biggr) \frac{d \hat{v}}{d \xi}= \frac{\eta}{\delta} \biggl[\frac{2a Z(a)}{m \Omega^{'}(a)} \biggr] \biggl(\hat{\sigma}+2i \xi+\frac{d}{d \xi} \biggr) \hat{p}.
\eeq
This suggests that the  incompressible limit  $\delta \rightarrow \infty$ which recovers the boundary layer equation (\ref{bl_asym})\,) still holds provided $\eta \ll \delta$ and this is confirmed by numerical computations: for example, see Table 3.

%
%
\begin{table}
\begin{center}
\begin{tabular}{@{}lrr@{}}
 $\delta$ & $\quad \Im m(\hat{\sigma})$ \quad & $\Re e(\hat{\sigma})$ \\
          &                                  &                       \\
$\infty$             &   -4.7795 & 0.9114\\
                     &           &       \\
$100  $              &   -4.7795 & 0.9114\\
$1 $                 &   -4.7796 & 0.9092\\
$10^{-1}$            &   -4.7784 & 0.9290\\
$10^{-2}$            &   -4.7793 & 0.9081\\
$10^{-3}$            &   -4.7748 & 0.9052\\
$10^{-4}$            &   -4.0979 & 1.1839\\
\end{tabular}
\end{center}
\caption{The most unstable eigenvalue $\hat{\sigma}$ for the eigenvalue problem (\ref{2D_1})-(\ref{2D_3}) with $\eta=10^{-4}$ and various degrees of compressibility. The presence of $\delta$ only becomes significant when it is $\leq O(\eta)$.}
\label{compressible}
\end{table}

%
%
%
%
\begin{figure}
\begin{center} 
\resizebox{0.7\textwidth}{!}{\includegraphics[angle=0]{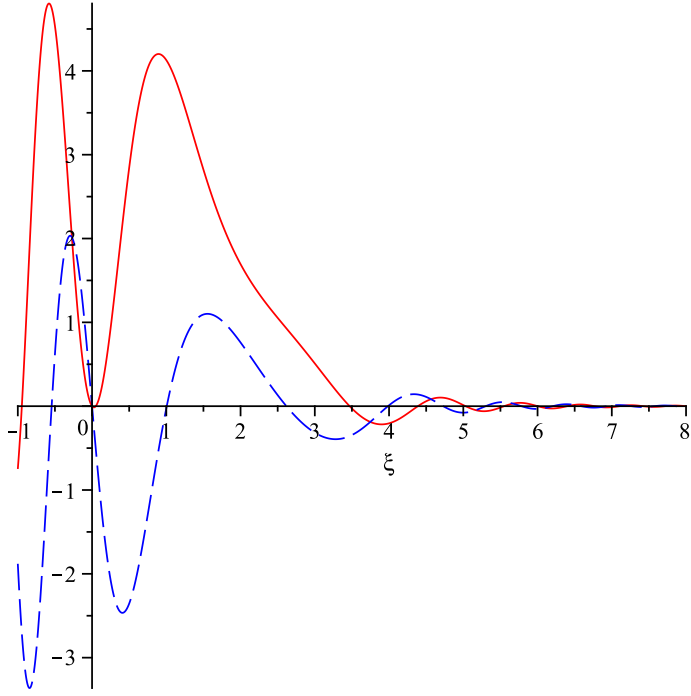}}
\end{center}
\caption{The leading growing $v$ eigenfunction calculated using (\ref{eigenfunction}) in Maple for $m=1$, $k=0.3$, $\alpha=-3/2$ corresponding to $\hat{\sigma}=0.64021-4.59526i$ (real/imaginary part is red solid/blue dashed line).}
\label{maple}
\end{figure}
%

%
%
%
\begin{figure}
\begin{center} 
\psfrag{X}{{\Large $\hat{\sigma}_r$}}
\psfrag{Y}{{\Large $\mu$}}
\resizebox{1.05\textwidth}{!}{\includegraphics[angle=0]{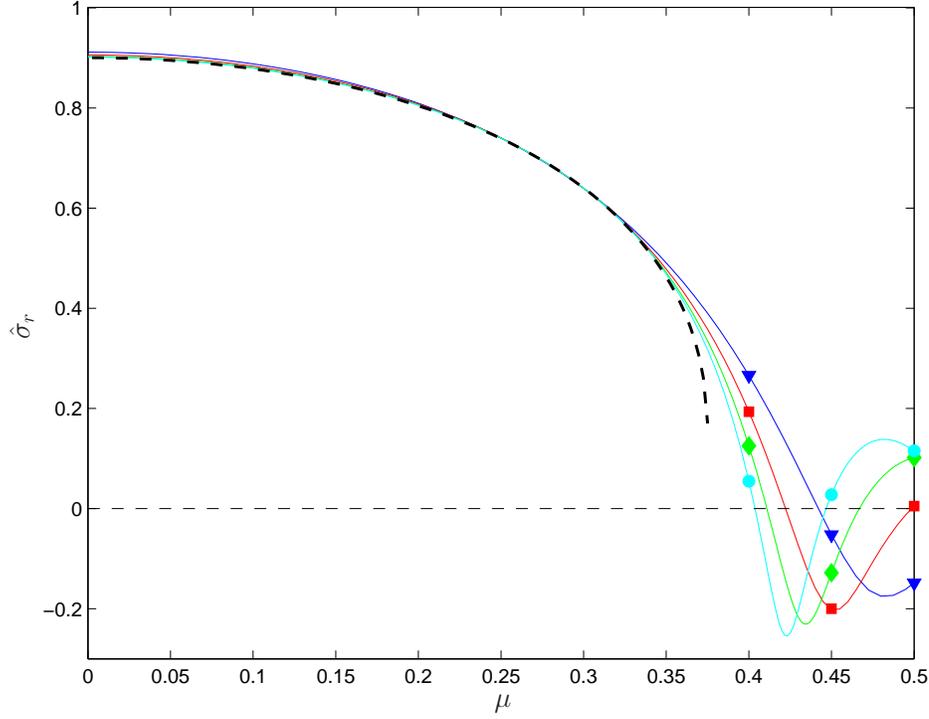}}
\end{center}
\caption{Leading inviscid 3D instability for $\alpha=-1.5$, $m=1$, $a=2$  with scaled growth rate $\hat{\sigma}_r=\sigma_r/\sqrt{- \half m \eta \Omega^{'}(a)/a}$ plotted against $\mu$, the degree of 3 dimensionality, as computed using  the full eigenvalue code - $\eta=10^{-4}$ uppermost (blue) line with triangles, $\eta=10^{-5}$ second uppermost (red) line withj squares, $\eta=10^{-6}$ third uppermost (green) line with diamonds and $\eta=10^{-7}$ lowest (cyan) line with filled circles - where $N=1000$ proves sufficient (The markers on each line actually show the $N=500$ stability results for $\mu= 0.4, 0.45$ and $0.5$ to demonstrate convergence). The leading instability is also shown for the boundary layer problem (\ref{1} \& (\ref{2}) as a dashed black line ($N=4000$). The boundary layer scaling works well - the full eigenvalue prediction smoothly converges to the boundary layer prediction as $\eta \rightarrow 0$ - for $\mu \lesssim 0.3$ but beyond this the $\sqrt{\eta}$ scaling is no longer correct. This plot demonstrates that increasing 3 dimensionality acts to suppress the 2D instability.}
\label{blayer}
\end{figure}

%
%
%
%
\section{3D Linear Instability \label{section_3D}}
%

We now study 3 dimensional disturbances. Adopting the ansatz  $\bu(r,\theta,z,t)=\bu(r) e^{i(m \theta+kz)+\sigma t}$, the linearized inviscid governing equations are 
\begin{align}
\biggl(\sigma+im\Omega(r) -\frac{\eta}{r} \frac{d}{d r} \biggr) u+\frac{\eta}{r^2} u-2 \Omega v+\frac{d p}{d r} &=0, \label{3D_1}\\
\biggl(\sigma+im\Omega(r)-\frac{\eta}{r} \frac{d}{d r} \biggr) v-\frac{\eta}{r^2} v +Z u+ \frac{im}{r}p &=0, \label{3D_2}\\
\biggl(\sigma+im\Omega(r)-\frac{\eta}{r} \frac{d}{d r} \biggr) w \hspace{2cm}+ik p &=0, \label{3D_3}\\
\frac{1}{r} \frac{d (ru)}{d r}+\frac{im}{r}v + ik w &= 0. \label{3D_4}
\end{align}
In contrast to the 2D situation, discrete neutral modes start to emerge for $k \neq 0$ in the absence of radial flow. For example, in the case of axisymmetric modes the equations (\ref{3D_1})-(\ref{3D_4})  boil down to the 2nd order ODE for $u$
\beq
\frac{d^2 u}{dr^2}+\frac{1}{r} \frac{du}{dr}-\biggl(\frac{2k^2}{\sigma^2}Z \Omega+k^2+\frac{1}{r^2} \biggr)u=0.
\eeq
There are two profiles $\Omega(r)$ which  make this just Bessel's equation: $\Omega=1$ $(\alpha=0$) and $\Omega=1/r$ ($\alpha=-1$). In the former, uniform rotation case, eigensolutions are  axisymmetric Poincar\'{e} modes \citep{Greenspan1968}. In the latter case, the general solution is $u=A J_{\nu}(ikr)+B Y_{\nu}(ikr)$
where $\nu:=\sqrt{1+2k^2/\sigma^2}$ with the dispersion relation (since $u(1)=u(a)=0$)
\beq
J_{\nu}(ik)\,Y_{\nu}(ika)
\,=\,
Y_{\nu}(ik)\,J_{\nu}(ika).
\eeq
This has purely imaginary eigenvalues $\sigma=i \lambda$ with $\lambda^2 <2k^2$. The issue is of course whether the emergence of discrete modes affect the instability. The answer is no as we now demonstrate again focussing on the stronger boundary layer instability.

%
%
%
\subsection{3D Extension of the Boundary Layer Instability}
Working with the primitive variables, the appropriate scalings to capture the boundary layer instability are
\beq
\sigma = -im \Omega(a)+\hat{\sigma} \sqrt{\frac{-m \eta \Omega^{'}(a)}{2a}}+\ldots, \qquad
\xi :=(a-r)\sqrt{\frac{-ma\Omega^{'}(a)}{2\eta}}  
\label{growthrate}
\eeq
\beq
(\,u,v,w,p\,) = \biggl(\,\sqrt{m \eta}\, \hat{u}, v, \hat{w}\sqrt{\frac{2\Omega(a)}{Z(a)}}, \sqrt{\frac{\eta}{m}}\, \hat{p}\,\biggr)  
\eeq
in the boundary layer $\xi=O(1)$ where 
\beq
\mu := \frac{k}{m}
\eeq
measures the degree of 3-dimensionality. With these, the equation set (\ref{3D_1})-(\ref{3D_4}) reduces to
\begin{align}
      0 & = 2\Omega v+\sqrt{-\half a \Omega^{'}(a)} \frac{d \hat{p}}{d \xi}, \label{bl_1}\\
\sqrt{\frac{-\Omega^{'}(a)}{2a}} 
\biggl(\hat{\sigma}+2 i \xi+\frac{d}{d \xi} \biggr)v &= -Z(a) \hat{u}-\frac{i}{a} \hat{p},\label{bl_2}\\
\sqrt{\frac{2\Omega(a)}{Z(a)}}\sqrt{\frac{-\Omega^{'}(a)}{2a}} \biggl(\hat{\sigma}+2 i \xi+\frac{d}{d \xi} \biggr) \hat{w} &= -i \mu \hat{p}, \label{bl_3}\\ 
-\sqrt{-\half a \Omega^{'}(a)} \frac{d \hat{u}}{d \xi}+\frac{iv}{a}+ i \mu\sqrt{\frac{2 \Omega(a)}{Z(a)}} \hat{w} &=0
\label{bl_4}
\end{align}
at leading order respectively for small $\sqrt{ m\eta}$. Eliminating $\hat{u}$ and $\hat{p}$ then gives
\begin{align}
\biggl(\hat{\sigma}+2 i \xi+\frac{d}{d \xi} \biggr)\frac{d       v}{d\xi} &=-i \gamma \hat{w},  \label{1}\\
\biggl(\hat{\sigma}+2 i \xi+\frac{d}{d \xi} \biggr)\frac{d \hat{w}}{d\xi} &= \,\,\,\,i \gamma v -2i\hat{w},       \label{2}
\end{align}
which is the 3D generalisation of the single boundary layer equation in 2D (see (\ref{bl_asym})\,)
where 
\beq
\gamma:=2a\mu \frac{\sqrt{2(\alpha+2)}}{(-\alpha)}.  \label{gamma}
\eeq
This reduced system only involves $m$ through $\mu$ and is valid for $m \ll 1/\eta$. So, for $\mu$ held fixed, the growth rate scales with  $\sqrt{m}$ (see the rescaling in (\ref{growthrate})\,) across the astrophysical regime of $1 \ll m \ll 1/\eta$.

In the particular case treated by \cite{ilin13} ($\alpha=-2$ so $\gamma=0$) adding an axial wavenumber does nothing to the 2D instability provided $\eta k$ stays small. In the general case of interest $-2 \leq \alpha < 0$ and $\mu >0$ (clearly the stability problem is symmetric under the transformation $(\mu,v,\hat{w}) \rightarrow (-\mu,v,-\hat{w})$ so only $\mu >0$ needs be considered) a solution is available in terms of Kummer functions.
The equations (\ref{1}) and (\ref{2}) can be combined to a single equation for $v$ 
\beq
(\hat{{\cal G}}^2+2i \hat{{\cal G}}-\gamma^2) v=0
\eeq
where 
\beq
\hat{{\cal G}}:=\frac{d^2}{d \xi^2}+(\hat{\sigma}+2 i \xi)\frac{d }{d\xi}.
\eeq
This can be factored as $(\hat{{\cal G}}+4i\Gamma_1)(\hat{{\cal G}}+4i\Gamma_2) v=0$ where 
$\Gamma_1:=\quart(1- \sqrt{1-\gamma^2})$ and $\Gamma_2:=\quart (1+\sqrt{1-\gamma^2})$. Under the change of variable
$z:=\quart i(\hat{\sigma}+2i \xi)^2$, each factor is just Kummer's differential equation,
\beq
z\frac{d^2 v}{d \xi^2}+(\half-z) \frac{dv}{d \xi}-\Gamma_j v=0  \qquad j=1,2
\eeq
%
%
%
%
\begin{figure}
\begin{center} 
\psfrag{X}{{\Large $\hat{\sigma}_r$}}
\psfrag{Y}{{\Large $\sigma_i$}}
\resizebox{0.9\textwidth}{!}{\includegraphics[angle=0]{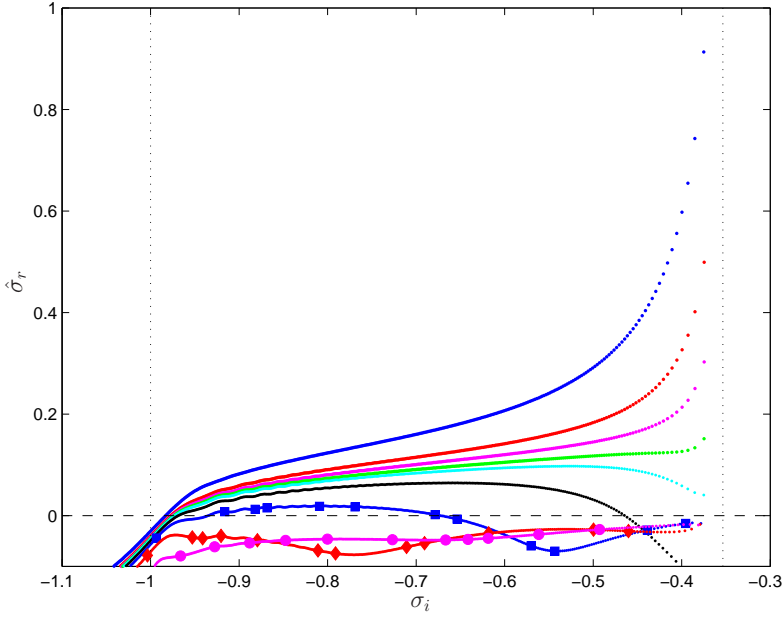}}
\end{center}
\caption{Inviscid 3D instabilities for $\alpha=-1.5$, $m=1$, $a=2$ and $\eta=3 \times 10^{-4}$  with scaled growth rate $\hat{\sigma}_r=\sigma_r/\sqrt{-m \eta \Omega^{'}(a)/(2a)}$ plotted against frequency $\sigma_i$ for various values of $\mu$. The right (left) vertical dashed line is $\sigma_i=-m\Omega(a)$ ($\sigma_i=-m\Omega(1)$).
The eigenvalues are calculated from a  full 3D eigenvalue calculation with $N=2400$ for (in order downwards from the uppermost 2D eigenvalues): $\mu=0$ (blue dots); $\mu=0.35$ (red dots); $\mu=0.4$ (magenta dots); $\mu=0.43$ (green dots); $\mu=0.45$ (cyan dots); $\mu=0.5$ (black dots); $\mu=0.6$ (blue dots with squares); $\mu=0.75$ (red dots with diamonds) and $\mu=1$ (magenta dots with circles). This plot demonstrates that increasing 3 dimensionality suppresses the 2D instability (in fact completely by $\mu=0.75$ here). Notice that numerical errors start to creep into the eigenvalues whose eigenfunctions have critical layers far from the inflow boundary (the developing corrugations in the curves) as $\mu$ increases (not unexpected as the numerical truncation $N$ is only $O(1/\eta)$). }
\label{plot_3e4}
\end{figure}

The solution for $v$ (and therefore indirectly $\hat{w}$) which decays exponentially for $\xi \rightarrow \infty$ and $\hat{\sigma}_r>0$ is
\beq
v=e^z({\cal A} U(\half-\Gamma_1,\half,-z)+{\cal B} U(\half-\Gamma_2,\half,-z)
\eeq
where $U$ is the multivalued Kummer's function  (see \S 13.2.25, \cite{DLMF}) and ${\cal A}$ and ${\cal B}$ are constants. If these are not both to vanish when the further boundary conditions  $v=\hat{w}=0$ are imposed  at $\xi=0$, then the product
\beq
U(\half-\Gamma_1,\half,-\quart i \hat{\sigma}^2) U(\half-\Gamma_2,\half, - \quart i \hat{\sigma}^2)=0
\eeq
which defines unstable ($\hat{\sigma}_r>0$) eigenfrequencies $\hat{\sigma}$.  
The eigenvalues for $\mu >0$ which smoothly connect with those at $\mu=0$ are given by $U(\half-\Gamma_1,\half,-\quart i \hat{\sigma}^2)=0$ since $\mu \rightarrow 0$ implies $\Gamma_1 \rightarrow 0$ and convergence to the 2D equation $\hat{{\cal G}} v =0$.
These eigenfunctions can be plotted using 
\beq
v=\begin{cases}
2\frac{{\rm Gamma}(\half)}{{\rm Gamma}(1-\Gamma_1)} e^z M(\half- \Gamma_1,\half,-z)-e^z U(\half-\Gamma_1,\half,-z) 
& 0\, \leq \, \xi\, <\, -\half( \hat{\sigma}_r+\hat{\sigma}_i)\\
\\
e^z U(\half-\Gamma_1,\half, -z) & -\half( \hat{\sigma}_r+\hat{\sigma}_i) \leq \xi \\
\end{cases}
\label{eigenfunction}
\eeq
and $\hat{w}=2 \Gamma_1 v/\gamma $ (\,with $z:=\quart i(\hat{\sigma}+2i \xi)^2$ and `Gamma' indicating the Gamma function\,) which compensates for the branch cut along the negative real axis in $U$ routinely imposed by packages such as Maple ($M$ is the single-valued Kummer function: \S 13.1.2, \cite{Abram}): see Figure \ref{maple}.


The equations (\ref{1}) and (\ref{2}) can also be directly treated numerically (see Appendix A). This boundary layer approach works well for the dominant instabilities as $\mu$ increases from zero showing how they are systematically suppressed until their growth rates are comparable with the interior critical layer modes: see figure \ref{blayer} which shows this for the leading instability. At this point (which is $\mu \approx 0.4$ for $a=2, \eta=10^{-4}, \alpha=-1.5$) the  full 3D eigenvalue (\ref{3D_1})-(\ref{3D_4}) must be solved (see Appendix A) which shows the complete suppression of the all the instabilities by $\mu=1$: see figure \ref{plot_3e4}. 

The general conclusion is that 3D disturbances are less unstable than 2D disturbances under boundary inflow. In fact since the boundary layer instability depends only on the shear at the boundary, this could have been anticipated by using a Squires tranformation to map a 3D disturbance to a more unstable 2D disturbance \citep{Squires1933}. Normally, such a transformation fails in a rotating system due to curvature terms but here these are marginalised by the dispersion relation being only sensitive to the shear at the boundary.

%
%
%
\begin{figure}
\begin{center} 
\psfrag{X}{{\Large $\eta$}}
\psfrag{Y}{{\Large $\sigma_r$}}
\resizebox{0.9\textwidth}{!}{\includegraphics[angle=0]{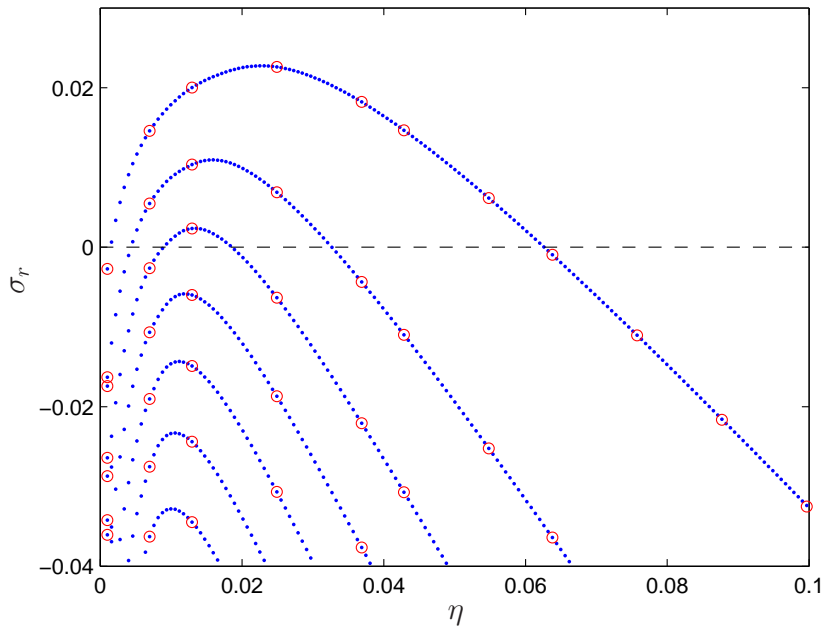}}
\end{center}
\caption{The growth rate $\sigma_r$ verses radial flow $\eta$ for linear 2D perturbations with $m_0=1$ on the 1D basic state (\ref{basic}) for $\alpha=-3/2$, $a=2$ and $Re=10^5$. The blue dots indicate the  eigenvalues using 200 radial modes to represent each component of the perturbation velocity field and the red circles are a sampling of results of using 400 to confirm convergence. The leading perturbation (top curve) becomes unstable - $\sigma_r>0$ - at about $Re=5 \times 10^3$ and remains the only instability at $Re=10^4$.}
\label{stab2D}
\end{figure}

%
%
%
\begin{figure}
\psfrag{X}{{\Large $\hat{\sigma}_r$}}
\psfrag{Y}{{\Large $\sigma_i$}}
\setlength{\unitlength}{1cm}                                                    
\begin{picture}(12,12) 
\put(2,4){\includegraphics[width=10cm]{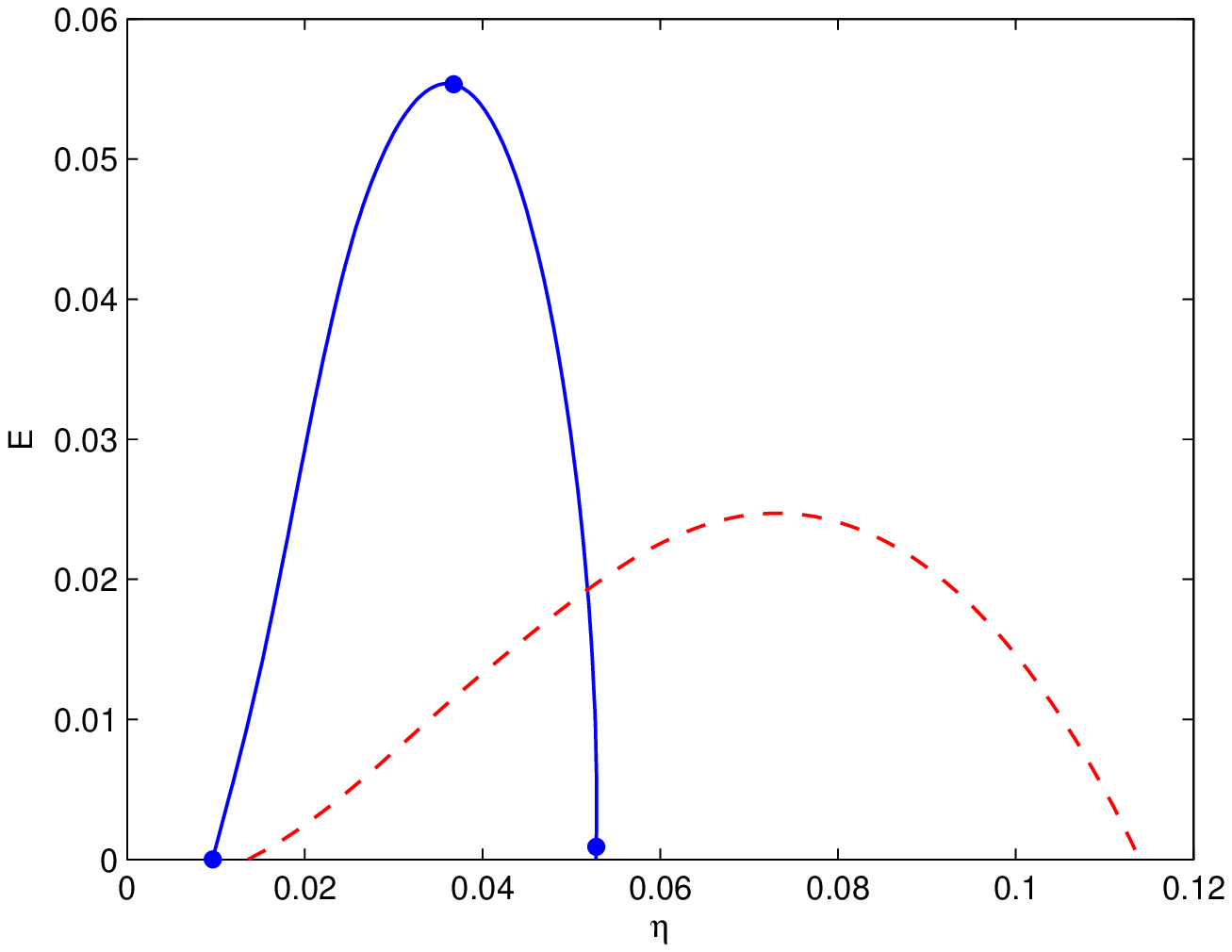}} 
\put(0.5,0){\includegraphics[width=4cm]{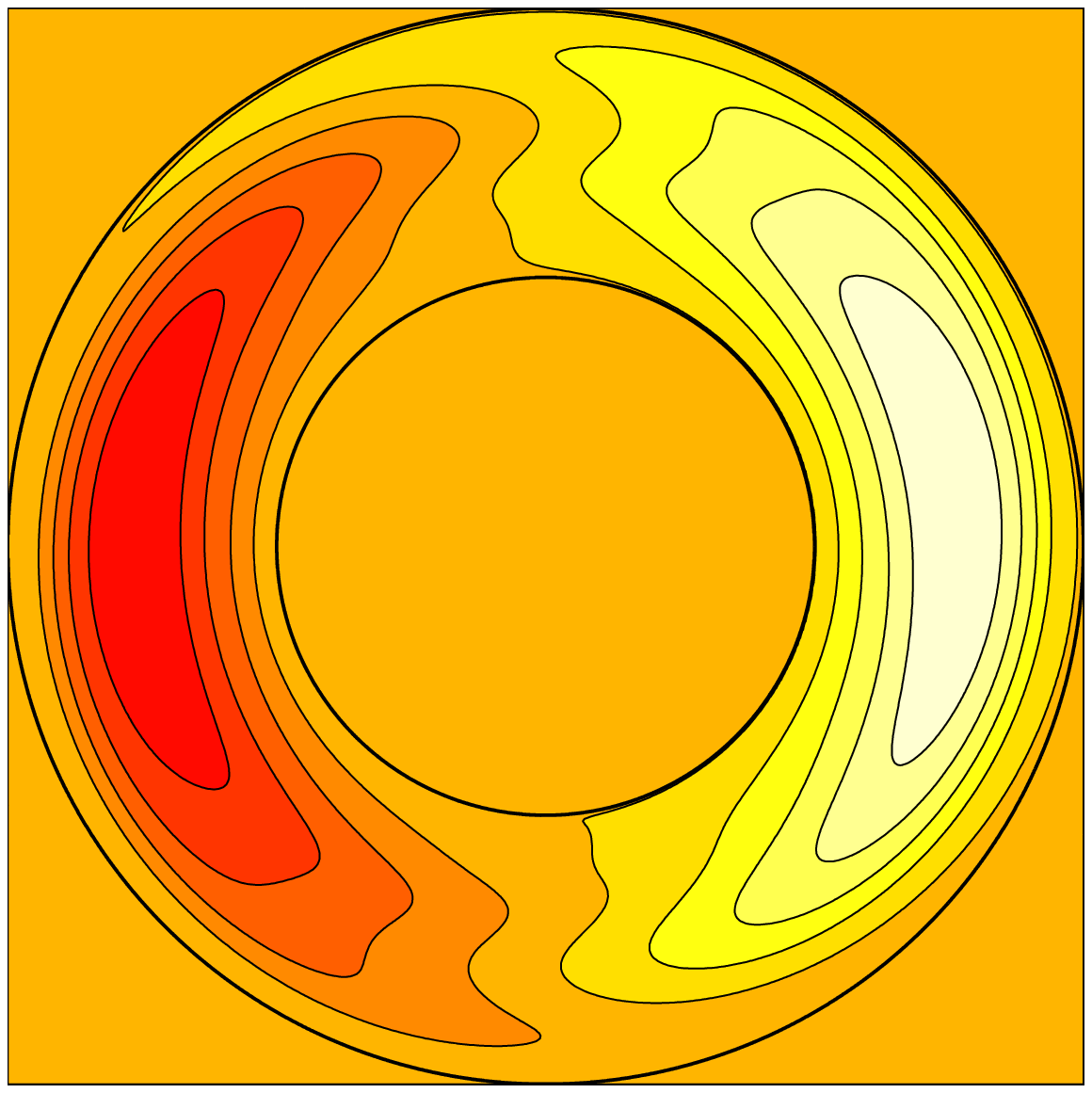}}                                                        
\put(5,0){\includegraphics[width=4cm]{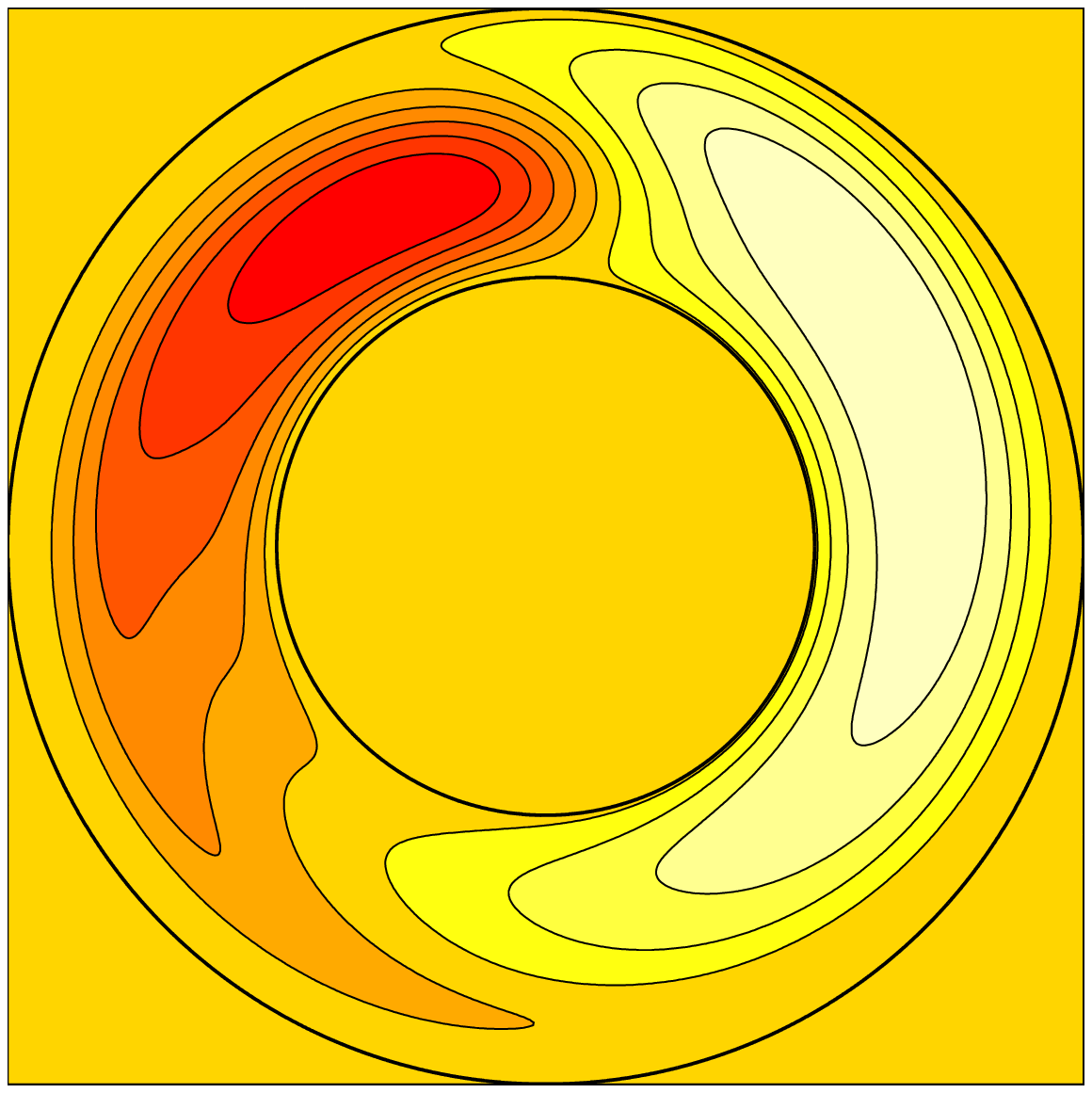}}
\put(9.5,0){\includegraphics[width=4cm]{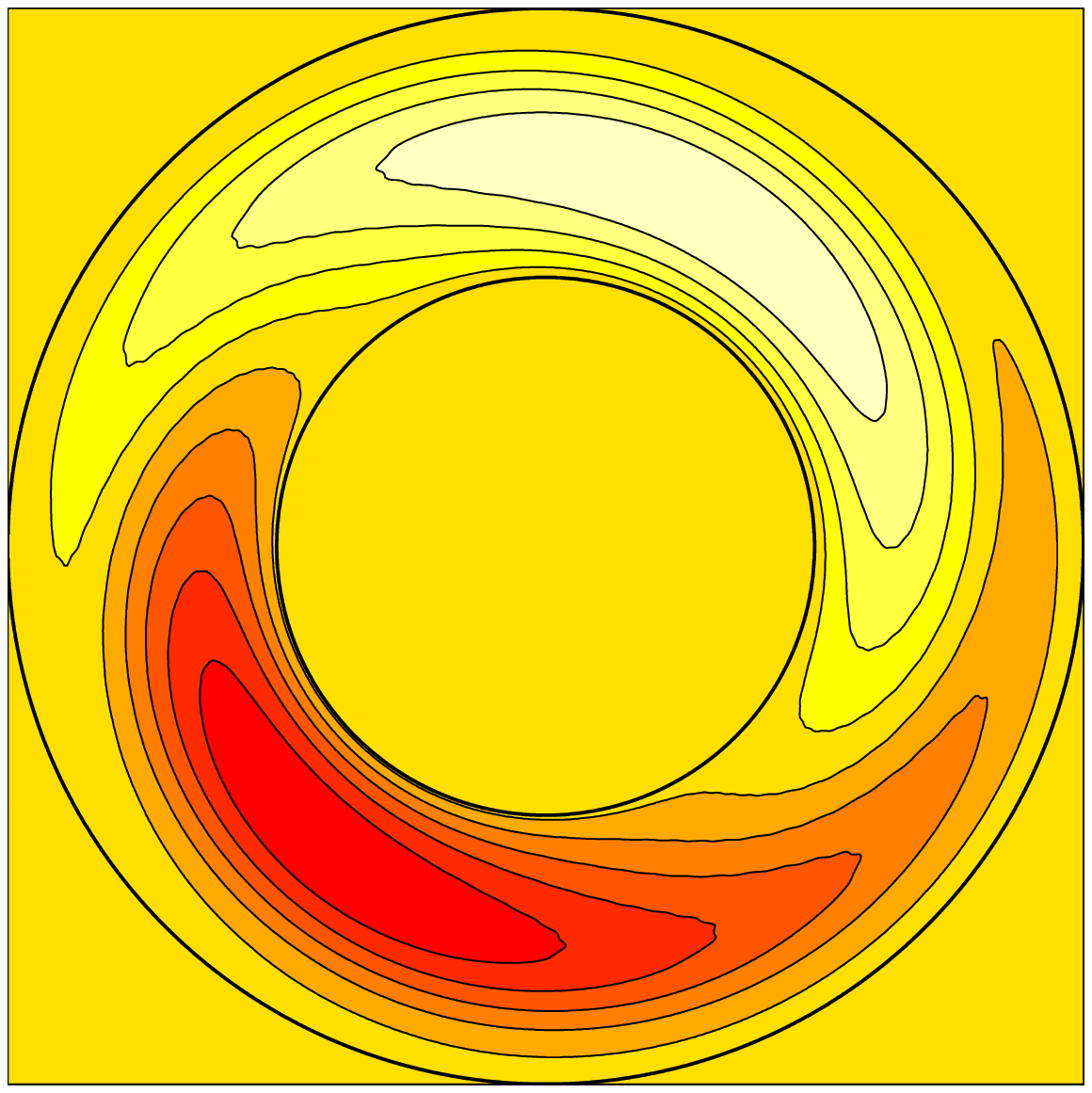}}
\end{picture}       
\caption{Upper: kinetic energy per unit length in $z$ of the 2D perturbation, $E$, verses the radial flow $\eta$ with the solid blue line representing the $m_0=1$ solution branch and the dashed red line the $m_0=3$ branch for $Re=10^4$, $\alpha=-3/2$ and $a=2$ (resolution $M=20$ and $N=60$). Lower: the 2D streamfunction for the $m_0=1$ solution branch plotted (left to right)  at $\eta=9.64 \times 10^{-3}$, $3.67 \times 10^{-2}$ and $5.28 \times 10^{-2}$ (shown as dots on the solution curve). In each, ten contours are plotted (dark/red-to-light/white being -ve to +ve) and at $\eta=3.67 \times 10^{-2}$, $\max \psi=0.0478$ and $\min \psi=-0.0662$ (in all, the outer colour/shading indicates the zero contour).
 }
\label{soln2D}
\end{figure}

%
%
%
\begin{figure}
\psfrag{X}{$\eta$}
\psfrag{r}{$r$}
\psfrag{z}{$\overline{v}$}
\setlength{\unitlength}{1cm}                                                    
\begin{picture}(12,6) 
\put(-0.3,0){\includegraphics[width=7.4cm]{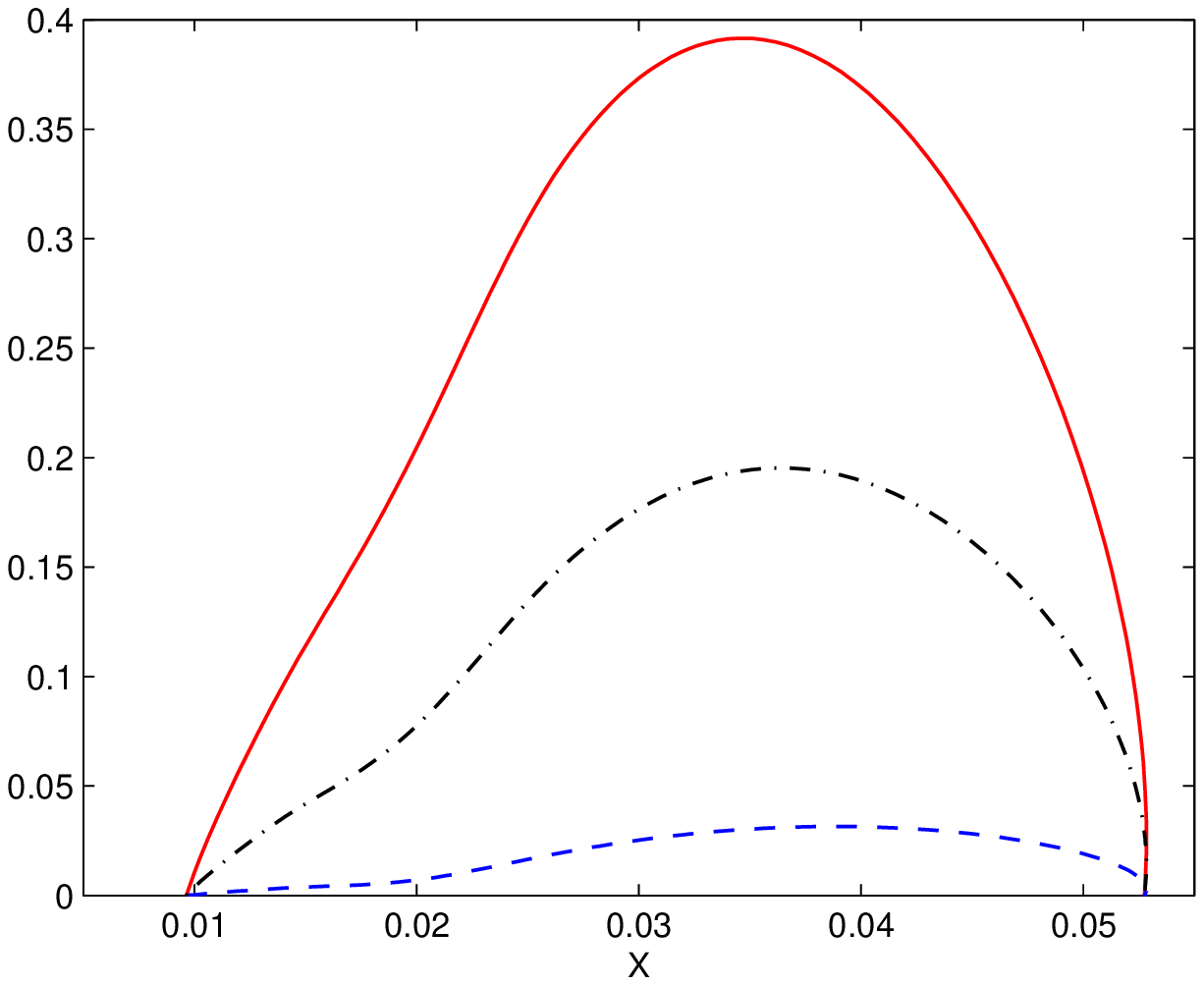}} 
\put(6.5,0){\includegraphics[width=7.4cm]{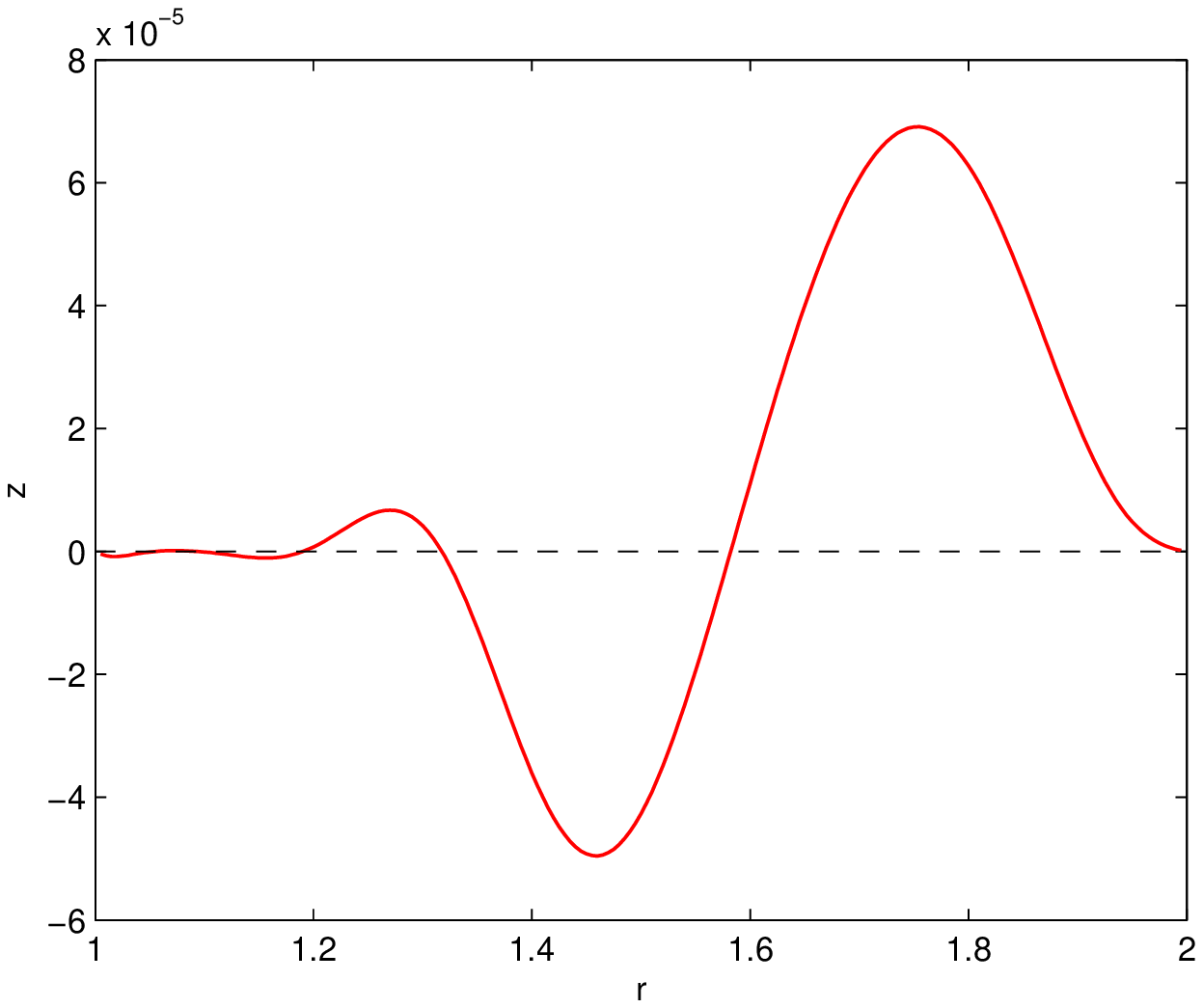}}                                                        
\end{picture}      
\caption{Left: surplus azimuthally-averaged  radial pressure drop $\delta p$ (blue dashed line), disturbance angular momentum $\delta I$ (red solid line) and surplus of 2D rotational energy above 1D rotational energy $\delta E$ (black dash-dot line) all plotted against $\eta$ for the 2D ($m_0=1$) solution branch of figure \ref{soln2D}. Right: the disturbance angular velocity profile for the 2D ($m_0=1$) solution at $\eta=9.64 \times 10^{-3}$ $Re=10^4$, $a=2$ and $\alpha=-3/2$. Notice this closely resembles the prediction of the nonlinear analysis given that the boundary layer thickness is relatively large ($10^{-4/3}$) and the boundary layer oscillation extends 6 boundary layer thicknesses outwards.}
\label{press}
\end{figure}

%
%
%
\begin{figure}
\psfrag{X}{$r$}
\psfrag{Y}{$\Omega$}
\psfrag{Z}{$ r^2 \overline{uv}$}
\setlength{\unitlength}{1cm}                                                    
\begin{picture}(12,6) 
\put(-0.3,0){\includegraphics[width=7.4cm]{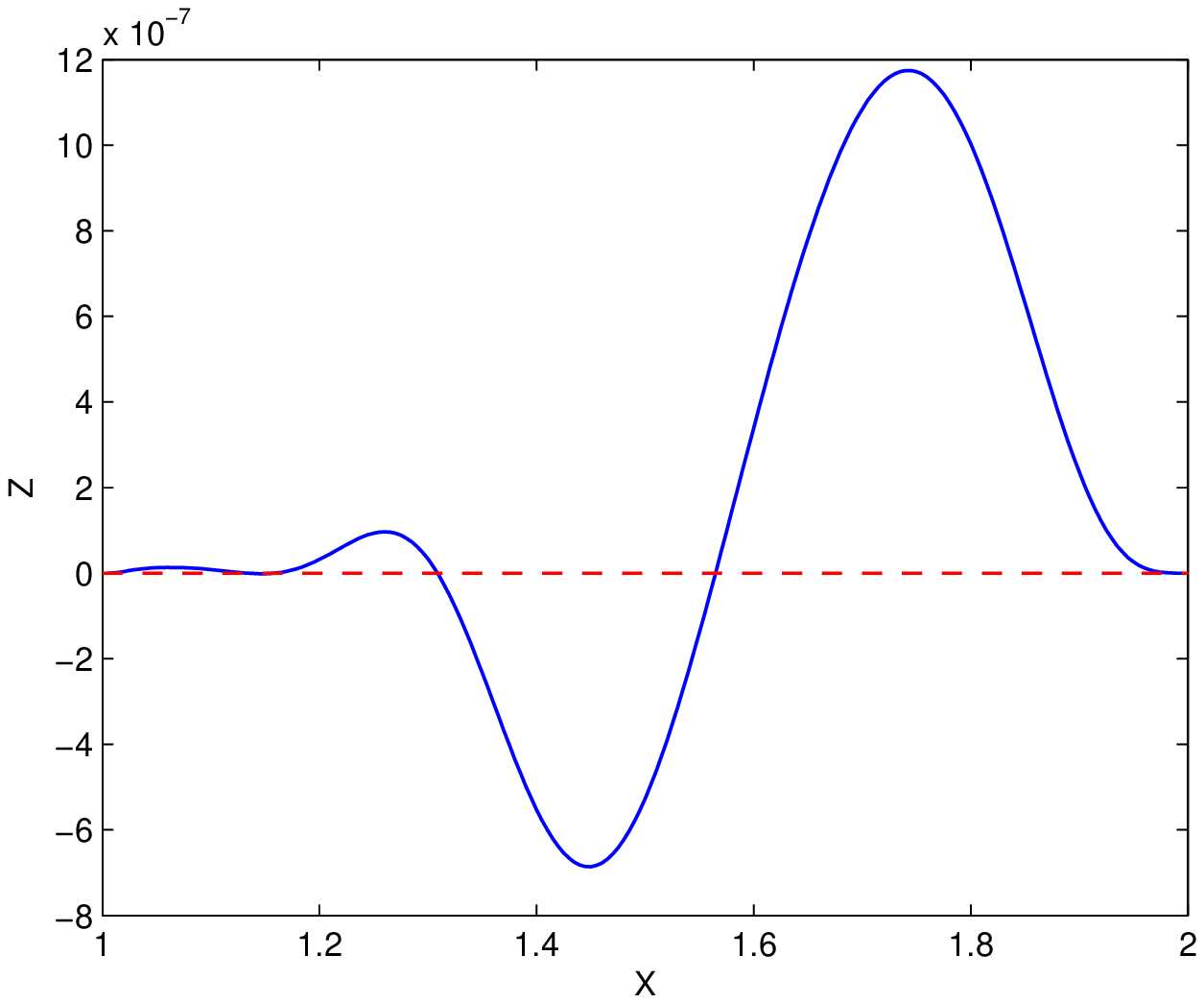}} 
\put(6.5,0){\includegraphics[width=7.4cm]{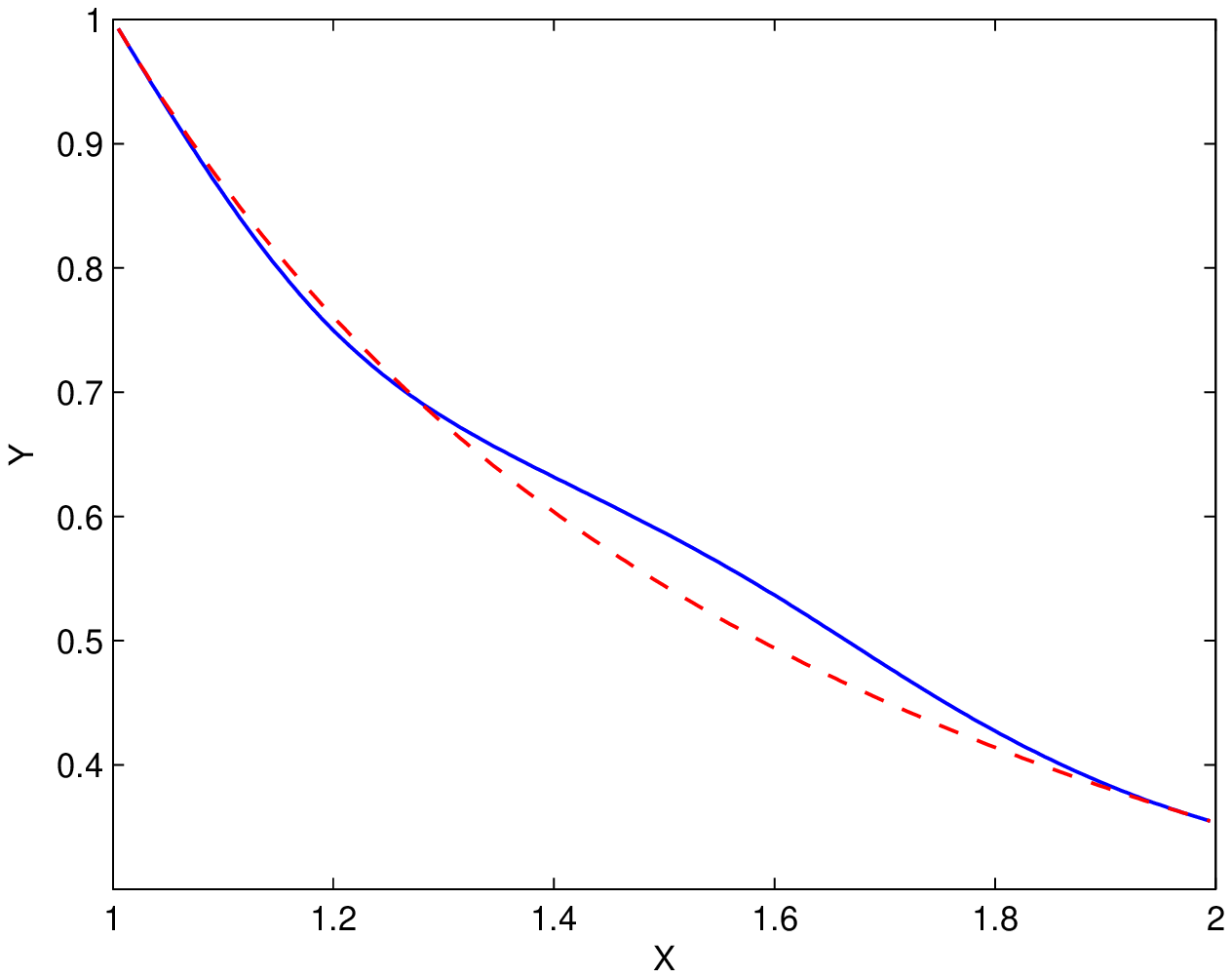}}                                                        
\end{picture}      
\caption{Left: $\int^{2\pi}_0 r^2 uv \,d \theta$ against $r$ for the bifurcating eigenfunction at $\eta=9.645 \times 10^{-3}$ (left dot on (upper) figure \ref{soln2D}). Right: the mean angular velocity profile for the $m_0=1$ 2D solution at $\eta=3.67 \times 10^{-2}$ $Re=10^4$, $a=2$ and $\alpha=-3/2$ (solid blue line) and the profile $\Omega=r^\alpha$ corresponding to the 1D basic state (red dashed). This plot clearly indicates that angular momentum has been transported outwards by the saturated instability.}
\label{Jflux}
\end{figure}

%
%
%
\begin{figure}
\psfrag{X}{{\Large $\eta$}}
\psfrag{Y}{{\Large $\Delta p $}}
\resizebox{0.95\textwidth}{!}{\includegraphics[angle=0]{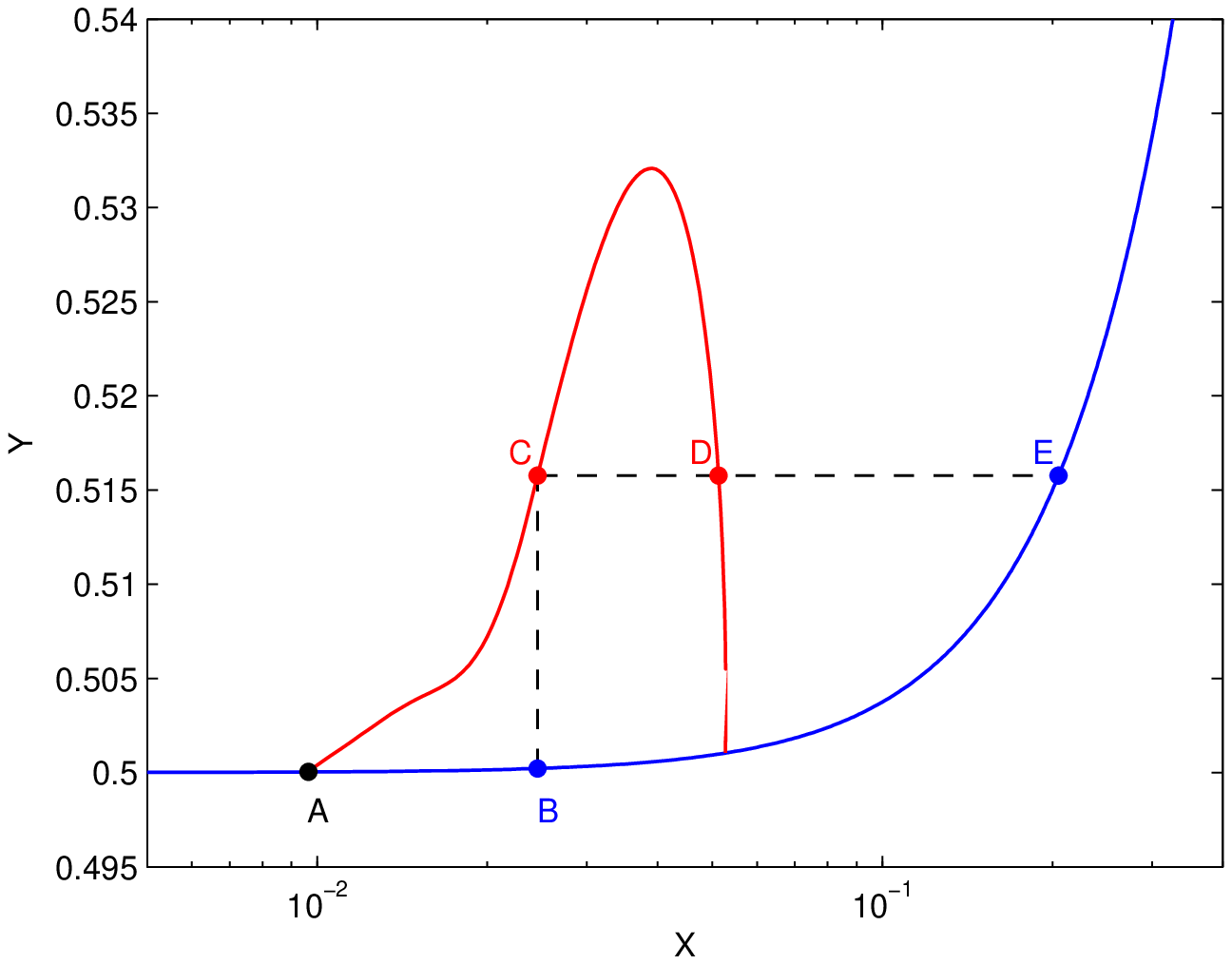}}                                               
\caption{The azimuthally-averaged radial pressure drop $\Delta p$ is plotted again the azimuthally-averaged radial flow $\eta$ for the undisturbed 1D basic state (blue solid line connecting points A,B and E; $\Delta p_{1D}:=(1-1/a)+\half \eta^2(1-1/a^2)$\,) and the 2D solutions (red solid line connecting points A,C and D) which emanate out of the bifurcation for $m_0=1$, $a=2$, $Re=10^4$ and $\alpha=-3/2$ (as shown in Figures \ref{soln2D}). The bifurcation point is point A and the plot shows that the bifurcation is supercritical for either  $\eta$ or $\Delta p$ being the control parameter.  Note that if $\Delta p$ is fixed, the bifurcation leads to states with lower radial flow (compare states C \& D with E).}
\label{bifurcation}
\end{figure}

%
%
\section{Nonlinearity \label{nonlinearity} }

In this section we examine the nonlinear aspects of the boundary inflow instability. The natural starting point is a weakly nonlinear analysis of the 2D boundary layer instability in the presence of viscosity so that there is a finite (radial flow) threshold for instability. Then the leading question is whether the new 2D solution branch exists for $\eta \leq \eta_{crit}$ (a subcritical bifurcation) or for $\eta \geq \eta_{crit}$ (a supercritical bifurcation). This question is most straightforwardly posed with the growing instability not permitted to change the (azimuthally-averaged) radial flow i.e. this is the control parameter of the flow. 

%
%
\subsection{Weakly Nonlinear Analysis of 2D Viscous Boundary Layer Instability}\label{wnl_anal}

The analysis revolves almost completely around the boundary layer and can be handled relatively straightforwardly: see Appendix B. The growing instability is found to be {\em supercritical} so that the branch of 2D solutions exists for $\eta \geq \eta_{crit}$. 
Along this branch of solutions, the azimuthally-averaged radial pressure drop, $\Delta p$, across the domain, which is a more natural control parameter for a laboratory experiment, changes. Furthermore, in an accretion disk, maintaining constant angular momentum of the flow is a more natural constraint under which to study flow bifurcations. These, however, just represent different perspectives of viewing the same bifurcation and the ensuing 2D branch of solutions. For example, solutions with constant $\Delta p$ and varying $\eta$ are the same as those with constant $\eta$ and varying $\Delta p$ provided the rest of the boundary conditions are consistent (e.g. azimuthally-asymmetric radial flow components vanish at the boundaries in both and the azimuthally-asymmetric pressure distribution on each boundary is unconstrained in both).  To make this clear and to be able to make statements away from the vicinity of the bifurcation point, we now compute the  fully nonlinear 2D solution branch.

%
%
\subsection{Fully Nonlinear 2D Solutions}

To study (non-helical) 2D bifurcations off the 1D steady state (\ref{basic}), a Reynolds number of $Re=10^4$ is chosen as a compromise, being hopefully large enough to be in the asymptotic regime but not too large that flows become too arduous to follow numerically. According to the asymptotics, $\eta_{crit}=a N_c^{2/3} \chi^{1/3} Re^{-2/3} \approx 0.0061\,(0.0087)$ for $m_0=1$ ($m_0=3$) whereas actually the thresholds $\eta_{crit}=0.009635\,(0.0136)$ are found numerically at $Re=10^4$ ($\alpha=-3/2$ and  $a=2$). 
Figure \ref{stab2D} shows how the spectrum of the linear operator depends on $\eta$ for $Re=10^5$. Instability is first possible  at $Re \approx 5 \times 10^3$ (not shown)  with the second mode of instability appearing for $Re > 10^4$ and by $Re=10^5$ there are 3 unstable modes. Further computations for $Re=10^6$ (not shown) show that further modes of instability emerge (now 8) and the persistent feature that each unstable mode is only unstable for a finite range of $\eta$. The lower threshold $\eta_l$ (which is the focus in this work and \cite{gal10})  must tend to $0$ as $Re \rightarrow \infty$ to be consistent with the inviscid analysis whereas the upper threshold $\eta_u$ must tend to a finite limit to be consistent with the analysis of \cite{ilin15}.

The bifurcation is a Hopf bifurcation and the oscillation can be made to look steady by going into a frame rotating at the phase speed $c_\theta:=\sigma_i/m_0$ at the bifurcation point. Subsequently, the 2D solution branch is traced out by using the representation
\begin{equation}
\left[ \begin{array}{c} 
u \\ v \\ p
\end{array}\right]:=\sum_{n=0}^{N-1} \sum_{m=0}^{M} 
\left[ \begin{array}{c} 
u_{mn}(t) (\,T_{n+2}(\xi)-T_{n}(\xi) \,)\\
v_{mn}(t) (\,T_{n+2}(\xi)-T_{n}(\xi) \,)\\ 
p_{mn}(t)  T_n(\xi)
\end{array}\right] e^{i m m_0 (\theta-c_{\theta}t)}+c.c.
\label{2Dsoln}
\end{equation}
where $\xi:=(2r-a-1)/(a-1)$, $m_0$ indicates the rotational symmetry of the bifurcating eigenfunction and $c_\theta$ is the azimuthal phase speed which is found as part of solution process. Figure \ref{soln2D}  shows the new branch of 2D solutions arising from the one unstable mode present at $Re=10^4$ for $m_0=1$ and $m_0=3$ found by a Newton-Raphson rooting finding algorithm with truncations  varying from $(M,N)=(20,60)$ to $(10,150)$ (so $O(10^4)$ degrees of freedom). Both bifurcations at the lower threshold $\eta_l$ are supercritical consistent with the weakly nonlinear analysis of \S \ref{wnl_anal} and the solution branches reconnect with the 1D solution (\ref{basic}) at the upper threshold $\eta_u$: see Figure \ref{soln2D}.


Along the 2D solution branch, the `surplus' azimuthally-averaged radial pressure drop,
\beq
\delta p:=(\Delta p)_{2D}-(\Delta p)_{1D}=\int^a_1 2\Omega(r)\overline{v}+\frac{\overline{v^2-u^2}}{r}\, dr,
\eeq
(\,where $\overline{( \cdot)}:=1/2\pi \int^{2 \pi}_0 (\cdot) d \theta$ is just an azimuthal average\,), the surplus of 2D rotational energy compared to the 1D solution,
\beq
\delta E:=2 \pi \int^a_1 r (\,\Omega \overline{v}+\half \overline{v}^2\,) \,r dr\,)
\eeq
and the disturbance angular momentum $\delta I$ all are positive quantities: see Figure \ref{press}.
The initial increase in the pressure drop and the total angular momentum indicates that if either were used as a control parameter instead of the azimuthally-averaged radial velocity, the bifurcation would remain supercritical. For example, Figure \ref{bifurcation} shows a plot of $\Delta p$ against the azimuthally-averaged radial flow parameter $\eta$ for the 2D solution branch already plotted in Figure \ref{soln2D}. The weakly nonlinear analysis in Appendix B takes $\eta$ as the control parameter 
and finds that the bifurcated 2D solution branch exists for $\eta> \eta_{crit}$ (marked as the point A in Figure \ref{bifurcation}) with $\Delta p$ larger for the 2D solutions than the equivalent (same $\eta$) 1D solution (see points B and C in Figure \ref{bifurcation}). If $\Delta p$ is the control parameter, the bifurcation is still supercritical as 2D solutions can only exist for $\Delta p \geq \Delta p$ at A but now the bifurcated 2D solutions give rise to 2 {\em smaller} radial inflow solutions (points C or D in Figure \ref{bifurcation}) compared to the 1D solution E. 

Another interesting issue for accretion disks is whether this instability acts to transfer angular momentum $I$ outwards. Forming the integral $\int^{2\pi}_0 r^2(\ref{azimuth}) \, d\theta$ gives the conservation equation
\begin{equation}
\biggl({\cal I}(r):=\int^{2 \pi}_0  r[r \Omega(r)+ v] \, \, rd \theta \biggr)_t+J_r=0,
\end{equation} 
where $I=\int^a_1 {\cal I}(r) dr$ and 
\begin{equation}
J:=\int^{2 \pi}_0  
\biggl\{r uv-\eta v-\frac{1}{Re}r^2 \frac{d}{dr}\biggl( \frac{v}{r} \biggr)
\, \,  \bigg\}rd \theta 
\end{equation}
is the associated radial flux of angular momentum (the flux vanishes for the 1D basic state (\ref{basic})).
The first (Reynolds-stress) term in $J$ is computable using the bifurcating eigenfunction alone and  can be used to determine the effect of the instability on the angular momentum distribution (the other two terms, which involve nonlinear aspects of the instability, subsequently balance the first term to give a finite amplitude 2D steady state: figure \ref{press} (right) actually shows that $r^2\overline{uv} \approx \eta r \overline{v}$).
Figure \ref{Jflux} plots this term over the radius for the $m_0=1$ state just after the bifurcation (shown in figure \ref{soln2D} as the leftmost point at $\eta=9.645 \times 10^{-3}$) and  shows that it alternates in sign but is mostly positive indicating outwards angular momentum transport. The angular velocity of the 2D state at its maximum amplitude  (the middle $m_0=1$ point in the upper plot and the middle cross-section in the lower plot of figure \ref{soln2D}) provides more definitive evidence of this outward transport: see figure \ref{Jflux}. 

%
%
%
\begin{figure}
\psfrag{X}{$\eta$}
\psfrag{Y}{$E$}
\setlength{\unitlength}{1cm}  
\begin{picture}(14,12) 
\put(-1.5,0)   {\includegraphics[width=16cm]{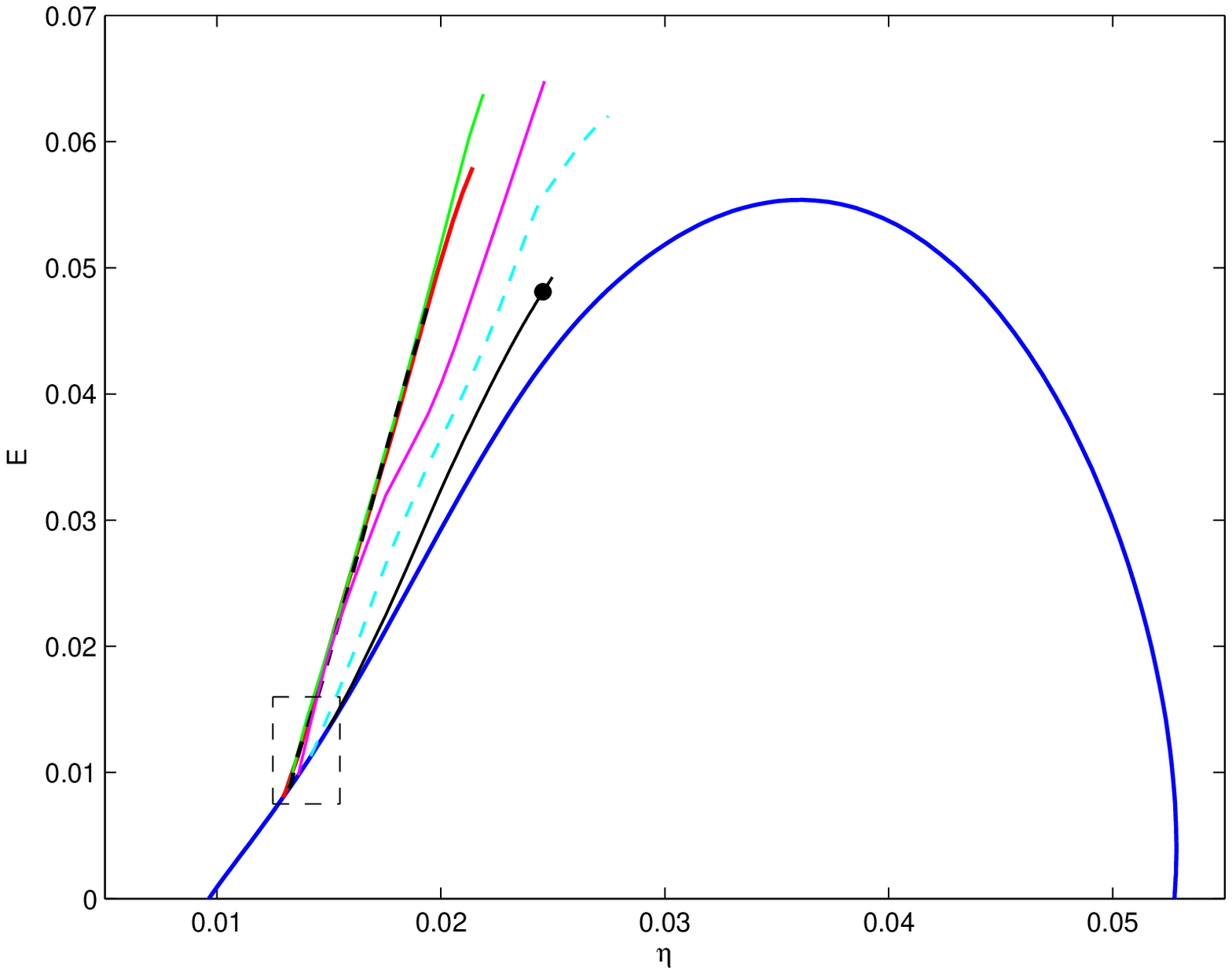}} 
\put(5.75,1.5)   {\includegraphics[width=7cm]{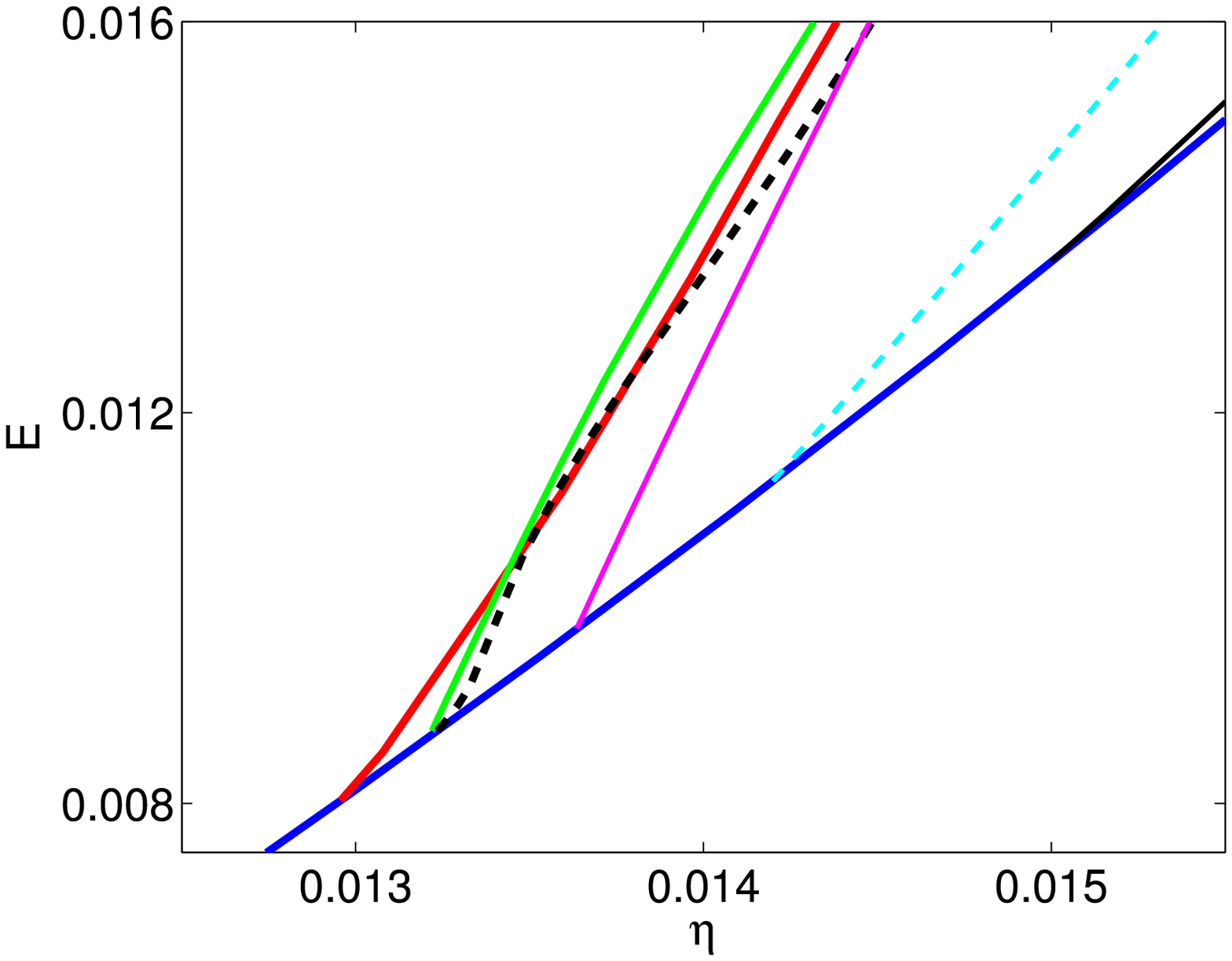}}
\end{picture}                                                         
\caption{The 3D solution branches continued out of the six Hopf bifurcations (see the inset which magnifies the dashed box for detail) from the 2D branch (thick blue loop as in figure \ref{soln2D}) found for $m_0=1$ at $Re=10^4$, $\alpha=-3/2$, $k=1$ and $a=2$. All the bifurcations are supercritical. The ordinate is the disturbance energy (2D and 3D) and the abscissa the radial flow $\eta$. The black dot indicates the particular 3D flow state shown in detail in figure \ref{3Dstate}. Typically a resolution of $(N,M,L)=(40,20,4)$ was used to follow these solutions with curve segments only shown if they are robust under truncation changes.}
\label{states3D}
\end{figure}

%
%
\subsection{Fully Nonlinear 3D Solutions}

\begin{figure}
\setlength{\unitlength}{1cm}                                                    
\begin{picture}(14,14) 
\put(0,6.75)   {\includegraphics[width=6.5cm]{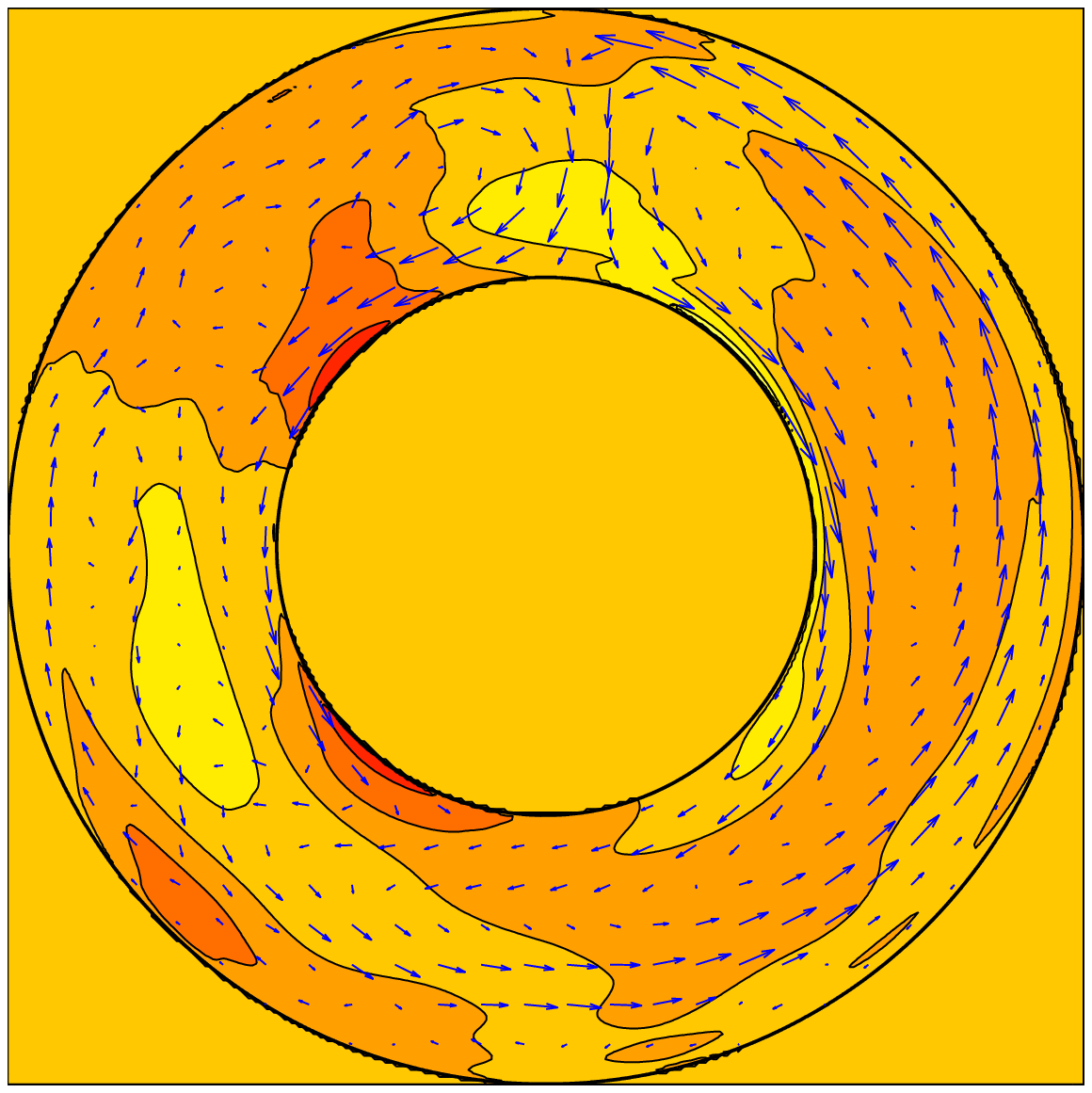}} 
\put(6.75,6.75){\includegraphics[width=6.5cm]{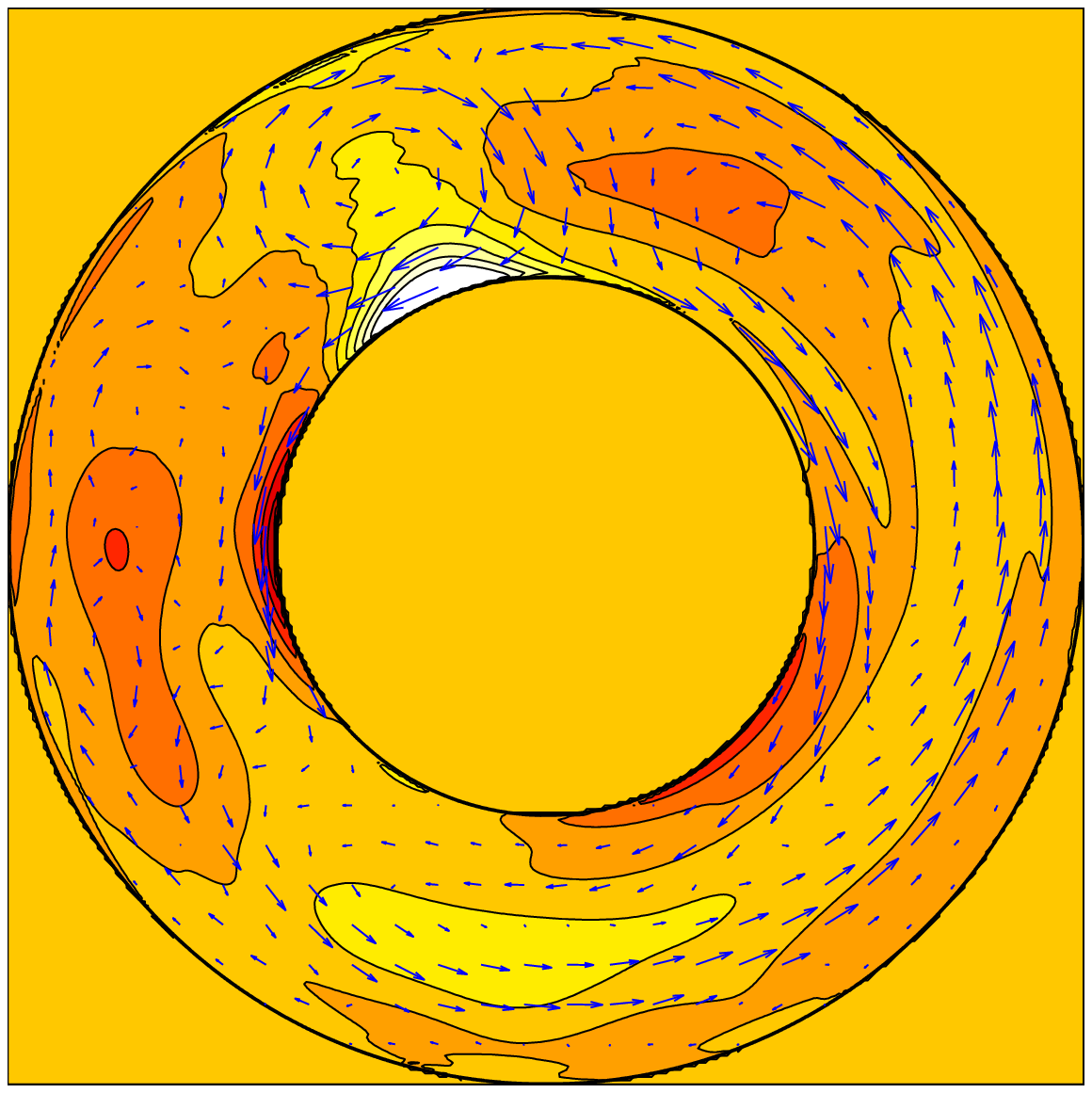}}                                                        
\put(0,0)      {\includegraphics[width=6.5cm]{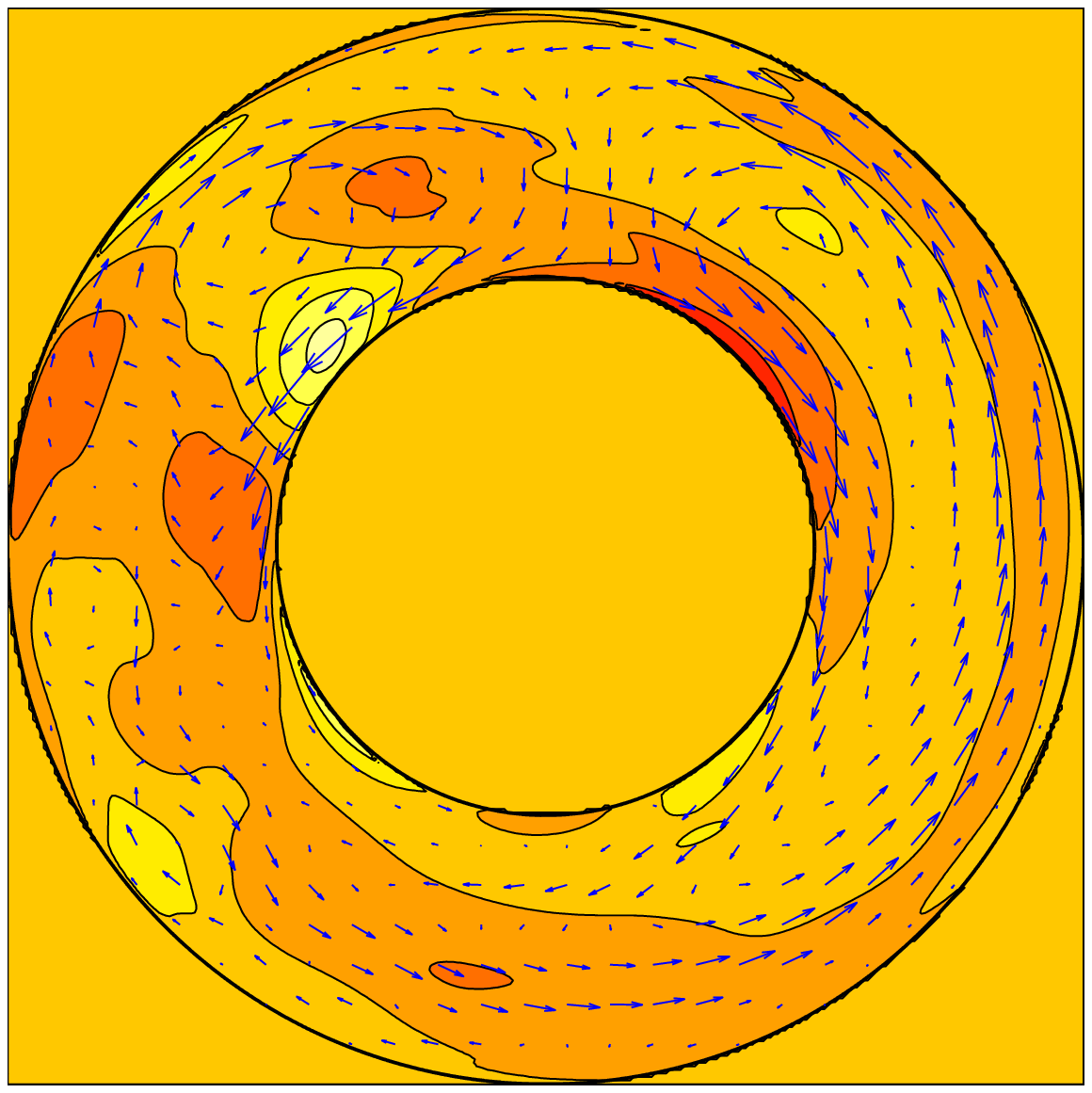}}
\put(6.75,0)   {\includegraphics[width=6.5cm]{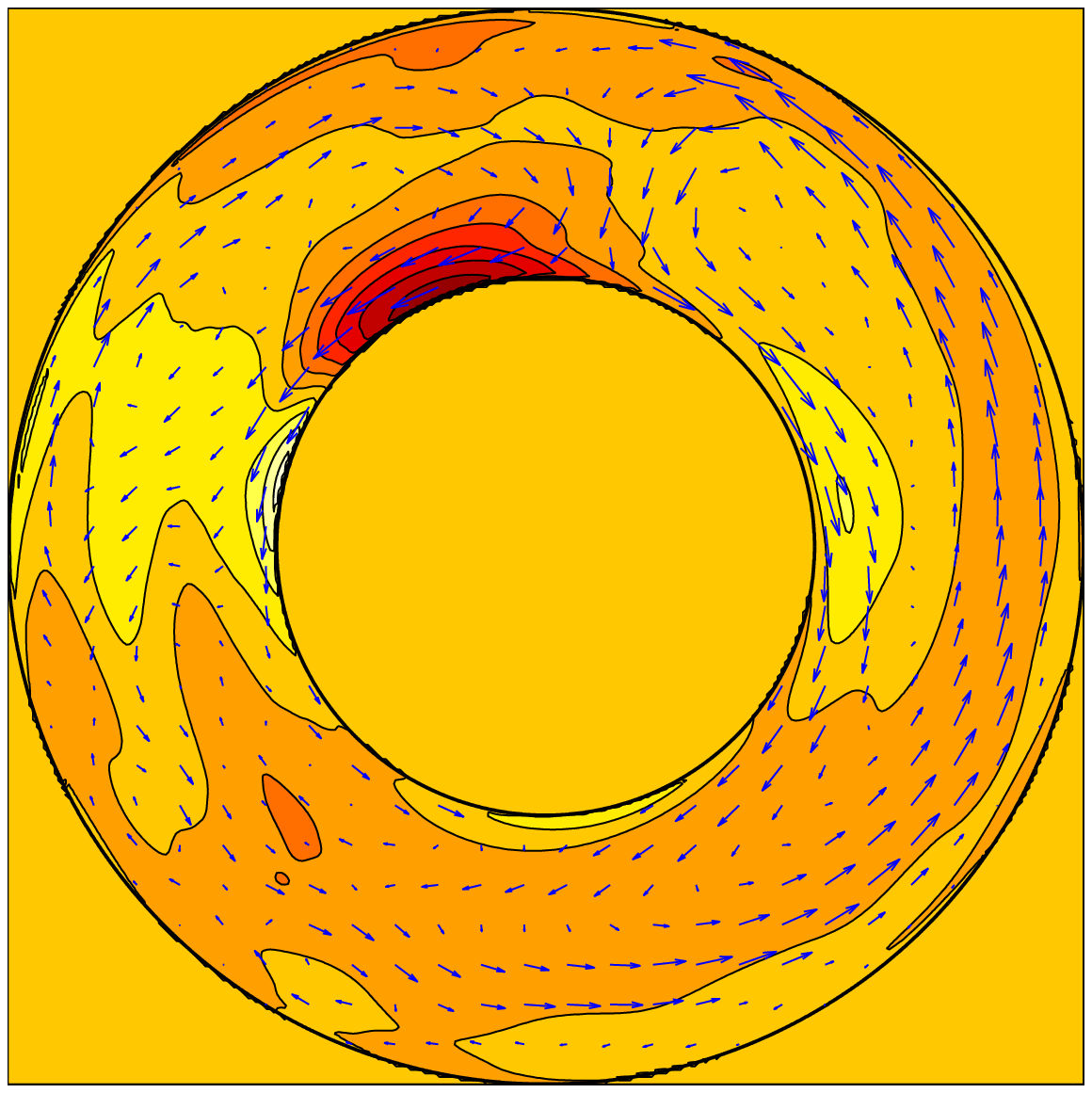}}
\end{picture}       
\caption{A typical 3D state (shown as a dot on figure \ref{states3D}) calculated using $(N,M,L)=(40,28,4)$ (82,003 d.o.f.) plotted at $z=0$ (top left), $z=\pi/2$ (top right), $z=\pi$ (bottom left) and $z=3\pi/2$ (axial wavelength = $2\pi$). The axial speed is shown by 10 contours going from -0.025 to 0.035 (the colour/shading of the zero contour is shown on the outside of the flow domain). The arrows indicate the speed and direction of the flow in the $(r, \theta)$ plane with the longest arrow representing a speed of 0.272.}
\label{3Dstate}
\end{figure}

To briefly explore the possibility of 3D states, 3D bifurcations were sought off the 2D branch of $m_0=1$ solutions. Concentrating on 3D states with an axial wavenumber $k=1$, six Hopf bifurcations were found between the initial 2D bifurcation point $\eta=\eta_{2D}=0.0096$ and $\eta=0.0152$. These were then traced using the fully nonlinear steady representation 
\begin{equation}
\left[ \begin{array}{r} 
u \\ v \\ w \\ p
\end{array}\right]:=\sum_{n=0}^{N-1} \sum_{m=-M}^{M} \sum_{l=0}^{L} 
\left[ \begin{array}{c} 
u_{lmn}(t) (\,T_{n+2}(\xi)-T_{n}(\xi) \,) \\
v_{lmn}(t) (\,T_{n+2}(\xi)-T_{n}(\xi) \,) \\ 
w_{lmn}(t) (\,T_{n+2}(\xi)-T_{n}(\xi) \,) \\ 
p_{lmn}(t)  T_n(\xi)
\end{array}\right] e^{i (m m_0 \theta^*+lk z^*)}+c.c.
\end{equation}
by moving into a frame moving with the phase speed $c_z$ in $z$ (initially $c_z:=\sigma_i/k$ at the bifurcation point where $\sigma_i$ is the instability frequency). Since $\theta^*:=\theta-c_\theta t$ already incorporates the phase speed of the initial 2D instability, these secondary 3D states are steady in a rotating  {\em and} translating frame. Figure \ref{states3D} shows that all the traced bifurcations are supercritical i.e. there is no evidence of solution branches reaching $\eta < \eta_{2D}$. Solution branches are shown as far as they are reproducible using different truncation levels. The maximum realistic resolution was $(N,M,L)=(40,20,4)$ giving 58,963 degrees of freedom since the branch continuation approach was a direct Newton-Raphson solver albeit with multithreaded linear algebra software. A typical 3D flow is shown in figure \ref{3Dstate} with the presence of vertical jets at the outflow boundary and the lack of any large scale  structure particularly noteworthy.

%
%
%
%

%
%
\section{Energetics}
%
%

The instability is inviscid in nature so we first consider the energetics of this simplest situation: note that in the absence of viscosity, $\alpha=-2$ gives the only consistent solution. If $\bu_{tot}=\bU+\bu$ is the total velocity field, then the scalar product of this with the governing Euler equations gives
\beq
\partial_t \langle \half \bu_{tot}^2 \rangle= -\oint \, \half \bu_{tot}^2 (\bu_{tot}.{\mathbf dS})- \oint \, p \,\bu_{tot}.{\mathbf dS}
\eeq
With no disturbance $\bu=0$, this amounts to 
\beq
-\oint \, p\, \bu_{tot}.{\mathbf dS}= \eta \int^{2\pi}_0 d \theta \int dz \, \biggl[\, p(a,\theta,z,t)-p(1,\theta,z,t)\, \biggr] 
= \pi \eta(1+\eta^2)\biggl(1-\frac{1}{a^2}\biggr)
\eeq
so a net pressure drop radially {\em inwards} across the domain drives the radial flow and works at a rate to replenish the net kinetic energy leaving the domain (the rightmost term using $\bU=-\eta/r \br+1/r \btheta$).

%
%
The need for interior shear to fuel the instability is apparent by looking at the disturbance energy balance.
Taking the scalar product of the disturbance velocity $\bu$ and the disturbance evolution equation, leads to
\beq
\frac{d}{dt}\biggl\langle \half \bu^2 \biggr\rangle = -\langle \, \bu \cdot \bnab \bU \cdot \bu\, \rangle
-\eta \int^{2\pi}_0 \half v^2 \biggl|_{r=1} \, d\theta
\eeq
or explicitly
\beq
\frac{d}{dt}\biggl\langle \half \bu^2 \biggr\rangle = \eta \biggl\langle \frac{v^2-u^2}{r^2} \biggr\rangle
- \biggl\langle ruv \frac{d \Omega}{dr} \biggr\rangle
-\eta \int^{2\pi}_0 \half v^2 \biggl|_{r=1} \, d\theta
\eeq
so the disturbance can only gain energy through the underlying shear field (the last term on the right hand side is the loss of kinetic energy through the outflow boundary). The second term on the rhs is the energy transfer from the swirl field and this has to be positive (i.e. the underlying swirl field supplies energy to the disturbance) for a growing disturbance to achieve growth rates $\gg O(\eta)$.

%
%
Adding viscosity complicates the (energetic) situation by introducing interior dissipation and the further possibility of viscous stresses at either or both radial boundaries inputting energy into the flow: explicitly
\begin{align}
\frac{d}{dt} \bigg\langle \half \bu_{tot}^2 \bigg\rangle=
\eta  \int^{2\pi}_0 d \theta \int dz \, 
\biggl[\, & \half \bu_{tot}^2(a,\theta,z,t)-\half \bu_{tot}^2(1,\theta,z,t) \nonumber \\
&+ p_{tot}(a,\theta,z,t)-p_{tot}(1,\theta,z,t)\, 
\biggr] \nonumber \\
&\hspace{0.5cm}
 +\frac{1}{Re} \oint \bu_{tot} \cdot 2\mathbf{e} \cdot \mathbf{dS}-\frac{1}{Re} \langle \,2 \mathbf{e:e} \, \rangle
\end{align}
where $\mathbf{e}$ is the rate of strain tensor for $\bu_{tot}$. However, non-slip boundary conditions and constant $\eta$ do at least fix the net advection of kinetic energy out of the domain. Then either an  increased radial pressure drop and/or increased work done by viscous stresses at the walls must offset the greater dissipation of a 2D bifurcated state. Certainly the nonlinear computations of \S \ref{nonlinearity} suggest that the radial pressure drop does go up for the bifurcated solutions. For an accretion disk, however, it may be more realistic to assume that the radial pressure drop is fixed but the radial flow $\eta$ can change instead. In this case, rather paradoxically, figure \ref{bifurcation} indicates that the 2D flow will adopt a smaller radial flow to accommodate the greater dissipation of the 2D state. This runs counter to the usual argument that a turbulent disk  should set up, via an enhanced eddy viscosity,  a more accreting flow travelling down the gravitational potential to power it. 

%
%
\section{Discussion}

In this paper we have revisited the 2D instability discussed separately by \cite{Nicoud97}, \cite{Doering00}, \cite{gal10} and \cite{ilin13} in various contexts to explore the interesting mathematical and physical issues surrounding it. A simple half-plane model indicates that the instability operates by the crossflow advecting vorticity introduced by the inflow boundary across the cross-stream shear. Imposing vanishing vorticity at the inflow boundary eliminates the  instability. This half-plane model also makes it clear that curvature or rotation is not important other that to restrict the allowed streamwise wavenumbers (azimuthal periodicity prevents very long wavelengths which are the most unstable in the rectilinear situation). It also highlights the fact that only inflow is destabilising suggesting that in situations where there is both an inflow and an outflow, the inflow boundary could be expected to destabilise the flow initially as the crossflow is turned on before the outflow (or `suction')  boundary eventually stabilises the flow at a higher crossflow value. This is clearly seen in circumstances where the underlying shear flow is not modified by the crossflow (e.g. \cite{Nicoud97}, \cite{ilin13} and here in \S \ref{hp}) but seems generally true even if it does (e.g. \cite{Doering00, Fransson03, gal10, Guha10, ilin15, Deguchi14} although note \cite{Hains71}). The identification of the instability with inflow also explains why this instability is relatively unstudied compared to outflow boundaries, long lauded as a reliable means to stabilise unidirectional flows (e.g \cite{Joslin1998}).

One of the most interesting aspects of this `boundary inflow instability' is that the growth rates scale as $\sqrt{\eta}$ for boundary layer modes and $\eta \log 1/\eta$ for interior critical layer modes when the perturbing crossflow is only $O(\eta)$. This means that the instability has to draw energy out of the underlying shear field. However, the 2D nonlinear solutions computed in the rotating situation do not show any despinning of the  swirl field but instead indicate that the radial pressure gradient which drives the crossflow more than replenishes this energy by working on the flow. 

The apparently delicate mathematical structure underlying the instability - the lack of any discrete normal modes for the inviscid, vanishing crossflow situation - suggests that additional physics may easily suppress it. However, the work by \cite{gal10} already indicated that it could survive the addition of viscosity (see also \cite{ilin15}) and here we have also explored different (more Rayleigh-stable) rotation profiles plus the addition of 3-dimensionality (see also \cite{ilin15b}) and compressibility. It is true that except for the rotation profiles, none of these have actually enhanced the instability (e.g. 2D modes are more unstable than 3D modes) but neither have they immediately suppressed it either (e.g. the viscous threshold for the instability is still only a very small crossflow of $O(Re^{-1})$ in a rectilinear geometry \citep{Nicoud97} and $O(Re^{-2/3})$ in the rotating situation; see \S \ref{swirl_viscous_bl}). The conclusion is therefore that the instability is relatively robust except to the exact boundary conditions imposed at the inflow boundary and one should therefore expect instability whenever there is boundary inflow together with shear.

Weakly nonlinear analysis of the primary bifurcation and branch continuation of the fully nonlinear 2D solutions indicate that the instability is supercritical in the crossflow. Furthermore, secondary bifurcations to 3D states also appear supercritical (6 were found and continued) suggesting a succession of bifurcations in which the flow gradually becomes more complicated spatially and temporally as the crossflow is increased. This all makes the boundary inflow instability an inviting target for an experimental study especially as the primary instability is oscillatory and hence clearly identifiable.


Astrophysically, we have established that a Keplerian rotation profile with crossflow of $-\eta/r \br$ and non-slip boundary conditions at the inflow boundary will be unstable if the crossflow $\eta \gtrsim O(Re^{-2/3})$ - hence Rayleigh's stability criterion can be circumvented by crossflow. However, in a quiescent disk, (molecular) viscous stresses only generate a crossflow of $O(Re^{-1})$ so the linear instability is not triggered. Moreover, both the (2D) primary and (3D) secondary instabilities are supercritical so there is also no apparent opportunity to reach the instability via finite amplitude perturbations at such low crossflows. Even if the crossflow was large enough, one would have to argue that the appropriate boundary conditions were in place at the inflow (outer) boundary of the disk.  Nevertheless, the analysis presented here does highlight the potential for mass entering a disk to disrupt the orbiting flow if this mass flux possesses vorticity and also showcases the complications of introducing inflow boundaries in disk models (e.g. \cite{Kersale04}).

\vspace{1cm}
\noindent
{\em Acknowledgements} The author thanks Jim Pringle for some informal discussions and a referee for their careful review of the manuscript.

%
%
\appendix

%
%

\section{Numerics}

The 3D boundary layer equations (\ref{1}) and (\ref{2}) can be
directly treated numerically by transforming the domain $\xi \in
[0,\infty)$ to $x \in [-1,1)$ via the definition $x:=(\xi-L)/(\xi+L)$
    where $L$ is a scale factor ($L=1$ works well here). $v$ and
    $\hat{w}$ are expanded as differences of consecutive even or odd
    Chebyshev polynomials 
\beq 
(v,\hat{w})= \sum_{n=0}^N (a_n,b_n) (\,T_{n+2}(X)-T_n(X)\, ), 
\qquad \qquad {\rm where}\, \, \,  T_n(x):=\cos(n\cos^{-1}x) 
\eeq 
so that $v$ and $\hat{w}$ are forced to vanish at $\xi=0$ and $\xi
\rightarrow \infty$, and the two equations (\ref{1}) and
(\ref{2}) are collocated across the $N$ zeros of $T_N(X)$. The resulting $2N \times
2N$ eigenvalue problem (with $N$ varying from 100 to 10,000) is then
solved using LAPACK routine ZGGEV (solving the 2D boundary layer
equation (\ref{bl_asym}) is just a $N \times N$ special case of the 3D
problem). The viscous boundary layer equation  (\ref{bl_eqn}) can be similarly solved. 

 
The  full 3D eigenvalue (\ref{3D_1})-(\ref{3D_4}) can be solved by expanding the primitive variables as follows
\beq
\biggl[\,u, \biggl(\begin{array}{c} v \\ w \end{array} \biggr),p \biggr]
:=\sum_{n=0}^N \biggl[ a_n (\,T_{n+2}(Y)-T_n(Y)\,), \biggl(\begin{array}{c} b_n \\c_n \end{array} \biggr) 
(\,T_{n+1}(Y)-T_n(Y)\, ), d_n T_n(Y) \biggr]
\eeq
where $Y:=(2r-a-1)/(a-1)$ so that the appropriate boundary
conditions - $u=0|_{r=1}, \quad u=v=w|_{r=a}=0$ - are automatically
imposed. This $4N \times 4N$ eigenvalue problem is again solved by
LAPACK routine ZGGEV typically with $N=2400$. An associated
inverse iteration code was developed where $N$ could reach $12,000$ on a
128GB machine to confirm eigenvalues.

%
%
%

\section{Weakly Nonlinear Analysis of 2D Viscous Boundary Layer Instability}

We introduce two small quantities: $\delta$ as the amplitude of the instability and $\eps=Re^{-1/3}$ as a measure of the viscosity. The analysis proceeds from the following expansion of the velocity field
\begin{align}
\psi &=  \del \biggl\{  
        (\,\psi^{(m)}_{10}(r)+{\color{red}\hat{\psi}^{(m)}_{10}(y)}\,)
+ \eps (\,\psi^{(m)}_{11}(r)+\hat{\psi}^{(m)}_{11}(y)\,) + \ldots \biggr\}E(\theta,t) \nonumber\\
      & +\del \biggl( \dele \biggr) \biggl( \frac{m}{a} \biggr)\biggl\{ 
 \frac{1}{\eps}\psi_{2,-1}^{(0)}(r)+N_c^{1/3} \chi^{-1/3}(\, \psi^{(0)}_{20}(r) +{\color{red}\hat{\psi}^{(0)}_{20}(y)}\,)+O(\eps))\biggr\} \nonumber\\
& +\del \biggl( \dele \biggr) \biggl( \frac{m}{a} \biggr) N_c^{1/3} \chi^{-1/3}\biggl\{ 
(\, \psi^{(2m)}_{20}(r) +{\color{red}\hat{\psi}^{(2m)}_{20}(y)}\, )+O(\eps)\biggr\}E(\theta,t)^2 \nonumber\\
      & +\del \biggl(\dele \biggr)^2  \biggl(\frac{m}{a}\biggr)^2 N_c^{2/3} \chi^{-2/3}
\biggl\{ (\,\psi^{(m)}_{30}(r) +\hat{\psi}^{(m)}_{30}(y)\,)   +O(\eps)\biggr\}E(\theta,t) \nonumber \\ 
&+ \del \biggl(\dele \biggr)^2  \biggl(\frac{m}{a}\biggr)^2 N_c^{2/3} \chi^{-2/3}
\biggl\{ (\,\psi^{(3m)}_{30}(r) +\hat{\psi}^{(3m)}_{30}(y)\, )+O(\eps) \biggr\} E(\theta,t)^3 \qquad \nonumber \\
& \hspace{5cm} +\ldots +c.c.
\end{align}
where
\begin{align}
E(\theta,t) &:= e^{im \theta+(\,-i m \Omega(a) +\eps \hat{\sigma}\,)t}, \nonumber \\
y           &:= N_c^{-1/3}\chi^{1/3}\biggl( \frac{a-r}{\eps} \biggr), \nonumber \\
\hat{\sigma}&:= N_c^{ 1/3}\chi^{2/3} \hat{\sigma}_0+\biggl(\dele \biggr)^2  \biggl(\frac{m}{a}\biggr)^2 \hat{\sigma}_2+\ldots, \nonumber \\
\eta        &:= \eps^2 \hat{\eta}_c+\eps^2 \biggl(\dele \biggr)^2  \frac{m^2}{a} N_c^{1/3}\chi^{-1/3}\hat{\eta}_2+\ldots  \biggr., \nonumber
\end{align}
with $N_c=4.57557$, $\hat{\sigma}_0=-5.63551i$ and  $\hat{\eta}_c= aN_c^{2/3} \chi^{1/3}$ being the critical values (see \S \ref{swirl_viscous_bl}) and $\chi=-\half m \Omega^{'}(a)$ (only flow components in {\color{red}red} need to be calculated).
The objective here is to look around the bifurcation point $\eta=\eps^2 \hat{\eta}_c$ where $\delta=0$ (fixed $Re$) to find a value of $\hat{\eta}_2$ consistent with $\delta>0$: $\hat{\eta}_2 <0(>0)$ indicates a subcritical (supercritical) bifurcation. We work with the full 2D disturbance equation (\ref{master})
\beq
\biggl( \frac{\partial}{\partial t}+\Omega(r) \frac{\partial}{\partial \theta} -\frac{\eta}{r}\frac{\partial}{\partial r} \biggr) \Delta \psi = \frac{1}{r}\frac{dZ}{dr} \frac{\partial \psi}{\partial \theta}+\frac{1}{Re} \Delta^2 \psi+\frac{1}{r} J(\psi,\Delta \psi).
\eeq

\subsection{Linear problem at $O(\del)$}
The boundary layer problem is  4th order in $y$
\beq
\biggl[
  \frac{d^2}{dy^2}-N_c\frac{d}{dy}-N_c(\hat{\sigma}_0 +2 i y)\biggr]
\frac{d^2 \hat{\psi}^{(m)}_{10}}{dy^2} =0
\eeq
with the boundary conditions  that $\hat{\psi}^{(m)}_{10}$, $d\hat{\psi}^{(m)}_{10}/dy$ and $d^2\hat{\psi}^{(m)}_{10}/dy^2$ all vanish at large $y$ and 
$d\hat{\psi}^{(m)}_{10}/dy=0$ at $y=0$.
The outer (leading order) problem is 2nd order in $r$
\beq
\biggl[\, \Omega(r)-\Omega(a) \, \biggr] \opL \psi_{10}^{(m)}=\frac{1}{r} \frac{dZ}{dr}\psi_{10}^{(m)}
\eeq
($m \neq 0$) to which the appropriate boundary conditions are $\psi_{10}^{(m)}(1)=0$ ($u(1)=0$) and $\psi_{10}^{(m)}(a)=-\hat{\psi}_{10}^{(m)}$ so $u(a)=0$.
(Formally there is a weak viscous layer at $r=1$ of thickness $O(1/Re)$ where the streamfunction boundary correction is $O(1/Re)$ to accommodate the non-slip condition $v(1)=0$ but this plays no role in the nonlinear equilibration of the instability centered on the other boundary and will be ignored here and below.)
The boundary layer problem alone identifies the bifurcation point (see \S \ref{swirl_viscous_bl}).

%
%
\subsection{Nonlinearity $O(\del(\del/\eps^2))$}

\begin{figure}
\begin{center} 
\psfrag{v}{{\Large $\tilde{v}$}}
\resizebox{1\textwidth}{!}{\includegraphics[angle=0]{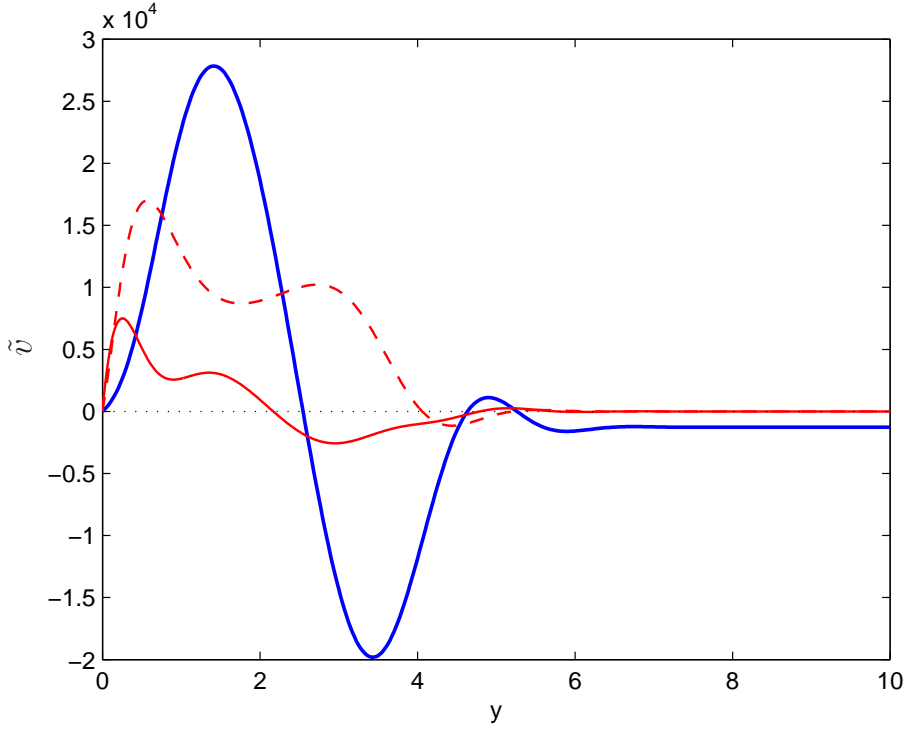}}
\end{center}
\caption{Azimuthal velocities for the (scaled) boundary layer mean flow at $O(m\delta^2/a\eps^3)$:  $\tilde{v}:=\bigg(d\hat{\psi}_{20}^{(0)}/dy- d\psi_{2,-1}^{(0)}/dr|_{r=a}\biggr)$ (thick solid blue line)  and the 2nd harmonic - $d\hat{\psi}_{20}^{(2m)}/dy$ (real/imaginary part solid/dashed red line) - plotted against $y$, the boundary layer variable. }
\label{wnla}
\end{figure}

Nonlinearity comes in at $O(\del(\del/\eps^2))$ generating boundary flows $\hat{\psi}^{(0)}_{20}$ and
$\hat{\psi}^{(2m)}_{20}$ as follows. For $\hat{\psi}^{(0)}_{20}$
%
%
%
\beq
\biggl[
  \frac{d^4}{dy^4}-N_c\frac{d^3}{dy^3} 
\biggr]\hat{\psi}^{(0)}_{20} =-i\biggl\{(\hat{\psi}_{10}^{(m)}+\psi_{10}^{(m)})\frac{d^3 \hat{\psi}_{10}^{*(m)}}{dy^3}
+\frac{ d \hat{\psi}_{10}^{(m)}}{dy} \frac{d^2 \hat{\psi}_{10}^{*(m)}}{dy^2}\biggr\} +c.c.
\label{B5}
\eeq
where the solution which vanishes as $y \rightarrow \infty$ is sought.
In the boundary layer, the outer solution $\psi_{10}^{(m)}$ is just the constant  $-\hat{\psi}_{10}^{(m)}(0)$ and since $\Re e (\,i |d \hat{\psi}^{(m)}_{10}/dy|^2\,)=0$, equation (\ref{B5}) can be integrated twice to
\beq
\biggl[
  \frac{d^2}{dy^2}-N_c\frac{d}{dy} 
\biggr]\hat{\psi}^{(0)}_{20} =f_0(y):=2\Re e \biggl\{-i(\,\hat{\psi}_{10}^{(m)}(y)-\hat{\psi}_{10}^{(m)}(0)\,)\frac{d \hat{\psi}_{10}^{*(m)}}{dy}\biggr\}
\eeq
and the required boundary layer solution is
\beq
\hat{\psi}^{(0)}_{20}(y)=\int^{\infty}_y \frac{1-e^{N_c(y-x)}}{N_c} f_0(x)\, dx.
\eeq
This cannot be made to  satisfy the no-slip condition $d \hat{\psi}_{20}^{(0)}/dy=0$ at $y=0$ and so drives an interior mean flow $\psi_{2,-1}^{(0)}$ via
\beq
\frac{d\psi_{2-1}^{(0)}}{dr}\biggl|_{r=a}-\frac{d \hat{\psi}_{20}^{(0)}}{dy}\biggl|_{y=0}=0.
\eeq 
to ensure $v=0$ at $r=a$.
%
%
The interior mean flow problem for $\psi_{2,-1}^{(0)}$ is just
\begin{align}
\frac{d}{dr} \biggl( \frac{d}{dr} + \frac{1}{r}\biggr) \frac{d \psi_{2,-1}^{(0)}}{dr} & =0 \label{psi_2-1}\\
{\rm with \,\,  b.c.s} \quad \frac{d\psi_{2,-1}^{(0)}}{dr}\biggl|_{r=1}=0, 
\qquad 
\frac{d\psi_{2,-1}^{(0)}}{dr}\biggr|_{r=a} &= -\int^{\infty}_0 e^{-N_cx} f_0(x)\, dx  \label{bc_psi_2-1}
\end{align}
since the leading interior nonlinear term $J(\psi_{10}^{(m)},\Delta \psi_{10}^{*(m)})+c.c.$ only drives a mean flow at an $O(\eps)$ smaller level. 
The solution is 
\beq
\frac{d \psi_{2,-1}^{(0)}}{dr}= -2 \frac{a(r^2-1)}{r(a^2-1)} \int^{\infty}_0 e^{-N_c x} f_0(x)\, dx \approx 2539\frac{a(r^2-1)}{r(a^2-1)}
\label{wnla_meanflow}
\eeq
so the mean flow ($\propto -d\psi_{2,-1}^{(0)}/dr$) is a decreasing function of the radius until the boundary layer is reached. There the mean flow undergoes an oscillation finally increasing sharply close to the boundary: see figure \ref{wnla}. \\

%
%
\noindent
For $\hat{\psi}^{(2m)}_{20}$,
\beq
\biggl[
  \frac{d^2}{dy^2}-N_c\frac{d}{dy}-2N_c(\hat{\sigma}_0+2i y)\biggr]
\frac{d^2 \hat{\psi}^{(2m)}_{20}}{dy^2} =\frac{d}{dy}\biggl[ -i \biggl(\frac{ d\hat{\psi}_{10}^{(m)}}{dy}\biggr)^2+i (\, \hat{\psi}_{10}^{m}(y)-\hat{\psi}_{10}^{m}(0)\, ) \frac{d^2 \hat{\psi}_{10}^{(m)}}{dy^2}\biggr]
\label{psi20_2}
\eeq
A particular integral for $d^2 \hat{\psi}_{20}^{(2m)}/dy^2$ can be generated by the variation of parameters method given that the complementary problem,
\beq
\biggl[
  \frac{d^2}{dy^2}-2 \alpha\frac{d}{dy}-(\beta \gamma^2 -\alpha^2+\gamma^3 y)\biggr]
\frac{d^2 \hat{\psi}^{(2m)}_{20}}{dy^2}=0 
\eeq
has solutions $ e^{\alpha y} Ai(\beta+\gamma y)$ and $e^{\alpha y} Bi(\beta+\gamma y) $
but the ensuing integral expressions become unwieldy when integrated twice to  get $\hat{\psi}_{20}^{(2m)}$ which is needed below. Instead, it is better to numerical solve (\ref{psi20_2}) directly. In fact (\ref{psi20_2}) can be integrated once to 
\begin{align}
\biggl[
  \frac{d^3}{dy^3}-N_c\frac{d^2}{dy^2}-2N_c( \hat{\sigma}_0 &+2 i y) \frac{d}{dy}
	+4iN
	\biggr]\hat{\psi}^{(2m)}_{20}\nonumber\\ 
	&=f_2(y):=-i \biggl(\frac{ d\hat{\psi}_{10}^{(m)}}{dy}\biggr)^2
	        +i (\, \hat{\psi}_{10}^{m}(y)-\hat{\psi}_{10}^{m}(0)\, ) \frac{d^2 \hat{\psi}_{10}^{(m)}}{dy^2}
\label{new_psi20_2}
\end{align}
where a vanishing solution is sought as $y \rightarrow \infty$.  The (numerical) solution strategy is to actually work over the domain $y \in[0,y_{max}]$ (with $y_{max}$ `large' but finite) which is transformed to $x \in [-1,1]$ by the definition $x:=2y/y_{max}-1$ and to expand $d \hat{\psi}_{20}^{(2m)}/dy$ as 
\beq
\frac{d \hat{\psi}_{20}^{(2m)}}{dy}= \sum_{n=0}^{N} a_n (\,T_{n+2}(x)-T_n(x)\,)
\label{expansion}
\eeq
rather than $\hat{\psi}_{20}^{(2m)}$ to improve the numerical conditioning. The expansion (\ref{expansion}) explicitly builds in the boundary conditions that $d \hat{\psi}_{20}^{(2m)}/dy$ vanishes at $y=0$ and $y=y_{max}$ and, on application of the further condition that $\hat{\psi}_{20}^{(2m)}(y_{max})=0$, means that the expansion can be integrated to give
\beq
\hat{\psi}_{20}^{(2m)}=\sum_{n=0}^N a_n \Theta_n(x) \quad {\rm where}\quad 
\Theta_n(x):=\left\{ \begin{array}{ll}
\frac{4}{3}-\frac{3T_1(x)}{2}+\frac{T_3(x)}{6}& n=0 \\
\frac{3}{8}-\frac{T_2(x)}{2}+\frac{T_4(x)}{8} & n=1 \\
\frac{1-T_{n+1}(x)}{n+1} +\frac{T_{n+3}(x)-1}{2(n+3)}+\frac{T_{n-1}(x)-1}{2(n-1)}  & n \geq 2
\end{array} \right.
\eeq
Using $N=100$ gives over 15 decades of drop off in $|a_n|$ and the solution is independent of $y_{max}$ well before the actual value $y_max=10$ chosen.

%
%
%
%

%
%
%
\subsection{Solvability condition at $O(\del^3/\eps^4)$} 

The problem for $\hat{\psi}_{30}^{(m)}$ is
\beq
\biggl[
  \frac{d^2}{dy^2}-N_c\frac{d}{dy}-N_c(\hat{\sigma}_0 +2 i y)\biggr]
\frac{d^2 \hat{\psi}_{30}^{(m)}}{dy^2} 
=-(J_0+J_2)
+  \hat{\eta}_2 \frac{d^3 \hat{\psi}_{10}^{(m)} }{dy^3}
+\hat{\sigma}_2 \frac{d^2 \hat{\psi}_{10}^{(m)} }{dy^2}
\label{psi_30}
\eeq
where the nonlinear terms are 
\begin{align}
J_0 &:= i[\,\hat{\psi}_{10}^{(m)}-\hat{\psi}_{10}^{(m)}(0)\,] \frac{d^3 \hat{\psi}_{20}^{(0)}}{dy^3}
-i \frac{d \hat{\psi}_{20}^{(0)}}{dy}\frac{d^2 \hat{\psi}_{10}^{(m)}}{dy^2} +c.c.\nonumber\\
J_2 &:= i \frac{d \hat{\psi}_{20}^{(2m)}}{dy}\frac{d^2 \hat{\psi}_{10}^{(m)*}}{dy^2}
      +2i[\,\hat{\psi}_{20}^{(2m)}-\hat{\psi}_{20}^{(2m)}(0)\,]\frac{d^3 \hat{\psi}_{10}^{(m)*}}{dy^3} \nonumber \\
			&\hspace{1.5cm} -2i\frac{d \hat{\psi}_{10}^{(m)*}}{dy} \frac{d^2 \hat{\psi}_{20}^{(2m)}}{dy^2}
			-i[\,\hat{\psi}_{10}^{(m)*}-\hat{\psi}_{10}^{(m)*}(0)\,] \frac{d^3 \hat{\psi}_{20}^{(2m)}}{dy^3}+c.c.
\end{align}
with boundary conditions that $d\hat{\psi}_{30}^{(m)}/dy(0)=0$ together with $\hat{\psi}_{30}^{(m)}$, $d\hat{\psi}_{30}^{(m)}/dy$, and $d^2\hat{\psi}_{30}^{(m)}/dy^2$ all vanishing as $y \rightarrow \infty$. The operator on the LHS of (\ref{psi_30}) is non-self adjoint so we need to find the appropriate adjoint operator to develop a solvability condition. Defining the operator  
\beq
{\cal L}_1:= \biggl[
  \frac{d^2}{dy^2}-N_c\frac{d}{dy}-N_c( \hat{\sigma}_0 +2 i y)\biggr] \frac{d}{
dy}
\eeq
then if $\Psi:=d \hat{\psi}_{30}^{(m)}/dy$ (and subscripts indicate 
derivatives)
\begin{align}
\int^\infty_0 \Phi {\cal L}_1 \Psi\, dy=
\biggl[
\Phi \Psi_{yy}-\Phi_y \Psi_y+\Phi_{yy}\Psi+N(\Phi_y \Psi-\Phi \Psi_y)&-N (
\hat{\sigma}_0 + 2 i y)\Phi \Psi \biggr]^{\infty}_0 \nonumber \\
& -\int^{\infty}_0 \Psi {\cal L}^{\dag}_1\,\Phi dy
\label{solvability}
\end{align}
where 
\beq
{\cal L}^{\dag}_1 \Phi:=-\frac{d^3 \Phi}{dy^3}-N_c\frac{d^2 \Phi}{dy^2}-N_c\frac{d}{dy} \biggl[(
\hat{\sigma}_0 + 2 i y)\Phi \biggr].
\eeq
We need now to generate a solution $\hat{\Phi}$ 
of ${\cal L}^{\dag}_1 \Phi=0$ such that all the boundary terms vanish.
The required solution (unique up to arbitrary renormalisation) is, via 
variation of parameters, 
\beq
\hat{\Phi}:= e^{-\half N_c y} Bi[\zeta(y)]\int^y_0 e^{\half N_c x} Ai [\zeta(x)] \, dx
            -e^{-\half N_c y} Ai[\zeta(y)]\int^y_0 e^{\half N_c x} Bi [\zeta(x)] \, dx	
\eeq
where
\beq
\zeta(y):=\frac{N_c^{1/3}\hat{\sigma}_0}{(2i)^{2/3}}+\frac{N_c^{4/3}}{4(2i)^{2/3}}
+(2i)^{1/3} N_c^{1/3} y.
\eeq
This has $\hat{\Phi}(0)=\hat{\Phi}_y(0)=0$ and $\hat{\Phi} \sim 1/y$ as $y\rightarrow \infty$ which ensures all the boundary terms 
in (\ref{solvability}) vanish (recall $\Psi(0)=d \psi_{30}^{(m)}/dy(0)=0$) and therefore means
\beq
\int^\infty_0 \hat{\Phi} {\cal L}_1 \frac{d 
\hat{\psi}_{30}^{(m)}}{dy}\, dy=0.
\eeq
The solvability condition on (\ref{psi_30}) 
is then
\beq
I_1 \,\hat{\sigma}_2+ I_2 \,\hat{\eta}_2= I_3+I_4
\eeq
where
\beq
I_1 := \int^\infty_0  \hat{\Phi} \frac{d^2 \hat{\psi}_{10}^{(m)}}{dy^2} \,dy, \quad
I_2 := \int^\infty_0  \hat{\Phi} \frac{d^3 \hat{\psi}_{10}^{(m)}}{dy^3} \,dy, \quad
I_3 := \int^\infty_0  \hat{\Phi} J_0 \, dy, \quad
I_4 := \int^\infty_0  \hat{\Phi} J_2 \, dy. \nonumber
\eeq
Since the frequency shift $\hat{\sigma}_2$ is an imaginary number and the radial flow adjustment is real, 
\beq
\hat{\eta}_2= \frac{ \Re e([I_3+I_4] I_1^*) }{ \Re e (I_2 I_1^*) } \, \approx \, 8.47 \times 10^{4} \, >\,  0
\eeq
as computations give $I_1=-24.17-75.26i$, $I_2=140.84+20.98i$, $I_3=(3.262+4.1265i)\times 10^6$
and $I_4=(2.949-0.5123i) \times 10^6$ (the contribution from the mean flow is an order of magnitude larger than that from the 2nd harmonic, both in the same sense). So the bifurcation is supercritical with the bifurcating solution branch existing at radial flows {\em larger} than the critical value.


\bibliographystyle{jfm}

\end{document}